\documentclass[]{aa}     
\pdfoutput=1
 \usepackage[usenames,dvipsnames]{color}          
\usepackage{graphicx}       
\usepackage{txfonts}        
\usepackage{lscape}

\usepackage{natbib}
\bibpunct{(}{)}{;}{a}{}{,} 
\begin{document}        
        
\def\ang{{\rm\AA}}        
        
\def\n{\noindent}        
        
\def\ph{\phantom}         
\def\blankline{\par\vskip 12pt\noindent}

\newcommand{\olevel}[4]{#1\,#2\,$^{#3}$#4$^{\scriptsize \rm  o}$}
\newcommand{\elevel}[4]{#1\,#2\,$^{#3}$#4}
 
\newcommand{\zetap}{$\zeta$~Puppis}
\newcommand{\lambcep}{$\lambda$~Cep}
\newcommand{\hdfourteen}{HD~14947}
\newcommand{\hdfifteen}{HD~15570}
\newcommand{\hdsixteen}{HD~16691}
\newcommand{\hdsixtysix}{HD~66811}
\newcommand{\hdonesixtythree}{HD~163758}
\newcommand{\hdoneninetwo}{HD~192639}
\newcommand{\hdtwohundten}{HD~210839}
\newcommand{\hdo}{HD~190429A}
\newcommand{\fuse}{\emph{FUSE}}
\newcommand{\iue}{\emph{IUE}}
\newcommand{\hst}{\emph{HST}}
\newcommand{\xmm}{\emph{XMM}}
\newcommand{\chandra}{\emph{Chandra}}
\newcommand{\copernicus}{\emph{Copernicus}}
\newcommand{\hipparcos}{\emph{Hipparcos}}
\newcommand{\rosat}{\emph{ROSAT}}
\newcommand{\ohp}{\emph{OHP}}
\newcommand{\eso}{\emph{ESO}}
\newcommand{\uves}{\emph{UVES}}
\newcommand{\elodie}{\emph{ELODIE}}
\newcommand{\feros}{\emph{FEROS}}
\newcommand{\narval}{\emph{NARVAL}}
\newcommand{\ergs}{ergs\,cm$^{-2}$\,s$^{-1}$\,\AA$^{-1}$}
\newcommand{\halpha}{\mbox{H$\alpha$}}
\newcommand{\hbeta}{\mbox{H$\beta$}}
\newcommand{\hgamma}{\mbox{H$\gamma$}}
\newcommand{\hdelta}{\mbox{H$\delta$}}
\newcommand{\bralpha}{\mbox{Br$\alpha$}}
\newcommand{\brgamma}{\mbox{Br$\gamma$}}
\newcommand{\lyalpha}{\mbox{Ly$\alpha$}}
\newcommand{\lybeta}{\mbox{Ly$\beta$}}
\newcommand{\lygamma}{\mbox{Ly$\gamma$}}
\newcommand{\lydelta}{\mbox{Ly$\delta$}}
\newcommand{\htwo}{\mbox{H$_{2}$}}
\newcommand{\lb}{$\lambda$}
\newcommand{\mdot}{$\dot{M}$}
\newcommand{\gcmcube}{g\,cm$^{-3}$}
\newcommand{\gcmcarre}{g\,cm$^{-2}$}
\newcommand{\vsini}{$v\sin\,i$}
\newcommand{\vrad}{$v_{\rm rad}$}
\newcommand{\vmc}{$v_{\rm MC}$}
\newcommand{\flux}{ergs\,cm$^{-2}$\,s$^{-1}$}
\newcommand{\teff}{$T_{\rm eff}$}
\newcommand{\logg}{$\log g$}
\newcommand{\loggc}{$\log g_{\rm c}$}
\newcommand{\met}{[Fe/H]}
\newcommand{\vturb}{$v_{\rm turb}$}
\newcommand{\vmax}{$v_{\rm max}$}
\newcommand{\vmac}{$v_{\rm mac}$}
\newcommand{\vmin}{$v_{\rm min}$}
\newcommand{\vcl}{$v_{\rm cl}$}
\newcommand{\vinf}{$\rm v_{\infty}$}
\newcommand{\finf}{$f_{\infty}$}
\newcommand{\vt}{$v_{turb}$}
\newcommand{\radius}{$R_{*}/R_{\odot}$}
\newcommand{\rstar}{$R_{*}$}
\newcommand{\mstar}{$M_{*}$}
\newcommand{\zsol}{$Z_{\odot}$}
\newcommand{\zzsol}{$Z/Z_{\odot}$}
\newcommand{\rsol}{$R_{\odot}$}
\newcommand{\msol}{$M_{\odot}$}
\newcommand{\tlusty}{{\sc Tlusty}}
\newcommand{\synspec}{{\sc Synspec}}
\newcommand{\cmfgen}{{\sc CMFGEN}}
\newcommand{\fastwind}{{\sc FASTWIND}}
\newcommand{\wmbasic}{{\sc WM-BASIC}}
\newcommand{\Av}{$A_{v}$}
\newcommand{\ebv}{$E(B-V)$}
\newcommand{\Rv}{$R_{v}$}
\newcommand{\Mv}{$M_{V}$}
\newcommand{\msp}{$M_{\rm spec}$}
\newcommand{\mevone}{$M_{\rm evol}^{t}$}
\newcommand{\mevtwo}{$M_{\rm evol}^{init}$}
\newcommand{\mic}{$\xi_{\rm t}$}
\newcommand{\vdag}{(v)^\dagger}
\newcommand{\grid}{{\sc Ostar2002}}
\newcommand{\kms}{km\,s$^{-1}$}
\newcommand{\cun}{cm$^{-1}$\ }
\newcommand{\cdeux}{cm$^{-2}$\ }
\newcommand{\ctrois}{cm$^{-3}$\ }
\newcommand{\ci}{\mbox{C~{\sc i}}}
\newcommand{\cii}{\mbox{C~{\sc ii}}}
\newcommand{\ciii}{\mbox{C~{\sc iii}}}
\newcommand{\civ}{\mbox{C~{\sc iv}}}
\newcommand{\cv}{\mbox{C~{\sc v}}}
\newcommand{\hi}{\mbox{H~{\sc i}}}
\newcommand{\hyd}{\mbox{H}}
\newcommand{\he}{\mbox{He}}
\newcommand{\hei}{\mbox{He~{\sc i}}}
\newcommand{\heii}{\mbox{He~{\sc ii}}}
\newcommand{\oii}{\mbox{O~{\sc ii}}}
\newcommand{\oiii}{\mbox{O~{\sc iii}}}
\newcommand{\oiv}{\mbox{O~{\sc iv}}}
\newcommand{\ov}{\mbox{O~{\sc v}}}
\newcommand{\ovi}{\mbox{O~{\sc vi}}}
\newcommand{\ovii}{\mbox{O~{\sc vii}}}
\newcommand{\nii}{\mbox{N~{\sc ii}}}
\newcommand{\niii}{\mbox{N~{\sc iii}}}
\newcommand{\niv}{\mbox{N~{\sc iv}}}
\newcommand{\nv}{\mbox{N~{\sc v}}}
\newcommand{\fe}{\mbox{Fe~}}
\newcommand{\neon}{\mbox{Ne~}}
\newcommand{\p}{\mbox{P~}}
\newcommand{\piv}{\mbox{P~{\sc iv}}}
\newcommand{\pv}{\mbox{P~{\sc v}}}
\newcommand{\pvi}{\mbox{P~{\sc vi}}}
\newcommand{\siii}{\mbox{Si~{\sc ii}}}
\newcommand{\siiv}{\mbox{Si~{\sc iv}}}
\newcommand{\siv}{\mbox{S~{\sc iv}}}
\newcommand{\sv}{\mbox{S~{\sc v}}}
\newcommand{\svi}{\mbox{S~{\sc vi}}}
\newcommand{\feiv}{\mbox{Fe~{\sc iv}}}
\newcommand{\fev}{\mbox{Fe~{\sc v}}}
\newcommand{\fevi}{\mbox{Fe~{\sc vi}}}
\newcommand{\msolyr}{$M_{\odot}$\,yr$^{-1}$}
\newcommand{\pcyg}{P~Cygni}
\newcommand{\vesc}{$v_{esc}$}
\newcommand{\xit}{$\xi_{t}$}
\newcommand{\xitmin}{$\xi_{t}^{min}$}
\newcommand{\xitmax}{$\xi_{t}^{max}$}
\newcommand{\us}{$\times\,10^{-7}$}
\newcommand{\un}{$\times\,10^{-9}$}
\newcommand{\epv}{$\times\,10^{5}$}
\newcommand{\ev}{$\times\,10^{-5}$}
\newcommand{\evi}{$\times\,10^{-6}$}
\newcommand{\evii}{$\times\,10^{-7}$}
\newcommand{\eviii}{$\times\,10^{-8}$}
\newcommand{\eix}{$\times\,10^{-9}$}
\newcommand{\logL}{$\log L/L_{\odot}$}
\newcommand{\logLf}{$\log \frac{L}{L_{\odot}}$}
\newcommand{\lognHI}{$\log N$(H~{\sc i})}
\newcommand{\lognHII}{$\log N({\rm H}_{2})$}
\newcommand{\logLX}{$\log L_{\rm X}/L_{\rm bol}$}

\title{Properties of Galactic early-type O-supergiants }
\subtitle{A combined FUV-UV and optical analysis 
\thanks{Based on observations made with the NASA-CNES-CSA {\sl Far Ultraviolet  
           Spectroscopic Explorer\/} (\fuse) and by the NASA-ESA-SERC {\sl International  
           Ultraviolet Explorer\/} (\iue), and retrieved from the Multimission Archive at  
           the Space Telescope Science Institute (MAST). Based on observations collected with the   
           ELODIE spectrograph on the 1.93-m telescope (Observatoire de Haute-Provence, France).  
           Based on observations collected with the FEROS instrument on the  
           ESO 2.2 m telescope, program 074.D-0300 and
           075.D-0061. } }
       
\author{J.-C. Bouret\inst{1},  D. J. Hillier\inst{2}, T. Lanz\inst{3}, A. W. Fullerton\inst{4}}         
 
 \offprints{J.-C. Bouret}

 \institute{Laboratoire d'Astrophysique de Marseille, Universit\'e d'Aix-Marseille \& CNRS, UMR7326, 38 rue F. Joliot-Curie, F-13388 Marseille Cedex 13, France, \email{Jean-Claude.Bouret@oamp.fr}
         \and
 Department of Physics and Astronomy, University of Pittsburgh,
    Pittsburgh, PA 15260, USA   
             \and
         Laboratoire J.-L. Lagrange, UMR 7293, Universit\'e de Nice-Sophia Antipolis, CNRS, Observatoire de
 la C\^ote d'Azur, B.P. 4229, F-06304 Nice Cedex 4, France
            \and
       Space Telescope Science Institute, 3700 San Martin Drive, Baltimore, MD 21218, USA     
    }

\date{Accepted May 3rd 2012}        
   
\abstract{}
{We aim to constrain the properties and evolutionary status of  early and mid-spectral type supergiants (from O4 to O7.5). These posses the highest mass-loss rates among the O stars, and exhibit conspicuous wind profiles.}
 {Using the non-LTE wind code \cmfgen\ we  simultaneously analyzed the FUV-UV and optical spectral range to determine the photospheric properties and wind parameters. We derived effective temperatures, luminosities, surface gravities, surface abundances, mass-loss rates, wind terminal velocities, and clumping filling factors.}
 {The supergiants define a very clear evolutionary sequence, in terms of ages and masses, from younger and more massive stars to older stars with lower initial masses. O4 supergiants cluster around the 3 Myr isochrone and are more massive than 60 \msol, while the O5 to O7.5 stars have masses  in the range 50 - 40 \msol\  and are 4$\pm 0.3$ Myr old. 
The surface chemical composition is typical of evolved O supergiants {\bf(}nitrogen-rich, carbon- and oxygen-poor). 
While the observed ranges of carbon and nitrogen mass-fractions are compatible with those expected from evolutionary models for the measured stellar masses, the N/C ratios as a function of age are inconsistent with
the theoretical predictions for the four earliest (O4 spectral type) stars of the sample. We question the efficiency of rotational mixing as a function of age for these stars and suggest that another mechanism may be needed to explain the observed abundance patterns. 
Mass-loss rates derived with clumped-models range within a factor of three of the theoretical mass-loss rates. 
The corresponding volume-filling factors associated with small-scale clumping are 0.05$\pm 0.02$. Clumping is found to start close to the photosphere for all but three stars, two of which are fast rotators.}
{}             


\authorrunning{Bouret et al.}        
        
\maketitle        
     
        
\section{Introduction}        
\label{intro_sect} 

Many fields of modern astrophysics are directly or indirectly related to the physics of massive stars. Locally, massive stars
are responsible for the ionization and expansion of their surrounding HII regions, and are believed to be responsible for triggering 
star formation \cite[e.g.][]{zavagno07,martins10}. More generally, they dominate the chemical and dynamical evolution of the global interstellar medium of their host galaxies. Starburst events, either in the local or the distant universe, are dominated by hot massive stars,
as revealed by the composite spectrum of distant, star-forming galaxies, which exhibit numerous features typical of O and B stars 
\cite[e.g.][]{steidel96}. Fast rotating massive stars are likely progenitors of some of the most energetic events in the Universe, such as core-collapse supernovae and gamma-ray-bursts (GRBs) \citep[e.g.][]{hjorth03, woosley06} and population III stars are believed to be very massive objects with extreme effective temperatures \citep{bromm99,nakamura99}. 

In the 1970s it was recognized that stellar evolution in the upper Hertzsprung-Russell (H-R) diagram is, to a large extent, governed by mass loss. Through the course of  their evolution, massive stars loose a significant fraction of their mass through their stellar winds \citep[see e.g., the review by][]{meynet07}, and through giant eruptions \citep[e.g.][]{smith06}. 
Although the line-driven wind theory has been successful in explaining the global behavior of stellar winds, the actual wind parameters remain subject to uncertainties, especially regarding the wind terminal velocity, mass-loss rate and the degree of clumpiness \citep[see e.g., the workshop held in 2007 in Potsdam,][]{hamann08}. 
The most striking uncertainty regarding the mass-loss rates is related to the intrinsically unstable nature of the line-driving mechanism \citep{lucy70, owocki88}, which leads to strong reverse shocks \citep[e.g.][]{owocki99} that separate fast low-density wind material from overdense clumps.  
Since many prominent mass-loss diagnostics are sensitive to the square of the density, they
overestimate the mass-loss rate in the presence of clumping. The uncertainty in the true value of the mass-loss rate translates in a corresponding uncertainty in the evolutionary tracks of massive stars.
  
Recent studies showed that winds are clumped, although there is no consensus concerning the amount, nature (optically thin versus optically thick) and stratification of clumping.  Initial studies  \citep{crowther02, hillier03, bouret03, bouret05, fullerton06} found a reduction in mass-loss rates by a factor of three to ten when using the optically thin assumption for clumping.  
Within the same framework, the radial distribution of the clumping factor was constrained by \cite{puls06} and \cite{najarro11}, based on a simultaneous modeling of \halpha\, {infrared, millimeter} and radio observations. These studies concluded that clumping is three to six times stronger in the lower wind, where \halpha\ forms, compared with the outer wind, where the radio continuum originates. These findings marginally agree with the theoretical predictions by \cite{runacres02, runacres05}. The aforementioned studies assumed a void-interclump medium, although \cite{zsargo08} demonstrated that the treatment of a non-void interclump medium is required to explain the global spectroscopic properties of massive stars.

 The optically thin approximation was subsequently challenged by \cite{oskinova07}, who claimed that significantly lower reduction (factor of three only) is found when using optically thick clumping. 
 When it is pronounced, this ÒmacroclumpingÓ (or ÒporosityÓ) affects the emerging spectrum for a given mass-loss by reducing the effective opacity of the wind, because optically thick clumps hide material. 
Another factor that might affect the mass-loss rate is the structure of winds in velocity space (hereafter called velocity porosity). This arises because the dense absorbing clumps along a given line of sight, only occur at discrete velocities, and  leads to a reduction in the strength of P~Cygni absorption components \citep{owocki08}. This velocity porosity has been implemented in 2D/3D stochastic wind models by \cite{sundqvist10a, sundqvist11}, who concluded that it could increase mass-loss rates derived from UV resonance lines by up to an order of magnitude.
  
Major progress has also been achieved in the last few years in stellar rotation theory, but more progress is needed. 
It has been known for a long time that the physical properties of stars are distorted by rotation, as predicted from the von Zeippel
theorem and indicated from recent interferometric studies \citep[e.g.][]{domiciano03}. Consequently, rotation is also expected to  cause a latitudinal dependence of the line force and hence mass loss  \citep{maeder00}.

From the view point of stellar evolution, rotation causes an increase in the mass-loss rate, a change in evolutionary tracks in the H-R diagram, a lowering of the effective gravity, an extension of the main-sequence phase, the transport of angular momentum and the mixing of CNO-cycle processed material up to the stellar surface \citep{maeder00, heger00}. 
The so-called helium discrepancy first identified by \cite{herrero92} in many galactic O-type stars and especially in fast rotators is naturally explained by evolution models with rotation --- higher helium (and nitrogen) abundances are predicted for higher masses and higher rotation velocities. Another consequence of rotation is that two stars at the same evolutionary phase and at the same location in the H-R diagram can have different initial masses.

Several other observational facts have been successfully accounted for using evolutionary models with rotation \citep{maeder00}. Models with rotation predict the correct blue-to-red supergiant ratio as observed in the Small Magellanic Cloud \citep{maeder01}, explain the observed population of Wolf-Rayet stars as a function of metallicity \citep{meynet03, meynet05}, and provide a consistent framework for the evolution of progenitors of soft-long GRBs \citep{yoon05, woosley06, meynet07}.

Despite these success, some observations remain unexplained. \cite{hunter07, hunter08, hunter09}, who investigated the link between surface abundance patterns and rotation in the Galaxy and in the Magellanic Clouds, concluded that up to 40\% of their sample presented nitrogen surface abundances that could not be explained by rotational mixing. These studies strongly suggest
that the global chemical evolution of massive stars, as probed from their surface abundances,
is an intricate combination of several physical parameters, including mass, metallicity, age (evolutionary status) and rotation \citep{maeder09}. 

Some of the spectral lines used to determine mass-loss rates and clumping factors originate from CNO ions. Because surface CNO abundance patterns are modified by rotational mixing during stellar evolution, it is very important to have reliable estimates of these patterns in order to fully
characterize the wind properties of O-type stars.   

For this study, we have analyzed a sample consisting of eight Galactic supergiants with spectral types ranging from O4 If to O7.5 If \citep{sota11}. 
We derive their properties, and address questions about as their evolutionary status and how their surface abundances relate to this. 
We also investigate whether a consistent picture for mass loss and clumping may be drawn from different diagnostics such as \pv\ \lb\lb 1118-1128, \ov\ \lb1371, \niv\ \lb1718, \heii\ \lb4686 
and \halpha. By using different diagnostics we can test the influence of the ionization structure, the adopted abundances, and to a limited extent, the clumping formulation on wind diagnostics.

In the following section, we present our sample selection and observational material. The modeling tools are presented
in Sect.\, \ref{modass_sect}, while diagnostics of the different stellar parameters are presented in Sect. \ref{proc_sect}. 
Results of the analysis for the individual objects are presented in Sect. \ref{result_sect} and a general discussion of the evolutionary status,
chemical properties and wind properties is given in Sect. \ref{disc_sect}, before our general conclusion in Sect \ref{conclu_sect}.


\section{Target selection and observations}
\label{obs_sect}
\subsection{Stellar sample}
\label{var_sect}

Table~\ref{tab_spec_types} presents the sample selected for this study and fundamental observational data.
The selected objects are considered to be representative of early-Of supergiants, whose surface properties are likely to show processed material from stellar evolution and whose strong winds are expected to exhibit conspicuous signatures of clumping. The spectral classification scheme of these stars was designed by \cite{walborn71a, walborn71b, walborn73} and was recently updated in \cite{sota11}. 
The scheme is meant to sort stars with different temperatures, luminosities, and abundances. CNO-cycled material must play a role in explaining the observed dispersion for a given  spectral type/luminosity class. 

 \begin{table*}[]
\centering \caption{Basic parameters of our targets.}
\begin{tabular}{lllccccccccc}
\hline
 Star			&	Other name	& Spetral Type		& $U$	&	$B$		&	$V$		&	$J$		&	$H$		&	$K$		& $E(B-V)$	& dist. (kpc)	& $M_{V}$   \\ \\
 \hline
\object{HD 16691}		& ...				& O4 If			& 8.620	&	9.182	&	8.702	&	7.653	&	7.444 	&	7.362	& 0.800	& 3.31$\pm 0.20$ 		& -6.38$^{+0.12}_{-0.14}$  \\	
\object{HD 66811}		& $\zeta$ Pup		& O4 I(n)fp		& 0.890	&	1.941	&	2.210	&	2.790	&	2.955	&	2.968	& 0.040	& 0.46$\pm 0.04$ 		& -6-23$^{+0.17}_{-0.20}$    \\  
 \object{HD 190429A}	& ...				& O4 If		 	&	....	&	7.201	&	7.088	&	6.189 	&	6.162	&	6.150	& 0.460	& 2.45$\pm 0.20$		& -6.28$^{+0.21}_{-0.24}$ \\	
\object{HD 15570}		& ...				& O4 If			& 8.391	&	8.796	&	8.110	& 	6.477	&	6.310	&	6.158	& 0.966	& 2.34$\pm 0.11$ 		& -6.73$^{+0.11}_{-0.12}$ \\
\object{HD 14947}		& ...				& O4.5 If			& 7.850	&	8.452	&	7.998	&	7.037	&	6.945	&	6.861	& 0.730	& 3.00$\pm 0.20$ 		& -6.43$^{+0.14}_{-0.15}$  \\
\object{HD 210839}	& $\lambda$ Cep	& O6 I(n)fp		& 4.620	&	5.242	&	5.050	&	5.053	&	4.618	&	4.500	& 0.513	& 0.95$\pm 0.10$ 		& -6.43$^{+0.11}_{-0.12}$  \\
\object{HD 163758}	& ...				& O6.5 If			& 6.458	&	7.346	&	7.318	&	7.194	&	7.163	&	7.157	& 0.300	& 3.60$\pm 0.20$ 		& -6.39$^{+0.12}_{-0.12}$ \\
\object{HD 192639}	& ...				& O7.5 Iabf		& 6.830	&	7.455	&	7.116	&	6.300	&	6.271	&	6.217	& 0.620	& 2.00$\pm 0.20$ 		& -6.31$^{+0.21}_{-0.23}$  \\
  \hline
\end{tabular}
 \label{tab_spec_types}
    \begin{list}{}{}
\item[Note :]    Spectral types are taken from \cite{sota11}, while the photometry is taken from the GOS catalog \citep{maiz04}. Reddening, absolute magnitude and distances are 
calculated using the procedure presented in Sect. 4.1. 
\end{list}
   \end{table*}
      
We checked that binarity is not a problem for our  spectroscopic analyses. \cite{debecker09} concluded that \object{\hdfifteen},  \object{\hdsixteen}, and  \object{\hdfourteen} are very likely single because the authors did not find signs of radial velocity variations that they could unambiguously relate to companions.   \object{\hdsixtysix} has been extensively observed but no companions have been detected for this star, even in 
high-sensitivity speckle surveys for binarity \cite[see e.g.][]{mason98}. 
The adaptive optics survey for faint companions led by \cite{turner08} also concluded that  \object{\hdtwohundten},  \object{\hdoneninetwo} and  \object{\hdonesixtythree} are single stars.
 \object{HD 190429} is a well-known binary but the spectrum (and the photometry) we used was for  \object{\hdo} (i.e., for the primary only).
Because the companion,  \object{HD 190429B}, has a spectral type of O9.5II, the luminosity difference is about 0.7 dex \citep{walborn00}, 
implying that its contribution to the total flux remains small.

The log of the spectroscopic observations is presented in Table~\ref{tab_spectra}.

\begin{table*}[]
\centering \caption{Star sample and observations logs.}
\begin{tabular}{lllclclc}
\hline
\hline
        &                          & \multicolumn{2}{c}{FUV} & \multicolumn{2}{c}{UV} & \multicolumn{2}{c}{Optical}  \\
Star			& Spectral Type				& Data Set$^{\mathrm{a}}$ & Date & Data Set & Date & Instrument & Date  \\
\hline
 \object{HD 16691}	& O4 If		& F - E8050101	& 2004-11-18 & SWP46545 				& 1992-12-21	& OHP/ELODIE		& 2005-11-05 \\
  \object{HD 66811}	& O4 I(n)fp	& C - C044-001	& 1973-02-22 & SWP15296 				& 1981-10-20	& ESO/FEROS 		& 2005-12-20 \\
   \object{HD 190429A}	& O4 If		& F - P1028401	& 2000-07-18 & SWP $^{\mathrm{b}}$ .....		& \dots		& OHP/ELODIE		& 2004-08-27 \\
  \object{HD 15570}	& O4 If		& F - E0820101	& 2005-11-08 & SWP11234 				&1981-02-04	& OHP/ELODIE		& 2004-11-17     \\
  \object{HD 14947}	& O4.5 If		& F - E0820201	& 2004-30-09 & SWP10724				& 1980-12-01	& OHP/ELODIE		& 2004-11-20 \\
  \object{HD 210839}	& O6 I(n)fp	& F- P1163101		& 2000-07-22 & SWP52623				& 1994-10-23	& TBL/NARVAL	& 2006-12-13 \\
  \object{HD 163758}	& O6.5 If    	& F - P1015901	& 2000-08-16 & SWP02892				& 1978-10-09	& ESO/UVES		& 2001-07-27 \\
  \object{object{object{HD 192639}}}	& O7.5 Iabf     	& F-C17101010	& 2002-09-04 & SWP09493				& 1980-07-11	& OHP/ELODIE		& 2001-08-11 \\
  \hline
\end{tabular}
\label{tab_spectra}
  \begin{list}{}{}
\item[$^{\mathrm{a}}$] Observatory used to obtain spectra: F \fuse ; C \copernicus
\item[$^{\mathrm{b}}$] Co-addition of several spectra; see Bouret et al. (2005)
\end{list}
\end{table*}

\subsection{FUV and UV data}
\label{fuv_sect}
We extracted short wavelength, high-resolution echelle (SWP) \iue\ spectra from MAST. The SWP spectra cover the spectral range, \lb\lb1150-2000\,\AA, at a resolving power $R = 10,000$. 
Only one SWP spectrum is available for  \object{\hdfourteen},  \object{\hdfifteen},  \object{\hdsixteen},  \object{\hdonesixtythree} and  \object{\hdoneninetwo}, respectively. 
For  \object{\hdo}, we used the co-added spectrum presented in \cite{bouret05}. 
For  \object{\lambcep} we selected  the only \iue\ SWP high-resolution spectrum that is not saturated \citep[see][]{bianchi02}.

As part of the \iue\ ``MEGA" campaign \citep{massa95, howarth95},  \object{\zetap} was observed over nineteen days, corresponding to slightly more than three rotation periods (5.2 days). 
We chose to stick to the MEGA campaign to avoid any problems with ``long-term" variations, which have not been studied extensively.
For the 107 spectra from this campaign, we constructed three mean spectra - one for each ``rotational" cycle. 
Because these three spectra were very similar, we averaged all of them to form the spectrum that is used in our analysis. 

We furthermore smoothed the spectra to a resolution of 40~\kms\  to
increase the signal-to-noise ratio. The spectra of all stars show many
narrow lines of interstellar origin. This interstellar contamination dominates the whole UV spectrum of
 \object{\hdfifteen}; only strong lines such as \civ\ \lb\lb1548-1550 or \niv\ \lb1718 can be used safely for this star.

All stars but  \object{\hdsixtysix} were observed with \fuse\ through the LWRS $30''\times30''$
aperture. The nominal spectral resolution is 20,000, or about 20~\kms, and the
wavelength range extends from 905 \AA\ to 1187 \AA. The processed \fuse\ spectra were 
retrieved from MAST and reprocessed with the final version  of the \fuse\ pipeline
CalFUSE 3.2.3. Individual sub-exposures were co-added for 
each segment and then merged to form a single spectrum, using Lindler's FUSE-REGISTER 
program. We avoided contamination by the ``worm'' artifact \citep{sahnow00}
by using only the LiF2A spectra on the long-wavelength side (\lb\lb1086-1183\,\AA)
of the spectrum. Finally, the co-added merged spectra were smoothed
to a resolution of 30~\kms\  to enhance the signal-to-noise ratio.\\
For  \object{\hdsixtysix}, we used the \copernicus\ spectra already discussed in \cite{pauldrach94}  
that were originally obtained by \cite{morton77}, to which we refer the reader for details concerning the observations and data reduction.

\subsection{Optical data}
For all but one of the northern stars, we used data obtained in November 2004 with the \elodie\ spectrograph, which was mounted on the 193cm 
telescope at Observatoire de Haute-Provence (\ohp).  The resolving power is 42,000 and the  wavelength coverage extends from 3895\,\AA\ to 6815\,\AA. 
The exposure times were chosen to ensure a signal-to-noise ratio of at least 100 at 5200 \AA. 
The reduction of spectroscopic data acquired with \elodie\ was performed using the standard reduction pipeline described in \cite{baranne96}.
Each order was normalized by a fit to the continuum, which was specified by 
fitting a smooth spline to carefully selected continuum windows. This step was then followed by a complete merging of successive orders to reconstruct the full spectrum for each star.

The spectrum of  \object{\lambcep} was collected with the NARVAL spectropolarimeter at Telescope Bernard Lyot (TBL) at Pic du Midi Observatory in December 2006.
The resolving power is R=65,000 for a wavelength coverage 3700-10050 \AA.  The set of four sub-exposures was processed using Libre~ESpRIT \citep{donati97}, a fully automatic reduction 
package installed at TBL for optimal extraction of NARVAL spectra.  The signal-to-noise ratios per 2.6~\kms\ velocity bin range from 400 to 1100 over the 3800-7000 \AA\ range that we used here.  

The spectrum of  \object{\hdonesixtythree} was extracted from the UVES Paranal Observatory Project, a program of acquisition, reduction, and public release of stellar spectra obtained with UVES at the VLT \citep{bagnulo03}.
Complete details concerning these data can be found at the $\it UVES\ POP$ URL \footnote{{\tt http://www.sc.eso.org/santiago/uvespop/}}. 
The spectrum was normalized by fitting a spline to the continuum, which was defined by the same windows used to normalize the \elodie\ spectra. 

The other southern star of this sample, namely  \object{\zetap}, was observed with the \feros\ spectrograph on the MPI 2.2m telescope at La Silla Observatory (prog. ID 074.D-0300(A)). The spectrum coverage is from 3800 to 8800 \AA, with a resolving power of 48,000. 
The spectrum was wavelength-calibrated and extracted using a modified FEROS pipeline 
that includes the modifications already described in Sana et al. (2006) and Sana (2009).

Some broad spectral lines like the \heii\ \lb4686 line in the \feros\ spectrum of  \object{\zetap} or the \halpha\ line in either \elodie\ or \feros\ spectra extend over two adjacent orders. 
In these cases, we found that adopting order shapes from adjacent orders gave the best continuum reduction. 

\subsection{Some considerations about spectroscopic variability}
\label{variability_sect}
The observations used for this study have been obtained at very different 
epochs for each star (see Table \ref{tab_spectra}). 
Therefore, we had to address the problem of the expected variability of the stellar winds, and particular how this would affect the physical quantities that we measured through spectral fitting.

Resonance line variations are well documented for several O-type stars in the FUV-UV range, including 
two of our targets, namely  \object{\zetap} and  \object{\lambcep}. For the other stars  only one FUV-UV spectrum has been obtained, and hence no information is available on the UV variability of their wind or photosphere. 
The most prominent signature of variable wind structures in hot stars is the presence of blueward-migrating ``discrete absorption components" (DACs), which are clearly visible in time-series IUE spectra \citep[e.g.][]{howarth95, massa95}. These data have established that the variations in the wind lines are not chaotic fluctuations, but are instead very systematic and probably associated with large-scale perturbations that arise from corotating interacting regions \citep[][]{cranmer96}. 

In the optical range short-timescale (of about hours)  stochastic spectral line variations of \halpha\ and \heii\ \lb4686  are well documented for O-type stars \cite[e.g.][]{prinja94, eversberg98, markova05, debecker09}. This variability is commonly attributed to the small-scale structures, or clumps,  which are predicted by time-dependent models of the line-driven instability \citep{owocki94, dessart03, dessart05}, though pulsations might also be involved at some level \cite[e.g.][]{markova02, prinja06}. The observed line profile variability indicates that the winds are not smooth, but perturbed. 
Observations furthermore showed that significant variations in \halpha\ start from the wind base up to velocities of a few hundreds of \kms, which corresponds to the maximum extension 
of the \halpha\ line formation region in dense enough winds such as those of O supergiants. Evidently, these clumps can coexist with the large-scale perturbations (DACs) observed in the UV domain.

For our modeling, we assumed that the clumping properties of  the wind are determined by small-scale structures that originate from the intrinsic instability of the line-driving mechanism.
Hence, the derived clumping properties should be (almost) independent of time, as long as the major wind characteristics remain largely constant. This is supported by observations that on longer time scales, the wind properties of most O stars appear to be constant. Additional support comes from a study showing that for early-type supergiants, variations observed in \halpha\ indicate mass-loss rate changes of about $\pm$ 4\% with respect to the mean value of \mdot\ \citep{markova05}.  
These changes are smaller than the uncertainties of our \mdot\ estimates, most of which are attributable to
uncertainties in the stellar luminosities.

\section{Model atmospheres}
\label{modass_sect} 
We performed the analyses using model atmospheres calculated with the code \cmfgen\  \citep{hillier98}. This code computes line-blanketed, non-LTE (NLTE) models by solving the radiative transfer and statistical equilibrium equations in the comoving frame of the fluid, for a spherically symmetric outflow. To help facilitate the inclusion of line blanketing, super-levels were used.  
This approach allows the inclusion of many energy levels from ions of many different species. After convergence of the model, a formal solution of the radiative transfer equation is computed  in the observer's frame \citep{busche05}, thus providing the synthetic spectrum for comparison to observations. 

For the photospheric density structure we initially used the hydrostatic density structure for a model computed with \tlusty\ \citep{hubeny95, lanz03}, while for the wind we used a standard $\beta$-velocity law that is connected to the hydrostatic density structure just above the sonic point  (at approximately 15 \kms). 
The mass-loss rate, density, and velocity are related via the continuity equation.
In more advanced models we iterated on the density structure so that the hydrostatic equation,
\begin{equation}
{dP/dr} = - g_{\rm grav} + g_{\rm rad}
\end{equation}
where
\begin{equation}
g_{\rm grav} = GM/r^{2}
\end{equation}
and the radiative acceleration $g_{\rm rad}$
\begin{equation}
g_{\rm rad} = \frac{4\pi}{c}\rho\int\chi_{\nu}H_{\nu}d\nu
\end{equation}

was satisfied. We solved for the density structure by integrating the hydrostatic equation using the Runge-Kutta method. To facilitate the integration, 
the Rosseland LTE opacity was computed as a function of temperature and electron number density for the model abundances prior to the \cmfgen\ run. 
In a typical \cmfgen\ model, a hydrostatic iteration is performed after the first model iteration and then every $n$ (with $n\sim 8$) iterations. Generally 3 to 4 hydrostatic iterations are needed to obtain convergence to better than 5\% everywhere. 
 
A depth-independent microturbulent velocity is included in the computation of the atmospheric structure (i.e., temperature and level populations). A value of 15 \kms\  is adopted as the default in our computations. 

We have computed clumped wind models with CMFGEN. Clumping is implemented through a parameteric, volume filling-factor approach, which assumes that the clumps are optically thin to radiation and the inter-clump medium is void. Under these assumptions, the clumped wind density, $\rho(r)$ is related to the  homogeneous (unclumped) wind density and the volume filling by  $\rho(r)=\bar{\rho}/f$. The filling factor decreases exponentially with increasing radius (or, equivalently, with increasing velocity):
$f = f_\infty + (1-f_\infty) \exp(-v/v_{\rm cl})$
where $v_{\rm cl}$ is the velocity at which clumping starts. We tuned $v_{\rm cl}$ to improve the fit to some observed lines \citep{bouret05}. 
A  limitation of this parameterization is that {\it f} is monotonic whereas theoretical simulations by \cite{runacres02} and observational studies \citep{puls06, najarro11} suggest it is non-monotonic. 


We have accounted for shock-generated X-ray emission in our models\footnote{In these stars the clumps are probably primarily responsible for producing the observed \nv\ resonance doublet, but in cooler stars the interclump medium is also likely to be very important \citep{zsargo08}}. An important consequence of the X-ray and EUV shock radiation is enhanced photoionization which results in ``wind super-ionization'' --- that is the presence of highly ionized ions (such as \ovi) that are not expected to be produced by the photospheric radiation field. These super-ions are primarily produced by the Auger process \citep{cassinelli79}. In this process a  X-ray photon causes the ejection of an inner shell electron from an ion. The ion, which is in a highly excited state, usually autoionizes. For \mbox{ CNO} elements, the net change in the charge of the X-ray-absorbing ion is generally $+2$.  Auger processes, and direct ionization by the shock radiation field, are accounted for in calculating the wind ionization. 
Only two stars in our sample have been observed with X-ray satellites, namely \zetap\ and \lambcep\ with \rosat; the X-ray luminosities are $\log L_{\rm X}=32.43$ and $\log L_{\rm X}=31.92$, respectively \citep[see][]{oskinova05}. The luminosity ratios, \logLX, for these stars were computed using the bolometric luminosities as determined from this study. For \hdo, \hdfourteen, \hdfifteen\ and \hdsixteen, we have adopted the X-ray luminosity  of \zetap\ , while the X-ray luminosity of \lambcep\ was chosen as a proxy for \hdonesixtythree\  and \hdoneninetwo. 

\section{Diagnosing stellar and wind parameters}
\label{proc_sect}

\subsection{Stellar luminosity}

The stellar luminosity $L$ is one of the key input parameters to \cmfgen. It is usually derived from the absolute magnitude, which requires accurate knowledge of the distance of a star.

Because  \object{\hdfifteen} is a member of the cluster IC1805,  we adopted the distance modulus of the cluster \citep{massey95}, $DM = 11.85\pm 0.1$ (which corresponds to a distance of 2.3 kpc), 
to derive \Mv, hence $L$. A significant absorption is also present toward this cluster, which translates into a fairly high reddening of the stellar spectra. 
Remarkably, the color excess we derive from fitting the observed SED differs by less than 
1\% from the \ebv\ computed from observed (B-V) and theoretical (B-V)$_{0}$ from \cite{martins06}.

For all other stars, the distance is poorly known, with sometimes fairly large differences in estimates of the distance modulus. 
This leads to significant uncertainties in the luminosity, which in turn translates into differences in some parameters such as the mass and the mass-loss rate, which depend on the stellar luminosity 
through the implied value of the stellar radius \citep[see, e.g.,][]{puls96}. We adopted the following procedure:
\begin{enumerate}
\item  We adopted an initial absolute magnitude (and corresponding distance) for each object from the literature. 
We used \Mv\ values listed in \cite{markova04} for  \object{\hdfourteen},  \object{\hdsixteen},  \object{\zetap},  \object{\hdoneninetwo} and  \object{\lambcep}, while \Mv\ was taken from \cite{repolust05} for  \object{\hdo}. For  \object{\hdonesixtythree}, we found no \Mv\ listed in the literature and we adopted \Mv\ from the calibration by \cite{martins06}.  

From \Mv, we derived L assuming a bolometric correction from \cite{martins05}. 
We then searched for the model that best fits the normalized UV and optical spectra. This model yields the photospheric and wind parameters. We derived \mdot\ by requiring that the model fits \halpha. 
Any value of \mdot\ and \rstar\ such that \mdot\ /\rstar$^{1.5}$ is constant provides a similar fit to \halpha\ (and \heii\  \lb4686). However, UV lines do not follow the same trend, and this allowed us to break the degeneracy between \mdot\ and \rstar\ to some extent..   

\item From the best-fit model, we fitted the flux calibrated \iue\ spectra and UBVJHK fluxes (for the photometry see Table 1) to constrain the distance and the extinction. 
The Galactic reddening law of \cite{cardelli89} was used. We varied the distance (global scaling) and color excess \ebv\ (wavelength-dependent leverage) to reproduce the UV-optical-near-infrared  spectral energy distribution (SED). 

\item The absolute magnitude \Mv\ was recomputed with the distance derived in step 2 from the fit to the observed fluxes. If the new value differed from the value initially assumed in step 1 by 0.05 dex or more, it was used to compute a new luminosity and a new model atmosphere. When the revised model atmosphere is calculated,
 \teff, \logg, \vinf, $\beta$ and the abundances were kept fixed to their respective values derived in step 1, but \mdot\ and $f$ (the clumping filling factor) were varied to determine a new best fit to the normalized spectra.

\item This set of three steps was repeated until we converged on a solution satisfying both the fit to the individual spectral wind features from the normalized spectra and the absolute UV + optical/near-IR (NIR) fluxes.
\end{enumerate}   

\begin{figure*}
\includegraphics[scale=0.52 , angle=0]{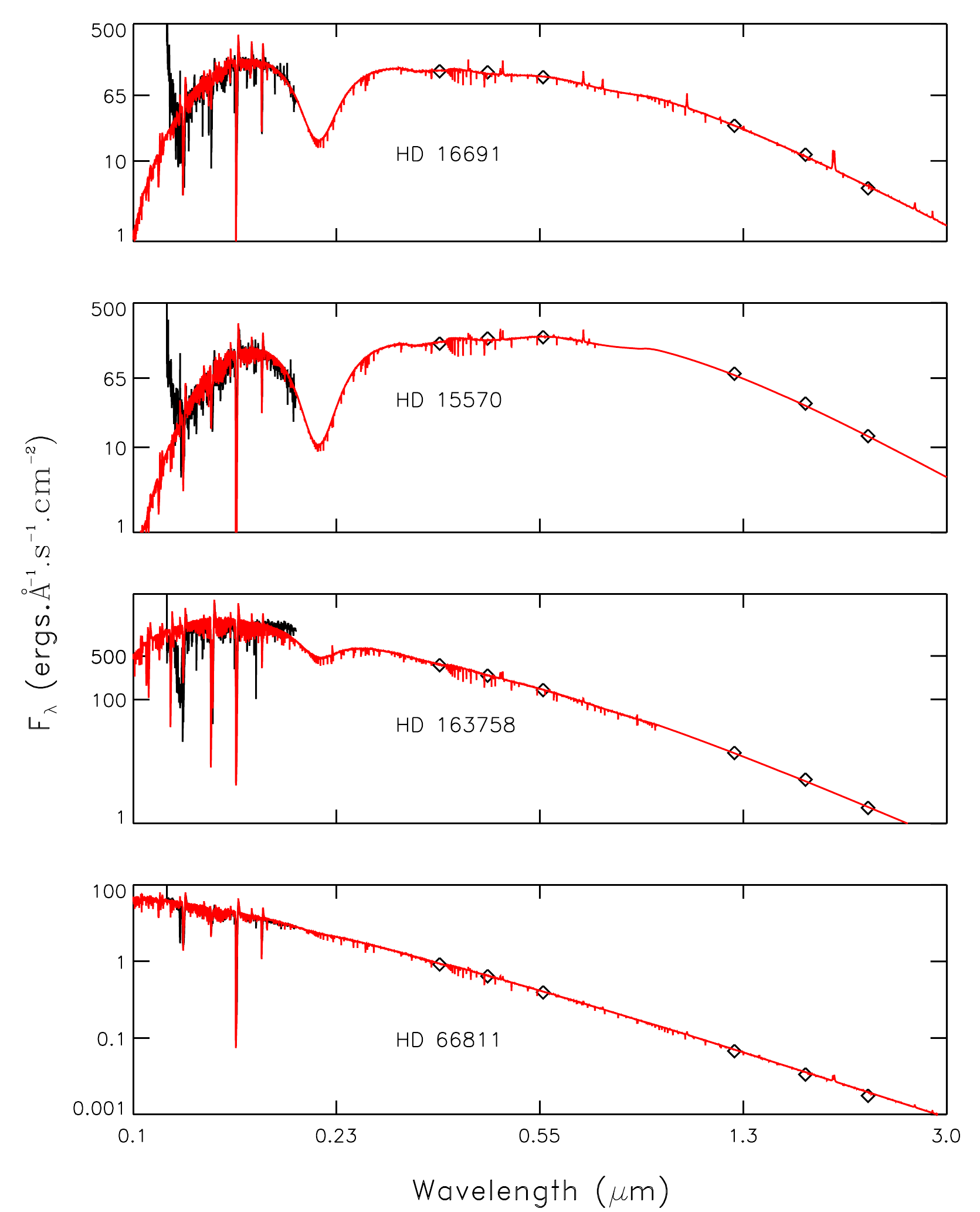}
\includegraphics[scale=0.52, angle=0]{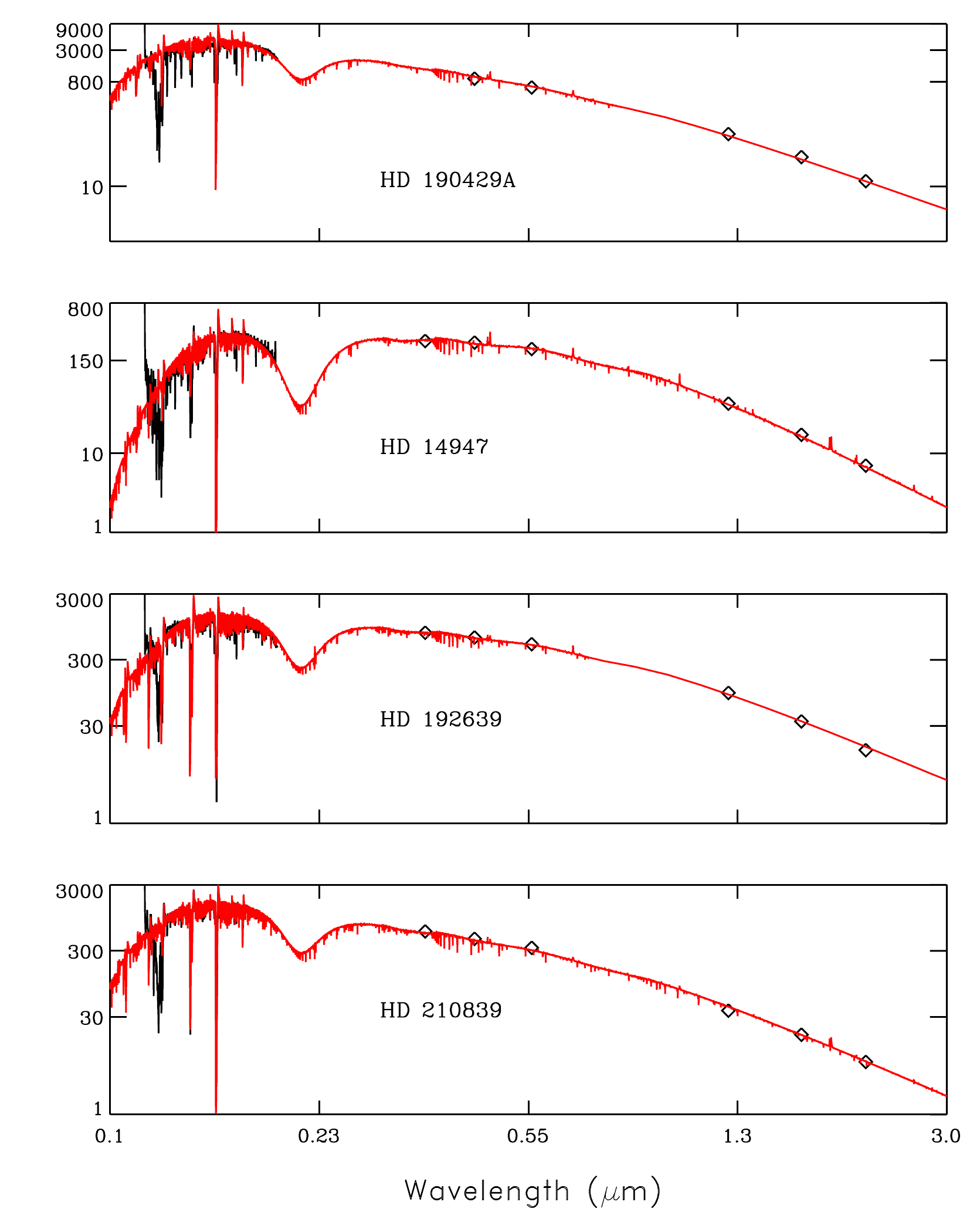}
      \caption[12cm]{Synthetic SEDs (in red) compared to flux calibrated + UBVJHK photometry for the target stars (in black). The distance, E(B-V) and luminosity were iterated to reach agreement between models and observations (see Sect. 4.1). For plotting purpose, the fluxes were scaled by a factor 10$^{14}$ for all stars but  \object{\hdsixtysix} and \object{\hdtwohundten}, where factors 10$^{9}$ and 10$^{13}$ were used, respectively.}
         \label{fig1}
\end{figure*}

Final values for the distances and absolute magnitudes are listed in Table \ref{tab_spec_types}. The SEDs for each star resulting from the process above are presented in Fig. \ref{fig1}.
 
\subsection{Effective temperature, \logg\ and helium abundance }
\label{teff_sect}
Our primary \teff\ diagnostics in the optical were the ratio of \hei\ \lb4471 to \heii\ \lb4542 equivalent widths, as well as line profile fitting. We used the other lines of \hei\ and \heii\ to refine the \teff\ determination. 
The optical helium lines were also used to constrain the relative \he\ to \hyd\ abundance. 
In addition to these classical helium diagnostics, we also used metal lines for estimating the effective temperature. This is particularly relevant for the earliest stars of the sample, which have weak \hei\ lines, and for which the validity of using the helium ionization balance to determine \teff\ accurately is questionable. In the UV the \feiv\ to \fev\ and \fev\ to \fevi\ ionization ratios provide  a sensitive temperature diagnostic. 
We found consistent estimates with the helium ion ratios within one thousand Kelvin for most objects, a slightly higher difference for  \object{\zetap} and  \object{\lambcep} is related to the higher \vsini\ of these objects. 
When possible ( \object{\hdo} and  \object{\hdsixteen}), we also used the \civ\ \lb1169 to \ciii\ \lb 1176 line ratio which has also been shown to provide a useful temperature diagnostic by \cite{heap06}. 

In some cases, effective temperature determinations based on the ionization ratio of optical helium lines are inconsistent with some temperature-sensitive metal lines, such as \nv\ \lb\lb 4605-4620. The presence of these lines in the observed spectra points to an enhancement in \teff\ of approximately 2000~K compared to the values derived from the helium lines. The relative strength of the \nv\ doublet, which is strongly dependent on the temperature, 
has a relatively weak dependence on gravity, while the absolute strength was better reproduced by tuning the nitrogen abundance.
Nevertheless, we could not achieve a good fit to these lines without degrading the agreement with the rest of the optical and UV spectrum.  The values of \teff\ and nitrogen abundance needed to fit these lines should be regarded as upper limits. 

The Stark-broadened wings of \hgamma\ provided the primary constraint on \logg\ (other Balmer lines were used as secondary indicators). The typical uncertainty in \logg\ is 0.1~dex. Despite the high S/N of our spectra it is difficult to do better than this --- \logg\ has
to be determined simultaneously with \teff, and there are slight inconsistencies with different lines (because of rectification errors and  residual uncertainties with the atomic physics).

\subsection{Micro-turbulence}
\label{microt_sect}
The formal solution of the equation of transfer was used with a radially dependent turbulent velocity to calculate the emergent spectrum \citep[see also][]{hillier98}. 
In this case, the microturbulent velocity follows the relation \xit$(r) =$ \xitmin + (\xitmax - \xitmin)$\frac{v(r)}{v_{inf}}$ 
where  \xitmin\ and \xitmax\ are the minimum and maximum microturbulent velocities. 
The microturbulence at the photosphere is chosen to best reproduce the strength of the \sv\ \lb1502 and iron lines (\feiv\ and \fev) in the \iue\ spectra. 
The adopted \vturb\ has little effect on the strength of most \hi\ lines. On the other hand, we found some significant changes for important \teff\ diagnostics such as \hei\ \lb4471 and some \heii\ lines \citep[see also][]{villamariz00}. 
We confirm findings by \cite{hillier03} that \heii\ \lb5411 is also sensitive to the microturbulent velocity, while \heii\ \lb4541 or \heii\ \lb4686 are not. 

In the outer region of the wind, the turbulence is adopted to best fit the shape and slope of the blue side of the absorption component of the \civ\ \lb\lb1548-1551 P~Cygni profile.  
 
\subsection{Surface abundance}
\label{surfabun_sect}
Accurate \mbox{CNO} abundances provide key constraints on both the
evolutionary state and the influence of rotation, and should preferably be determined from photospheric lines of these elements. 

We adopted the solar abundances from \cite{asplund05} as a reference. 
The new values better agree with surface abundances of B-type stars in the solar neighborhood \citep[see][]{cunha06, przybilla08, simon10, nieva11}. Some caution may be required when comparing these baseline abundances to those we derives since only one star ( \object{\zetap}) is located within 1 kpc, which is the limit usually set for solar neighborhood abundance studies 
\citep[but see the recent work by][]{nieva12}.   

\subsubsection{Carbon abundance}
Most carbon lines in the  FUV/UV range suffer from wind contamination and we chose not to use them as prime indicators for carbon abundances. 
For stars with \teff\ $\leq$ 35,000~K, \ciii\ \lb1247 appears as an emission line with a clear response to abundance variations. This is the only diagnostic for carbon abundance in the UV for these stars.

In the optical, available lines depend on the temperature range (except for \civ\ \lb\lb5801-5812 which are present for all stars).
In a few stars, \ciii\ \lb\lb4647-50-52 lines are clearly detected and were used.  For stars cooler than 38,000~K, \ciii\ \lb5696 was used.  
For stars with \teff\ $\leq$ 35,000~K, \ciii\ \lb4068 is present as a non-saturated absorption line and was also used to measure the carbon abundance.
Finally, we checked for consistency with the other \ciii\ diagnostics listed above using the \ciii\  emission lines at \lb6721, \ciii\ \lb\lb6727-6730.

\subsubsection{Nitrogen abundance}
In the UV, the photospheric lines of \niii\ \lb\lb1748-1752 were used.  However, these features saturate when the nitrogen abundance exceeds about five times the solar value,
 at which point they are no longer useful abundance diagnostics.

In the optical, \niii\ \lb4379, \niii\ \lb\lb4634-4640-42, \niii\ \lb\lb4510-4515, \niii\ \lb5320-5327, \niv \lb\lb5204-5205, \niv \lb6380, \nv\ \lb4944 and  $\niv$ \lb 6212-6220 are available. 
We furthemore used the emission lines of \nv\ \lb\lb6716-6718 to check for consistency only, because these lines are often blended with \ciii\  lines (\lb6721, \lb\lb6727-6730).

We found that the  \niv\ \lb5205 line provided the most reliable abundances in optical spectra of O4 -- O5 supergiants.
It is sensitive to abundance changes, with no sign of saturation within the limits we tried. 
While there may be problems with the absolute calibration of abundances based on this line, relative abundances should be much more accurate.
 
\subsubsection{Oxygen abundance}
The \oiii\ \lb5592 line provides the best constraint on the oxygen abundance in optical spectra.
Indeed, this line is not contaminated by wind contribution and is observed as an absorption line (though often weak), in a region of the spectrum where the continuum is well defined.  
We also used the \oiii\ triplet (\lb\lb1150-1154\,\AA) in the \fuse\ spectrum, when fast rotation does not blend them with other metal lines.
UV lines (\oiv\ \lb\lb 1338-1343 and \ov\ \lb1371) indicate a (sometime significant) contribution from the wind and were not used.

\subsubsection{Formation of some specific lines}
Before proceeding further it is worth making some general comments about photospheric lines, which sometimes appear in emission in O stars. If we assume the Eddington-Barbier relation holds, these transitions are in emission because the source function is larger than the neighboring continuum source function. This can occur because the population of the upper level is enhanced, or because the population of the lower level is depleted. Many of the transitions occur between the n=3 levels of CNO ions (n=4 for \siiv), and involve one, or more levels, coupled to low-lying levels by transitions in the EUV (i.e., shortwards of 900\,\AA). 
 Depending on the EUV radiation field, these transitions can drain or populate a particular level. 
 These transitions can also be influenced by overlapping lines from other species \citep[e.g.][]{najarro06}, and by the adopted abundances (which directly alters the optical depth of the line). 
 In \mbox{CNO} elements, dielectronic recombination may play a role by selectively enhancing the populations of some levels relative to others \citep{rivero12}.
 
In WN stars, \cite{Hillier88} found that the strong N\,{\sc iv} emission lines (4058\,\AA, 3478-3485\,\AA, 7013-7129\,\AA) are influenced by continuum fluorescence and dielectronic recombination. The \civ\ \lb\lb5801-5812 doublet is also strongly influenced by continuum fluorescence, which is pumped by a transition at 312\,\AA. On the other hand, in WC stars, with their large carbon mass-fraction ($> 0.1$), radiative and dielectronic recombination dominate the formation of optical carbon lines \citep{Hillier89}. For O stars the situation is likely to be more complicated (the densities are higher and more processes are likely to be important) because LTE must be recovered close to the line formation region.  
Fits to these lines significantly improved when we computed models with a much larger number of atomic elements, which were originally included to improve the calculation of the line radiative force. We found that the change induced in the total EUV line-blanketing yielded \civ\ \lb\lb5801-5812 profiles that better agreed with observations.   

\ciii\ \lb\lb4647-50-52 lines form very near the stellar surface and are marginally sensitive to wind parameters. 
In addition, the lower level of \ciii\ \lb\lb4647-50-52 ($1s^{2} 2s3s - 1s^{2} 2s3p$) corresponds to the upper level of \ciii\ transitions 
around 538\,\AA\ ($1s^{2} 2s2p - 1s^{2} 2s3s$). Around this wavelength, \feiv\ lines are present and overlap with the \ciii\ line. 
These \feiv\ lines drain and depopulate the lower level of the \ciii\ \lb\lb4647-50 transition \citep[see also][in preparation]{martins1213}. We had to reduce significantly the oscillator strengths of \feiv\ lines  around 538 \AA\ to obtain  satisfactory fits to these \ciii\ lines.  
This sensitivity to atomic model assumptions must be kept in mind when using these lines for abundance determinations. 
 
\ciii\ \lb5696 requires many energy levels in our model atom to form in emission as observed. However, in some cases we could not reach a satisfactory fit of the lines, indicating that some mechanism or process that is affecting the formation of this line has not been correctly accounted for in our modeling. 
To our knowledge, systematic studies of the formation of \ciii\ emission lines have been performed for WC stars only.  
We note first that \ciii\ \lb5696 behaves differently from other \ciii\ emission lines, even though it varies smoothly along the WC spectral sequence, 
\citep[e.g.][]{Torres86}. \ciii\ lines such as $\lambda2296$ and \ciii\ \lb6741 are strong in WC4 stars, while $\lambda5696$ is very weak or absent. On the other hand, in late WC stars, $\lambda5696$ is the strongest optical \ciii\ transition. This behavior occurs because the upper level of $\lambda$5696 (2s\,3d\, $^1D$) can also decay to the 2s\,2p\,$^1$P$^{\scriptsize \rm o}$ via a transition at 574\,\AA\ --- only when the transition is optically thick will $\lambda5696$ be strongly in emission (A($\lambda$574)/A($\lambda$5696)=150, Storey [private communication]) \citep{Hillier89}. The lower level is coupled to the 2p$^2$\,$^1$D state by a transition at 884\AA, highlighting the importance of EUV transitions on the source function for the $\lambda$5696 transition.
Note that $\lambda 4058$ is the \niv\ isoelectronic equivalent transition to \ciii\ $\lambda5696$.

\cite{mihalas72} showed that the \niii\ \lb\lb4643-4640-4642 feature can be driven into emission by dielectronic recombination, and highlighted the importance of two-electron transition for draining the 3p state. They also noted that the Swings mechanism (i.e., continuum fluorescence) could be important in stars with winds. Very recently, \cite{rivero12} showed that in stars with stellar winds, the Swings mechanism is indeed more important, although they also note that in O stars cooler than 35,000\,K the interaction between the \oiii\ and \niii\ resonance lines could also be important.

 $\niv$ \lb\lb6212-6220 (\elevel{2s}{4}{3}{S} --\olevel{2s}{4p}{3}{P}) 
appears as an emission feature in the earlier supergiants. Our original models failed to  reproduce this line, which was traced to a problem with the theoretical model atom. In the atomic data calculations the levels designations
of the \olevel{2s}{4s}{3}{S} with the \olevel{2p}{3p}{3}{P} level \citep[as previously noted by][]{Lau90NIV, AAL91_NIV} are inverted relative to the spectroscopic designations. When we rectified this inversion, emission in $\niv$ \lb 6212-6220 was correctly reproduced.
 
In our models for some of the O4 supergiants, the \niv\ \lb4058  line shows up as an intense emission feature, which is not observed. 
The \nv\ 4605-4620 lines are present as photospheric absorptions in O4 supergiant spectra only. Since they are also very sensitive to effective temperature we did not use them to constrain the nitrogen abundance.
\cite{walborn01} describes an apparent connection between the \nv\ \lb\lb 4605-4620 and the \niv\ \lb 
4058 lines: the former appear in absorption in early-type supergiants, while the latter are seen in emission of ``similar inverse intensity". 
Although the presence and relative intensity of these lines is clearly defined in Of supergiants, there is no obvious physical connection that would explain their observed proportionality (other than they both depend on \teff).

\subsection{Wind parameters}
The wind terminal velocities, \vinf, were estimated from the blueward extension of the absorption part of UV P-Cygni profiles, which occurs up to  \vinf\ +  \vmax, where \vmax\ is the maximum microturbulent velocity described above. Fits of the UV P-Cygni profiles using the above relation for microturbulent velocity allows a direct determination of \vinf. The typical uncertainty in our determination of \vinf\ is 100 \kms (depending on the maximum microturbulent velocity we adopt).

Mass-loss rates are derived from the simultaneous analysis of strong UV lines such as \civ \lb1169 + \ciii \lb1176, \nv\ \lb1240, \civ\ \lb\lb1548, 1551, \siiv\ \lb\lb1394, 1403, \ov\ \lb1371 and \niv \lb1718. In the optical the strength of \halpha\ and \heii\ \lb4686 is sensitive to \mdot, while the shape of their emission profile is sensitive to the $\beta$ exponent of the wind velocity law. 
 
Following \cite{bouret05}, clumping-related quantities (\mdot, $f_\infty$ and $v_{\rm cl}$) were derived from \pv\ \lb\lb1118, 1128, \ov\ \lb1371, and \niv \lb1718 in the FUV/UV domain, to which we added clumping sensitive lines from the optical, primarily \halpha\ and \heii\ \lb4686. We emphasize here that some photospheric lines in the UV and optical are also sensitive to the adopted filling factor (and scaled \mdot). For photospheric \hi\ and \he\ lines for instance, this is essentially caused by a weaker wind contribution (emission) in clumped models, which produces deeper absorption 
compared to smooth-wind models. For a more detailed discussion concerning the sensitivity of line to the wind and clumping,
see \cite{hillier03}.

\section{Results of the spectroscopic analysis}
\label{result_sect}
The synthetic spectra were convolved with a representative instrumental profile and a rotational profile calculated for a trial value of the projected rotational velocity (\vsini) to permit direct comparison with observations.
We varied \vsini\ until 
we achieved a good fit to the observed photospheric profiles.

As an independent check, we used the Fourier transform method described in \citet[][and reference therein]{simon07}
to determine \vsini\ from the observed spectra.

To apply the Fourier transform method, we selected lines with as weak a wind contribution and as symmetric a profile 
as possible. 
We identified several isolated photospheric lines (which included \nv\ \lb\lb4604 - 4620 for  \object{\hdo} and  \object{\zetap}, \oiii\ \lb5592 for cooler stars), which we used to 
derive individual values that we averaged to yield the values listed in Table \ref{tab_results}. 
For the fast rotators, namely  \object{\zetap} and  \object{\lambcep}, \vsini\ from the two method agree better than within 10 \kms, which is well within the typical error bars of each method.
For the other moderately rotating stars, the difference is smaller than 20 \kms,
which is slightly smaller than the difference found 
for other O-type supergiants by \cite{simon07}.  The final values used in our models are listed in Table \ref{tab_results}. 

The quality of the fits was then additionally improved by convolving the rotationally broadened profiles with a Gaussian of FWHM \vmac\ to account for (isotropic) 
macro-turbulence. More complex description of the macro-turbulence exists, e.g. the so-called radial-tangential broadening.
In the specific case of massive stars, \cite{aerts09} recently interpreted the missing line broadening in terms of the collective effect of numerous, low-amplitude, non-radial gravity-mode oscillations. 
These authors emphasized that using the Gaussian formulation of macro-turbulence to fit line profiles could lead to significant underestimation of the actual
rotational velocity and warned that appropriate expression for the pulsational velocities should be used instead. 
Therefore, the values quoted in Table \ref{tab_results} should be regarded as lower limits, but we do not expect significant modifications of our conclusions 
concerning surface abundances. First, rotation and macro-turbulence in the adopted description are not expected to affect line equivalent widths,
only the line profiles (contrary to micro-turbulence).
Second, differences between rotational broadening from line fitting, which should be regarded as an upper limit, or from Fourier transform are quite small (less
than 20 \kms, see above). Third, in the total broadening we considered, the rotational term is always dominant with respect to macro-turbulence.

Given the complexity of the fitting process over the FUV to optical spectral range, we chose not to derive formal statistical uncertainties,
nor did we try to estimate the correlation of errors between parameters. Instead, we varied the different parameters
until we obtained the solution that provided the best fit Òby eyeÓ to the observations, with more weight being given to primary 
diagnostics (cf Sect. \ref{proc_sect}).
Then, for a more quantitative estimate, we varied each parameter independently around this solution and computed the residual (i.e. the difference) between the observed and the synthetic spectrum. We imposed the condition that this residual must remain within $\pm$10\% and adopted the range of values that fulfilled this 
condition as error bars on each parameter. The stellar and wind parameters we derived are gathered in Table \ref{tab_results}.
We quote all chemical abundances by number and mass fraction, except for \hei, which is given by number relative to hydrogen (see Table \ref{cno_tab}). 
As for the uncertainty in the \mbox{CNO} abundance measurements, error bars are based on the fit quality of the lines listed above, as estimated from a classical $\chi^{2}$ procedure.
Our conclusions for individual objects are summarized in the remainder of the section. The best-model fits to the data are presented in Appendix A.
\subsection{ \object{\hdo} - {\rm O4~If}}
\label{hdo_sect}
 \object{\hdo} was studied extensively in the UV in \cite{bouret05}. 
We recomputed their best-fit model, using the new \cite{asplund05} abundances. We scaled the \mbox{CNO} as well as \mbox{P} and \mbox{S} abundances to the same relative values as in \cite{bouret05}. The best fit to the  \pv\ \lb\lb1118-1128 resonance doublet is obtained for P/P$_\odot$ = 0.7, further confirming the {\it apparent} depletion in phosphorus of this star. 
Only a limited tuning of the model parameters was needed to improve the fit to the optical lines, which demonstrates the reliability of our FUV/UV analysis. The best fit to FUV-UV and optical spectra is presented in Appendix \ref{bestfit_sect}.
Note however that the presence of strong lines of \nv\ \lb\lb 4605-4620 would require significantly higher temperature to be reproduced (\teff\ $\simeq$ 43,000 K), which would then make the fit
worse everywhere else in the spectrum (see also comments on Sect. \ref{teff_sect}).

The \ciii\ \lb\lb4647-50-52 lines are weak, especially compared to the strong neighboring \niii\ \lb\lb 4634-4640-4642 lines. This was recognized \cite[e.g. by][]{walborn10} as due to an abundance effect, 
as \mbox{CNO} processed material is expected at the surface in supergiants. We found that the atmosphere of  \object{\hdo} is highly depleted in carbon (X(C) = 0.009\% by mass), and is enriched in nitrogen (X(N) =  0.73\%). 
Although derived from fewer lines, the oxygen abundance corresponds to a mass-fraction X(O) = 0.067\%. Together with the \he\ enrichment $y = $ 0.2, this is qualitatively consistent with the general trend predicted by evolutionary models with rotation.  
\subsection{ \object{\zetap} - {\rm O4~I(n)fp}}

Although a model with \teff\ = 39,000~K provides an excellent fit to helium lines, raising \teff\ up to 41,000~K better agrees with the observed \nv\ \lb\lb4605-4620 lines, which are otherwise too weak. 
Deriving the actual value proved to be elusive, but we expect that  \teff\ = 40,000 $\pm 1000$ K is a realistic estimate of the effective temperature of  \object{\zetap}.
The wind properties agree very well with those presented in \cite{najarro11} based on the modeling of the near-infrared spectrum of  \object{\zetap}. 
Only clumped-wind models matched the whole set of wind-sensitive lines. The best fit to \pv\lb\lb1118-1128 lines was nevertheless obtained only when using sub-solar
phosphorus abundance, P/P$_\odot$ = 0.5. 
We needed to start clumping at \vcl\  = 100 \kms, significantly above the sonic point to obtain a reasonable fit of the central trough of \halpha.
This indicates that more absorption is needed in the line formation region, which is missing if clumping starts right above the photosphere. 
It is remarkable that this behavior is observed on the other fast rotator of our sample  \object{\lambcep}, as well as in  \object{\hdsixteen}, whose emission lines present strong similarities with
those of  \object{\zetap} (see below): \niii\ \lb\lb4634-4640-4642 presents a box-shaped profile, while \heii\  \lb4686 is clearly double-peaked. 
Based on preliminary models using a realistic 2D treatment of rotation as described in \cite{busche05} and \cite{zsargo06}, we concluded that 
the shapes of these emission lines are most easily interpreted in terms of an effect of high rotational velocity. These conclusions will be presented in
a forthcoming paper. 

Despite the high \vsini\ of this star, many lines of \mbox{CNO} ions are present in the spectra to derive reliable estimates of the surface abundances (see Fig. \ref{Fig_hd66}). 
One such line worth mentioning is the \niv\ 
\lb\lb7103-7129 multiplet, which is the triplet equivalent of \niv\ \lb4058. 
Our \feros\ spectrum of  \object{\zetap} shows an emission feature at these wavelengths whose strength is very sensitive to abundance variations and is well-matched in our final model.  
The total of \mbox{CNO} abundances is clearly super-solar in \zetap\ (see Table \ref{cno_tab}). 
It has been argued that this could be the result of a binary/merger evolution \citep{pauldrach11}, in line with the runaway status often invoked for this star \citep{sahu93}.
We stress, however, that in \cite{pauldrach11}, this conclusion is based on abundances for several elements, which are very different from what we derive. 
Their high \mbox{C} abundance for instance seems impossible to reconcile with the observed optical spectrum when modeled with \cmfgen. Detailed comparisons between 
the two analyses are needed to elucidate this problem.

We also note that the luminosity we adopted differs significantly from those quoted in \cite{najarro11} or \cite{pauldrach11}. 
These two studies chose the higher value, often related to its runaway status, while we adopted a lower value for the distance, compatible with the \hipparcos\ distance quoted for  \object{\zetap}. 
On the other hand, \cite{schilbach08} found \zetap\ to be even closer (300 pc) than given by \hipparcos\ (430 pc). These authors suggested that the star originally was in the open cluster Trumpler 10, 
which it left about 2.5 Myrs ago. A direct consequence of this lower distance would be to revise the luminosity downward to \logL\ = 5.42. The stellar mass would go down accordingly ($\approx$ 17 \msol), 
which starkly contradicts the usual parameters of an O4 supergiant.

\subsection{ \object{\hdsixteen} - {\rm O4~If}}
\label{hdsixt_sect}

A model with \teff\ = 41,000\,K $\pm$ 1000\,K best fits the \he\ lines as well as \nv\ \lb\lb4605-4620 in  \object{\hdsixteen}. 
Carbon lines in the optical are weak, which suggests a high depletion of this element (but see Sect. \ref{surfabun_sect} for comments 
about the sensitivity of \civ\ \lb\lb5801-5812 doublet to the treatment of UV blanketing). This is also supported by the \civ\ \lb1169 and \ciii\ \lb1176 lines, which show up as weak absorptions. 
Nitrogen lines are strong, indicating that this element is altogether highly enriched. We note that despite several tests we performed, \niv\ \lb4058 is predicted to be strongly in emission compared 
to the observed profile. Fitting this line would require a much lower nitrogen abundance, inconsistent with the rest of the nitrogen lines.
In accordance with its Of/WN transition status \citep{conti89},  \object{\hdsixteen} furthermore presents a clear enrichment in \he.  

Only clumped models agree well with the observed wind sensitive lines. The best fit to \pv\lb\lb1118-1128 lines was obtained using sub-solar
phosphorus abundance P/P$_\odot$ = 0.5. 

UV  and optical spectra of  \object{\hdsixteen} and  \object{\hdo} show strong similarities, reminiscent of those seen in their IR $K$-band spectra (e.g. the subordinate \heii\ 2.189$\mu$m in emission). 
\cite{conti95} noted that both stars would have been classified as WN stars had they been observed only in the $K$-band. These spectral similarities strongly suggest a common evolutionary status for these objects. 

 \object{\hdsixteen} and  \object{\zetap} also share some striking similarities (see Table \ref{tab_results}), especially regarding the emission lines of their optical spectra, such as \niii\ \lb\lb4634-4640-4642 and \heii\ \lb4686.
We argue that a high rotation rate is able to explain the observed emission profiles of \object{\hdsixteen}, and there is no need to invoke an extended corotating
region that would be caused by a putative magnetic field, as proposed by \cite{debecker09}. 
Furthermore, we needed to start clumping at \vcl\ = 130 \kms\ in our models to best reproduce the observed spectra. This behavior is again very reminiscent of that of the other fast rotators of our sample (see below)
and might indicate that the wind of  \object{\hdsixteen} is indeed rotating faster than suggested from the photospheric lines.   

\subsection{ \object{\hdfifteen} - {\rm O4~If}}
 \object{\hdfifteen} is one of the brightest early-type stars in the young open cluster  \object{IC 1805}. 
The FUV-UV spectra are heavily masked by interstellar absorption lines (mostly \htwo\ in the \fuse\ spectrum), consistent with the large \ebv\ we derive (see Sect 4.1).
The microturbulent velocity, which is usually best determined from \sv\ \lb1502 and the iron line strengths (\feiv\ and \fev) could not be accurately measured and we used optical lines only, to derive a value \xit\ = 15\,\kms.

We derived \teff\ = 38,000 K, \logL\ = 5.94 for  \object{\hdfifteen}, which implies that it is somewhat cooler than expected for its spectral type, although at the expected luminosity according to \cite{martins05}.
Usually the observed differences/dispersion in this calibration can account for a large part of the uncertainty in the luminosity, i.e., the distance 
to this object. This should not be the case for  \object{\hdfifteen} because the distance modulus IC1805 cluster is well determined. 

\begin{figure}[h]
  \centering
   \includegraphics[scale=0.5,angle=0]{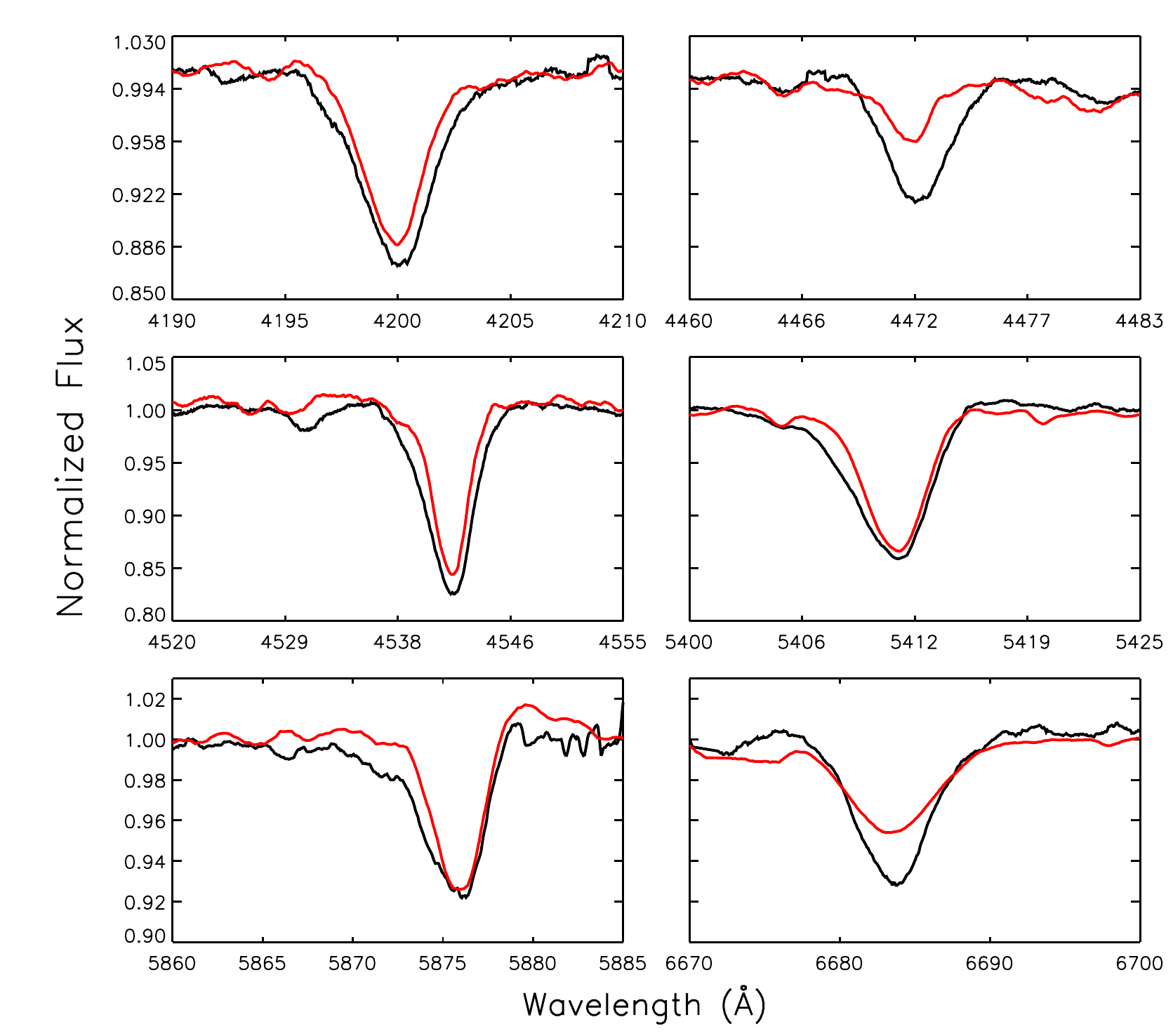}
      \caption[12cm]{Helium line profiles from the optical spectra of  \object{\hdo} (black) and  \object{\hdfifteen} (red).  \object{\hdo} has \vsini\ = 150 \kms, while it is 97 \kms\ for  \object{\hdfifteen}.}
         \label{he_prof}
\end{figure}

More puzzling, the \he\ lines of  \object{\hdfifteen} are weak for its spectral type. 
Since the photospheric and wind parameters of  \object{\hdfifteen}  (see Table \ref{tab_results}) are very close to those of  \object{\hdo}, we compared the optical spectra of both stars. 
The unusually weak helium lines of  \object{\hdfifteen} are clearly apparent in Fig. \ref{he_prof}, where \he\ lines are stronger and broader in  \object{\hdo} (\vsini\ for  \object{\hdo} is higher than for  
\object{\hdfifteen}).

The high-resolution spectrum of  \object{\hdfifteen} presented in \cite{debecker06}\footnote{{Obtained with the echelle spectrograph at the 2.1 m telescope at the Observatorio 
Astronomico Nacional of San Pedro Martir (SPM) in Mexico}} was kindly provided to us by G. Rauw for comparison to our spectrum.
Both spectra are remarkably similar throughout the common spectral range (4000 - 6800 \AA), including the weak and narrow(er) helium lines with respect to the star's photospheric properties. This confirms that our finding is not an observational artifact caused by incorrect data reduction or spectrum rectification. 
Spectra retrieved from the \elodie\ archives\footnote{{\tt http://atlas.obs-hp.fr/elodie/}}, which cover the epoch from 1996 to 2005 and have a signal-to-noise ratio of 100 or higher, also display the same narrower, weaker helium lines. This confirms that the weakness of \he\ lines is not time-dependent, at least on this timescale. 

Quantitatively, we found that $y$(He/H) = 0.07 (by number) is needed to best fit the helium lines in the optical domain. Such a low \he\ abundance is not supported by stellar evolution considerations, but we could not find any problems with the observations that would affect the analysis of \he\ lines. It is unlikely that this lower helium abundance is an indication of the initial composition of the molecular cloud  from which \object{\hdfifteen} formed. 
\cite{martins05} found $y = 0.1$ for  \object{HD 15629}, a main-sequence O5V((f)) star in  \object{IC 1805}, which suggests that the initial helium content of this cluster is normal.  
In the rest of this work, we use $y$ = 0.1, which should be considered as an upper limit on the helium abundance. 

Only models strongly depleted in carbon match the carbon lines, while all nitrogen lines in the optical suggest an enrichment in nitrogen.
As for  \object{\hdsixteen}, \niv\lb4054 appears as a strong emission in our models, which we could not reconcile with the observed profile (a very weak emission). 
The \oiii\ \lb5592 line shows that oxygen is also depleted. Qualitatively, the \mbox{CNO} abundances are consistent with the evolved nature of  \object{\hdfifteen}, 
as inferred from its position in the H-R diagram. 

The best agreement for wind-sensitive lines was obtained with clumped models, still requiring a sub-solar phosphorus abundance to match \pv\lb\lb1118-1128 (see Table \ref{cno_tab}).
\subsection{ \object{\hdfourteen} - {\rm O5~If}}
\label{hdfour_sect}
The best fit was obatined for \teff\ = 37000 $\pm$ 1000~K, which is 1500~K cooler than expected from the \teff\ - spectral type calibration by \cite{martins05}. 
This \teff\ is low enough that the \ciii\ \lb5696 line is present in the spectrum and can be used to measure the carbon abundance, which is mildly depleted. 
The adopted value also gives a consistent fit  to \ciii\ \lb\lb4647-50-52 and \civ\ \lb\lb5801-5812. 
We obtained a consistent fit to several nitrogen lines, including \niii\ \lb\lb4634-4640-42, \niii\ \lb\lb4510-4515, \niv \lb\lb5204-5205 and \niv \lb6380 by adopting
a surface enrichment comparable to those we measured in other stars of this sample. 
Similarly, \oiii\ FUV triplet (\lb\lb1150-1154) and \oiii\ \lb5592 indicate a depletion in oxygen. 
Qualitatively, the \mbox{CNO} patterns are compatible with both the evolutionary status (indicated by the position of  \object{\hdfourteen} in the H-R diagram), 
and the relatively minor \he\ enhancement ($y$  = 0.12).

The UV and optical wind lines could only be fitted with very highly clumped models. The best fit was obtained for a model with \finf\ = 0.03, implying 
a reduction of a factor of six for the mass-loss rate compared to the homogeneous model. This is by far the highest reduction of the whole sample. 
This highly clumped model is dictated by the \pv\ resonance doublet, for which we could achieve a very good fit by adopting a normal solar abundance for phosphorus. 
Using this low filling factor also maintains a good fit to all other wind/clumping-sensitive lines (see Sect. \ref{proc_sect}). 
At the temperature we derive for  \object{\hdfourteen}, the effect of introducing clumping is to strongly recombine P$^{+4}$ ion to P$^{+3}$ and the former ion is no longer dominant. The opacity in the line is greatly reduced and 
thus there is no additional need to vary the abundance to improve the fit. 
\subsection{ \object{\lambcep} - {\rm O6~I(n)fp}}
\label{lambcep_sect}
This star is another famous rapid rotator, with conspicuous spectral variability \citep[e.g.][]{howarth96, kaper97}. 
The best fit we could achieve to the FUV/UV and optical spectra was for 
\teff\ = 36,000~K, \logg\ = 3.5. 
Despite the high rotation rate, several lines are available for accurate abundance determinations. We found that  \object{\lambcep} is only moderately enriched in nitrogen, and depleted in carbon and oxygen
(Table \ref{cno_tab}), which is quite surprising because more extreme patterns would be expected from rotational mixing in such a fast rotator. We also found that $y$ = 0.12, in line with this moderate chemical evolution. 

The introduction of clumping in the model dramatically improved the fit to the spectrum, including \halpha.
Here again, we needed a lower-than-solar phosphorus abundance to achieve the fit to the \pv\ FUV lines presented in Fig. \ref{Fig_hd210}.  
As discussed in Sect. \ref{wind_section}, this indicates that the model optical depth for the resonance doublet is too high for solar abundance, which likely implies that either the ionization balance of phosphorus 
is computed incorrectly by \cmfgen, or that its treatment of clumping is incomplete.

The most severe disagreement between the model and the observations is for \heii\ \lb4686  -- with or without clumps we fail to reproduce the double-peaked profile, which is typical of a high rotation rate.
In this context it is no surprise that the \niii\ \lb\lb4634-4640-4642 lines present the same profile as already observed in  \object{\zetap} and  \object{\hdsixteen}. 

\subsection{ \object{\hdonesixtythree} - {\rm O6.5~Iaf}}
\label{hdonesixtythree_sect}
A low effective temperature \teff\ $=$ 32,500 -- 33,000 K would provide a very good fit to the forest of photospheric iron ions lines in the FUV-UV  domain \citep[see also][]{bianchi02}.
However, such a cool effective temperature is ruled out from the analysis of the optical spectrum and \teff\ = 34,500 $\pm$ 1000~K is preferred to fit the \hei\ and \heii\ lines in the optical and the FUV-UV spectrum as well.  
Although significantly higher than the UV-based \teff, it is also more compatible with the spectral type of the star \citep{martins05}.    
We found no documented variability that would explain the contradictory conclusions reached from the \fuse\ - \iue\ spectra on one hand and optical spectra on the other hand.  

The carbon lines in  \object{\hdonesixtythree} are stronger than in other O supergiants of similar spectral type \citep{gomez87}. 
This is the case for lines in emission (e.g. \ciii\ \lb1247, \ciii\  \lb\lb4647-50-52 and \ciii\ \lb5696) or in absorption (e.g. \ciii\ \lb4068 or \civ\ \lb\lb5801-5812). 
All these suggest that this star is carbon-rich. 
Our analysis shows that  \object{\hdonesixtythree} is indeed the most carbon-rich star of our sample, with an abundance of one and a half times the solar carbon abundance.  
While this abundance allows model spectra to match most photospheric \ciii\ and \civ\ lines, it still does not reproduce the strong \ciii\ \lb5696 emission line.

\object{\hdonesixtythree} is also one of the most nitrogen-rich stars star of the sample, as indicated by the strong emission lines of \niii\ \lb\lb4634-4640-4642 (the \niii\ \lb\lb1748, 1752 absorption lines are saturated). 
Because nitrogen is produced at the expense of carbon, and assuming mixing processes are unchanged when super-solar element abundance are present, this suggests that the initial carbon abundance 
of this object was very high. On the other hand, the oxygen abundance is sub-solar, as indicated by the very good fits to the photospheric \oiii\ lines in the FUV (\lb\lb1150, 54) and optical (\lb5592) spectrum.This seems
at odds with the conclusion that the initial \mbox{CNO} mixture of this star was super-solar. 

Whatever parameters we chose to fit the photospheric features, the whole set of wind lines could only be matched with a clumped models. The improvement is especially striking for the \siiv\ \lb\lb1394-1403 resonance doublet; this suggests that for this range of effective temperature, \siiv\, in addition to \pv,  can be used to probe clumping, thus extending the conclusions reached by \cite{prinja10} for B-type supergiants up to significantly higher temperatures
(or different ionization fractions). 

\subsection{\object{\hdoneninetwo} - {\rm O7.5~Iabf}}
\label{hdoneninetwo_sect}
The best fit was obtained for \teff\ = 33,500 $\pm1000$~K. In constrast to \object{\hdonesixtythree}, the analysis of both FUV-UV and optical spectra yields consistent results for  \object{\hdoneninetwo}. 
Given the strong line variability  of this star, especially in the \heii\ \lb4686 line \citep[e.g.][]{rauw01}, and the long time span separating the FUV, UV and optical spectra, this is quite remarkable.

Our analysis indicates that  \object{\hdoneninetwo} is moderately depleted in carbon and oxygen, although its nitrogen enrichment is relatively stronger. 
Whatever carbon abundance we used, we could not reproduce the \ciii\ \lb5696  or \civ\ \lb\lb5801-5812 lines with the correct intensity (see Fig. \ref{Fig_hd192}). Note that the latter would be 
better reproduced using higher \teff, which is ruled out from the fit of \he\ lines. The carbon abundance we quote is the upper limit above which \ciii\ \lb4068 and \ciii\  \lb\lb4647-50-52 become stronger than observed. 
The nitrogen abundance is constrained from the strong \niii\ lines in the optical at \lb\lb4510-4515 and \niii\ \lb\lb4634-4640-42. The \niii\ \lb\lb1748-1752 lines are saturated and were not used.  
This abundance is consistent with other \niii\ diagnostics, including \niii\ \lb5320-5327 in emission, but fail to reproduce  the weak absorption of \niv \lb\lb5204-5205. 
Oxygen lines are altogether very well defined and responsive to variations of abundances, either in the UV or in the optical. 

Finally, clumped models are needed to reproduce the wind lines, together with a factor of two depletion of the phosphorus abundance, compared to solar, to fit the \pv\ FUV lines. 
As for  \object{\hdonesixtythree}, the improvement in the fit resulting from clumped models is striking for the \siiv\ \lb\lb1394-1403 doublet. 
\begin{table*}[]
\caption{Fundamental parameters and wind properties for the stars in our sample. 
  }
  \begin{tabular}{lcccccccccccccc}
\hline
\hline
Star			& \teff	& \loggc	& \logLf	& log(\mdot)	& \vinf	&$\beta$	&\finf		& \vcl		& \vsini	& \vmac	& \msp					&\mevone & \mevtwo	& Age  \\
HD                      & [kK]          &                &                &                         & [\kms]     &                &                & [\kms]       & [\kms]     & [\kms]    & [\msol]                                         & [\msol] & [\msol] & [Myr] \\
\hline
16691		& 41.0	& 3.66	& 5.94	& -5.52 	& 2300	& 1.2		& 0.06	& 130			& 135	& 37		& 56.6$^{+14.6}_{-11.7}$		& 53.4	 & 65.2$^{+9.5}_{-8.6}$	& 3.1$^{+0.5}_{-0.1}$ \\ 
66811		& 40.0	& 3.64	& 5.91	& -5.70	& 2300	& 0.9		& 0.05	& 100		 	& 210	& 90		& 56.1$^{+14.5}_{-11.6}$		& 51.1	 & 62.3$^{+8.1}_{-7.8}$ 	& 3.2$^{+0.4}_{-0.2}$ \\
190429A	& 39.0	& 3.62	& 5.96	& -5.68	& 2300	& 1.0		& 0.04	& 30				& 150	& 57		& 66.0$^{+17.4}_{-13.4}$		& 54.6	 & 68.0$^{+8.3}_{-6.3}$	& 3.3$^{+0.4}_{-0.2}$ \\				 
15570		& 38.0	& 3.51	& 5.94	& -5.66	& 2200	& 1.1		& 0.05	& 30				& 97		& 40		& 54.0$^{+14.0}_{-11.1}$		& 48.7	 & 59.5$^{+6.6}_{-6.7}$	& 3.4$^{+0.8}_{-0.1}$ \\
14947		& 37.0	& 3.52	& 5.83	& -5.85	& 2300	& 1.3		& 0.03	& 30				& 130	& 36		& 48.0$^{+12.4}_{-9.8}$		& 43.4	 & 51.2$^{+6.0}_{-5.1}$ 	& 3.9$^{+0.5}_{-0.5}$ \\    
210839	& 36.0	& 3.54	& 5.80	& -5.85	& 2100	& 1.0		& 0.05	& 120			& 210	& 80		& 51.4$^{+15.2}_{-12.0}$		& 42.3	 & 50.5$^{+4.4}_{-4.5}$	& 4.0$^{+0.2}_{-0.2}$ \\
163758 	& 34.5	& 3.41	& 5.76	& -5.80	& 2100	& 1.1		& 0.05	& 30				& 94		& 34		& 41.6$^{+10.8}_{-8.5}$		& 41.2	 & 48	0$^{+5.0}_{-4.2}$	& 4.1$^{+0.3}_{-0.3}$ \\
192639	& 33.5	& 3.42	& 5.68	& -5.92	& 1900	& 1.3		& 0.05	& 30			& 90		& 43		& 38.3$^{+9.4}_{-8.2}$		& 39.1	 & 43.	0$^{+4.6}_{-3.7}$	& 4.2$^{+0.3}_{-0.2}$ \\
  \hline
\end{tabular}
 \label{tab_results}
   \begin{list}{}{}
\item \msp, \mevtwo, \mevone\ refer to spectroscopic mass, initial mass and mass at the corresponding stellar age, respectively.  
Uncertainties for \teff\ are $\pm$ 1000~K,  $\pm $ 0.1 dex for \logg\ and  $\pm $ 0.1 dex on \logL\ (reflecting the uncertainty on the distance). 
An uncertainty of $\pm$ 0.2\,\evi\ \msolyr\ was estimated for the mass-loss rates. \vinf\ is measured within $\pm$ 100 \kms.  
\end{list}
 \end{table*}

 \begin{table*}[htbp]
\centering
\caption{He, C, N, O and P abundances. 
}
\begin{tabular}{lcccccclll}
\hline
\hline
Star 			& y(He/H)		& $\epsilon$(C)	& $\epsilon$(N)	& $\epsilon$(O)	&  $\epsilon$(P)	& $\epsilon\ (\Sigma$ CNO)	& $X(\mathrm{C})$		& $X(\mathrm{N})$		& $X(\mathrm{O})$  \\
\hline
\object{HD 16691}		& 0.15		& 6.52$\pm0.18$	& 9.00$\pm0.17$	& 7.83$\pm0.25$	& 5.05$\pm0.24$	& 9.03 					& 1.00E-04	& 8.63E-03	& 6.67E-04 \\
\object{HD 66811}		& 0.16		& 7.60$\pm0.25$	& 9.10$\pm0.17$	& 8.13$\pm0.30$	& 5.05$\pm0.26$	& 9.16					& 2.86E-04	& 1.05E-02	& 1.30E-03 \\
\object{HD 190429A}	& 0.15		& 7.09$\pm0.18$	& 8.92$\pm0.16$	& 7.83$\pm0.26$	& 5.26$\pm0.20$	& 8.96					& 9.10E-05 	& 7.27E-03	& 6.69E-04 \\
\object{HD 15570}		& 0.10		& 7.52$\pm0.16$	& 8.62$\pm0.15$	& 8.30$\pm0.24$	& 5.06$\pm0.18$	& 8.81					& 3.27E-04	& 4.79E-03	& 2.63E-03 \\
\object{HD 14947}		& 0.12		& 8.30$\pm0.16$	& 8.78$\pm0.12$	& 8.13$\pm0.24$	& 5.25$\pm0.15$	& 8.97					& 1.66E-03	& 5.00E-03	& 1.44E-03 \\
\object{HD 210839}	& 0.12		& 8.22$\pm0.21$	& 8.70$\pm0.15$	& 8.48$\pm0.14$	& 4.70$\pm0.18$	& 8.99					& 1.32E-03	& 4.67E-03	& 3.23E-03 \\
\object{HD 163758}	& 0.15		& 8.56$\pm0.16$	& 8.78$\pm0.16$	& 8.36$\pm0.19$	& 5.02$\pm0.21$	& 9.08					& 2.72E-03	& 5.19E-03	& 2.25E-03 \\
\object{HD 192639}	& 0.15		& 8.17$\pm0.17$	& 8.76$\pm0.15$	& 8.61$\pm0.21$	& 5.07$\pm0.17$	& 9.05					& 1.09E-03	& 5.01E-03	& 4.01E-03 \\
Ori OB1 (B stars)		& \dots		& 8.35$\pm0.03$	& 7.82$\pm0.07$	& 8.77$\pm0.03$	& \dots			& 8.92					& 1.96E-03	& 6.77E-04	& 6.43E-03 \\
Sun 3D 		& 0.09		& 8.39$\pm0.05$	& 7.78$\pm0.05$	& 8.66$\pm0.05$	& 5.36$\pm0.03$  	& 8.88					& 2.16E-03	& 6.18E-04	& 5.35E-03 \\
\hline
\end{tabular}
\label{cno_tab}
   \begin{list}{}{}
\item By convention log $\epsilon$(X) = 12 + log [X/H]. For comparison, we give the solar chemical composition determined by \citet[][Sun 3D]{asplund05} and the chemical composition of B stars of the Ori OB1 association determined by \cite{simon10} and \cite{nieva11}. Mass-fractions are also indicated for \mbox{CNO}.
\end{list}
\end{table*}


\section{Discussion}
\label{disc_sect}
\subsection{Evolutionary status}
\label{hr_diag}
\begin{figure}[h]
   \centering
\includegraphics[scale=0.6, angle=0]{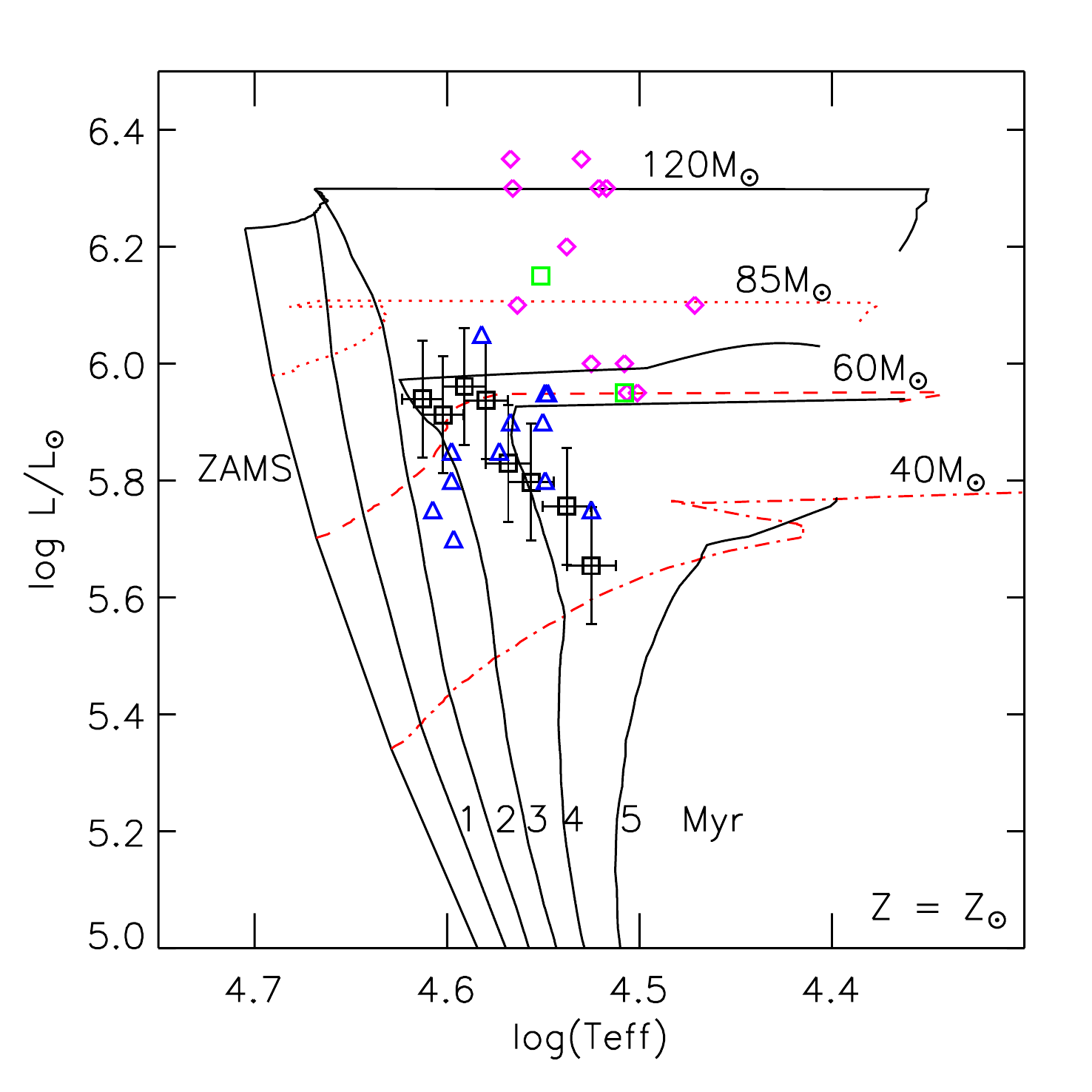}
    \vspace{-0.5cm}
      \caption[12cm]{H-R diagram showing the program stars (black squares) and the stars from \cite{martins08} for comparison. Blue triangles are ``normal" O4-6 supergiants, 
      green squares are ``extreme"  O4-6If$^{+}$ stars,  pink diamonds are WN7-9h objects.  
 Evolutionary tracks and isochrones are from \cite{meynet03} with an initial rotation of 300 \kms.}.
         \label{fig2}
   \end{figure}

We constructed the H-R diagram of the sample using effective temperatures and luminosities derived from the modeling, as well as
evolutionary tracks and isochrones from \cite{meynet03} (see Fig. \ref{fig2}), with an initial rotational velocity of  300 \kms\ and solar metallicity. 
The most remarkable feature of this diagram is how well it distinguishes the different spectral types in terms of masses and ages. 
The O4 supergiants cluster tightly around the 60 \msol\ track, 3 Myr isochrone. 
Later spectral types probe lower masses or more evolved regions of this diagram, but in any case, the stars are more massive than 40 \msol\ and younger than 5 Myr.

Alternately, the position of the O4 supergiants in this diagram might be explained by assuming that they are in fact much older, and
located on the 60 \msol\ track in the range 4.3 Myr to 4.4 Myr. 
However, the predicted properties of this track in this age range, like the hydrogen mass fraction for instance, are incompatible 
with the observed properties of the O4 supergiants. 
Indeed, we find $X(\mathrm{H}) = 55\%$ or higher while $X(\mathrm{H}) \approx 4\%$ or lower should be expected for a star with initial
mass 60 \msol\ after 4.3 Myr. 
This definitively rules out this possibility. For the same reason, the O4 stars of this sample could 
not be stars with initial masses slightly above 85 \msol, which would be around 3.7 Myr old, because the expected $X(\mathrm{H})$ is $5\%$ or less. 
The same comment holds for the later-type stars though for different initial masses. 

The luminosities were derived for the sample stars under the assumption that they are single. If they turn out to be binaries, the luminosity
of each component would have to drop, thus translating into lower initial masses and a somewhat different ages for the stars. 
As already stated (Sect. \ref{var_sect}), seven of the eight stars in our sample are most likely single and the known binary is well separated, 
therefore we were able to use photometry of the primary  for the luminosity calculation. 

When studying the most massive stars in the Arches cluster, \cite{martins08} found a significant overlap between the positions of 
the faintest WN7-9h stars and the most luminous, ``extreme", O supergiants in their sample, which turned out to have spectral types O4-6If$^{+}$. 
The positions of the stars of our sample in the H-R diagram overlap with the stars studied by \cite{martins08} of the same spectral class.
 We note that  \object{\hdo} and \object{\hdsixteen} show a late-type WN spectrum in the infrared $K$ band \citep{conti95} and 
 could be regarded as ``transitional" objects in terms of evolutionary properties, much like the extreme OIf$^{+}$ found by 
 \cite{martins08}, although they are systematically less luminous. 
 When merged with those from \cite{martins08}, our results confirm that the position of WN7-9h stars and O supergiants 
is distinct in the H-R diagram. 

\begin{figure}[h]
   \centering
   \includegraphics[scale=0.6, angle=0]{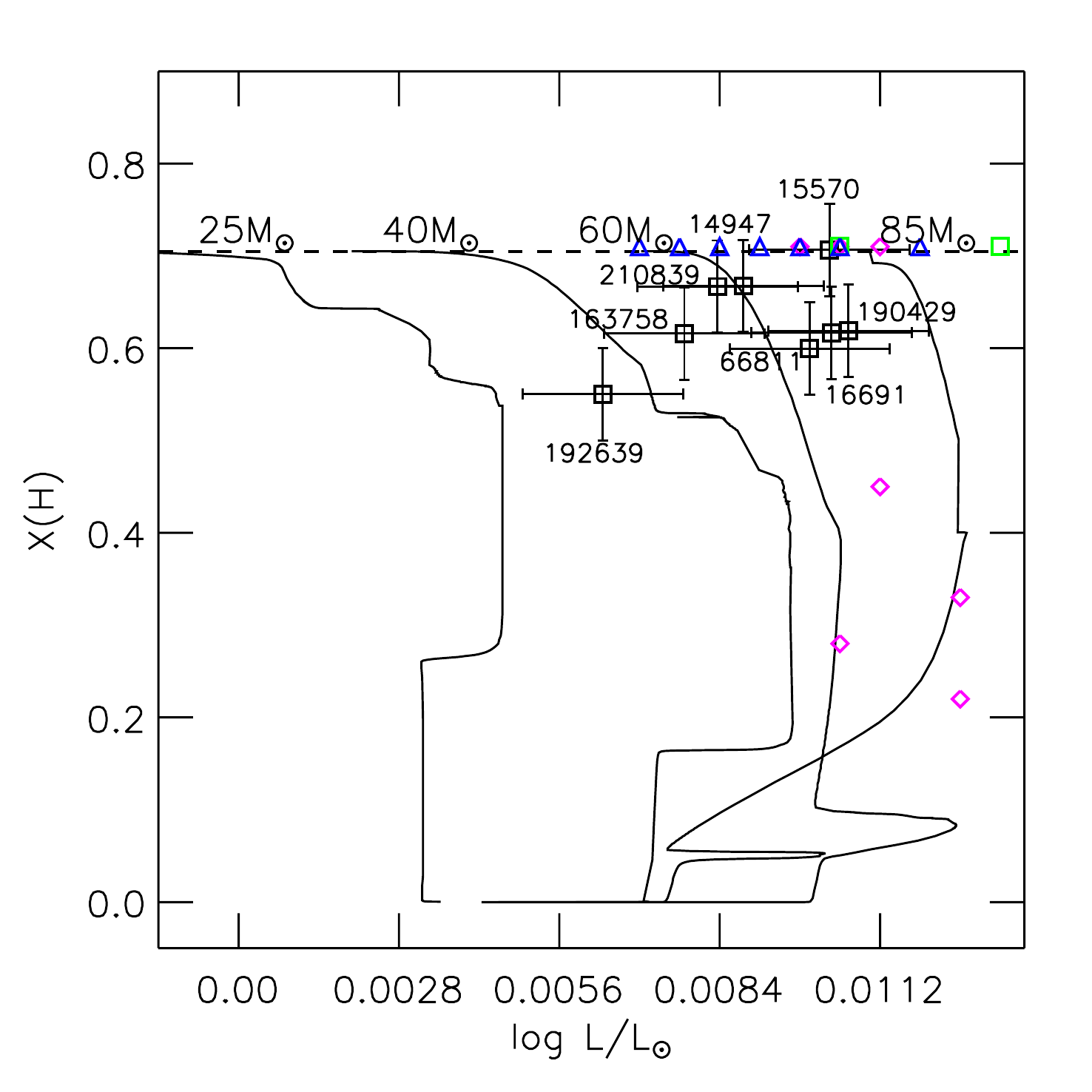}
   \vspace{-0.5cm}
      \caption[12cm]{Hydrogen mass-fraction as a function of luminosity. Symbols are the same as in Fig \ref{fig2}. Full lines are
      for  Geneva models with an initial rotational velocity of 300 \kms\  \citep{meynet03}. The dashed line indicates the initial value X(H)=0.705}
         \label{fig3}
   \end{figure}

Another insight into the evolutionary status of the stars is shown in Fig. \ref{fig3}, where the hydrogen mass fraction is plotted against the stellar luminosity. 
It is clear from Fig. \ref{fig3} that all stars in our sample are on the early part of their evolution away from the zero-age main-sequence, even the most extreme of them ( \object{\hdo}). 
Compared to the results from \cite{martins08} our stars appear to be more evolved (more hydrogen-depleted or equivalently helium-enriched) than their sample of supergiants, either normal or extreme. The reason for this difference is unclear, because helium lines in the  near-infrared are as sensitive to variations of helium content as lines in the optical range. A younger age of these cluster stars and possibly a higher metallicity could influence both their evolution and spectra.   

\subsection{Masses}
\label{mass_sect}
The spectroscopic masses were derived after correcting the effective gravity derived from the analysis by the effect of centrifugal forces caused by rotation (see Table \ref{tab_results}),
following the approach outlined in \cite{repolust04}. Assuming a random distribution of orientations for the rotational axis, the true surface gravity  $g_{c}$ is the sum of the effective gravity $g_{eff}$ and a 
correction term.
To a good approximation, the centrifugal acceleration averaged over the stellar disk is given by the projected rotational velocity, whence $g_{c} = g_{eff} + (v\sin\,i)^{2}/R_{*}$ \\
For the stars in our study, the corrections are small, with $\Delta \log g \leq 0.04$, which corresponds to a correction lower than 10\% for the actual mass. 

Evolutionary masses are obtained by bilinear interpolation between the tracks and isochrones in the H-R diagram. In Table \ref{tab_results}, we list masses at the corresponding stellar age and the original mass.
Spectroscopic masses and evolutionary masses (at the star age) generally agree well. 
The difference is below $10\%$ for  \object{\hdsixteen},  \object{\zetap},  \object{\hdo},  \object{\hdfourteen},  \object{\hdonesixtythree}, and  \object{\hdoneninetwo}, while it is around 20\% for 
 \object{\hdfifteen} and  \object{\lambcep}. 
For the most part, the observed differences could result from uncertainties in \logg. 
Assuming the stellar distance is known, the error in \logg\ is the dominant source of error in the spectroscopic mass, hence in the ratio of the evolutionary mass to the spectroscopic mass.
These uncertainties are partially attributable to the difficulties associated with rectifying echelle spectra, 
especially in the vicinity of broad \hyd\ lines.  
If accurate surface gravities are to be obtained, it is crucial that observing procedures are adopted to facilitate rectification, and to allow the accuracy of the rectification to be verified 
(e.g., measurement of multiple standards). Ideally, it would be useful to check the rectification against lower resolution (but still high-signal-to-noise) spectra. 
Obviously, distance determinations to Galactic O stars are quite uncertain (at least for field stars) and we cannot discard just yet the contribution of distance uncertainties as an important potential 
source of errors in spectroscopic masses.

Likewise, continued progress in stellar evolution theory confirms that stellar rotation is another fundamental physical process that affects the evolution of massive stars. 
The global agreement reached with spectroscopic masses furthermore suggests that, along with the introduction of NLTE line-blanketed stellar atmospheres, the assumptions adopted in the evolution models provide a realistic description of the actual conditions in young early-type massive objects.

 \object{\zetap} is very interesting in this context. It is likely seen near equator-on \citep[][and refs. therein]{harries00}, which implies that its true rotation rate should be 210 \kms\ 
or slightly higher. 
Applying the centrifugal correction to the measured surface gravity, we derive an actual (spectroscopic) mass of 56 \msol, while the initial mass, as derived from the star's location in the H-R diagram 
is 62 \msol\ (cf. Table \ref{tab_results}). 
The evolutionary model with an initial mass of 60 \msol\ and an initial rotation rate of \vsini\ = 300 \kms\ that best represents the position of  \object{\zetap} shows that after 3.2 Myr, the stellar mass should be $\approx$ 50 \msol\ and the rotation rate $\approx$ 108 \kms, more than 100 \kms\ lower than the measured equatorial velocity of the star. 
Either the initial rotation velocity of  \object{\zetap} was higher than 300 \kms\ or the star has been spun up during its evolution.
Estimating the real value for the initial rotation rate is not possible because too many other quantities are involved in the evolution of the star's angular momentum.
For instance the mass-loss rates, as measured from this study, are a factor two to three times lower than the values from \cite{vink00}, which were adopted in the evolution models of \cite{meynet03}. 
After 3.2 Myr a
60 \msol\ (initial mass) star should have lost only 3 \msol\ rather than 8 \msol\ (assuming the other parameters would still evolve according to the higher mass-loss scenario). 

\subsection{Surface abundances}
\label{abund_sect}

\subsubsection{Helium and nitrogen}
To assess the chemical evolutionary status of the stars in our sample, we compared the nitrogen to helium mass-fraction ratio as a function of effective temperature to predictions from stellar evolution models.
According to stellar evolution, nitrogen and helium are produced during the hydrogen burning phase but the relative increase of abundances is larger for nitrogen than for helium, since the latter is already abundant
in the atmosphere while the former is not. The N/He ratio quickly evolves from the initial value to a value corresponding to the CNO equilibrium, and stars for which the equilibrium has not been reached will present intermediate values of N/He. 

\begin{figure}[h]
   \centering
   \includegraphics[scale=0.6, angle=0]{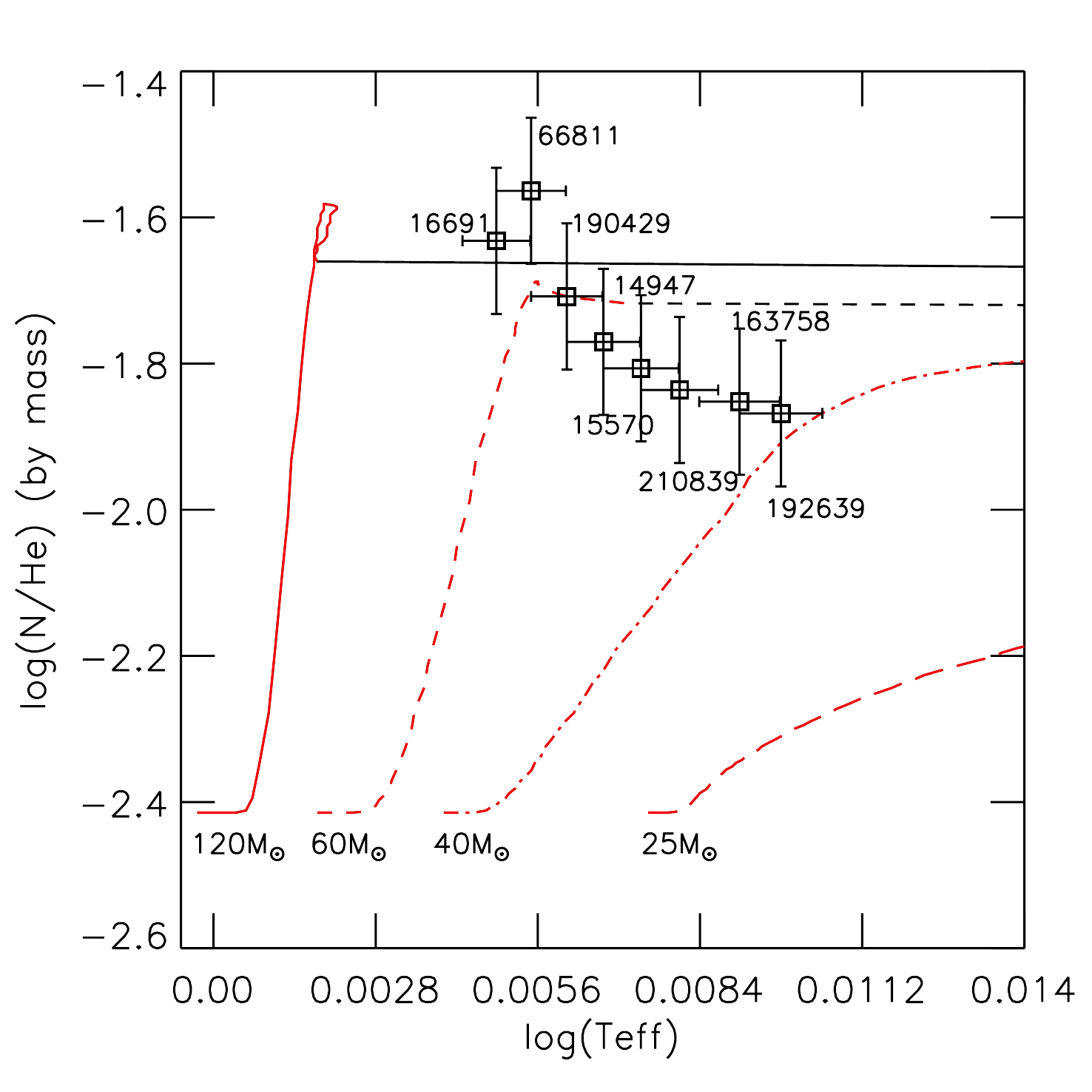}
   \vspace{-0.6cm}
      \caption[12cm]{Ratio of nitrogen-to-helium mass-fraction as a function of effective temperature, as predicted in Geneva models 
      for stars with an initial rotation rate of 300 \kms\  \citep{meynet03}, and as  derived from this study (symbols). 
      The red part of each line corresponds to the part of evolutionary tracks where X(H) $\geq$ 0.4. }
         \label{fig4}
   \end{figure}

Two of the O4 supergiants, \object{\hdsixteen} and  \object{\zetap}, seem slightly offset in this diagram, and masses higher than inferred from the H-R diagram would be needed to account for their position. 
In both cases, this is a consequence of the strong nitrogen enrichment we derive rather than from too small a helium mass-fraction, which we find to be normal for this range of masses and ages.
The case of  \object{\hdfifteen} is the exact opposite: a moderate nitrogen enrichment and low helium abundance shifts the star up in the diagram. 
Although \hdo\ is a binary, its properties seem well described by standard evolution models.
The location of these two stars in Fig.\,\ref{fig4} also indicates that they will soon reach the \mbox{CNO} equilibrium (possibly the case for  \object{\hdfourteen} as well). 

Similarly, the location of the later-type stars (cooler \teff) in this diagram is consistent, both qualitatively and quantitatively, with the predicted N/He fraction from stellar evolution models for objects in the mass range 50 - 40 \msol. 
These stars lie in a region of the plot where the N/He ratio is still rising, which indicates that the \mbox{CNO} equilibrium has not been attained. 

\subsubsection{Carbon versus nitrogen}
Together with the surface enrichment of nitrogen, the stars in our sample reveal carbon depletion, resulting in significant increases of the N/C ratios. 
This ratio is an excellent tracer of the chemical evolution of massive stars, because it increases substantially when CNO cycle-processed material appears at the
stellar surface.

\begin{figure}[h]
   \centering
   \includegraphics[scale=0.6, angle=0]{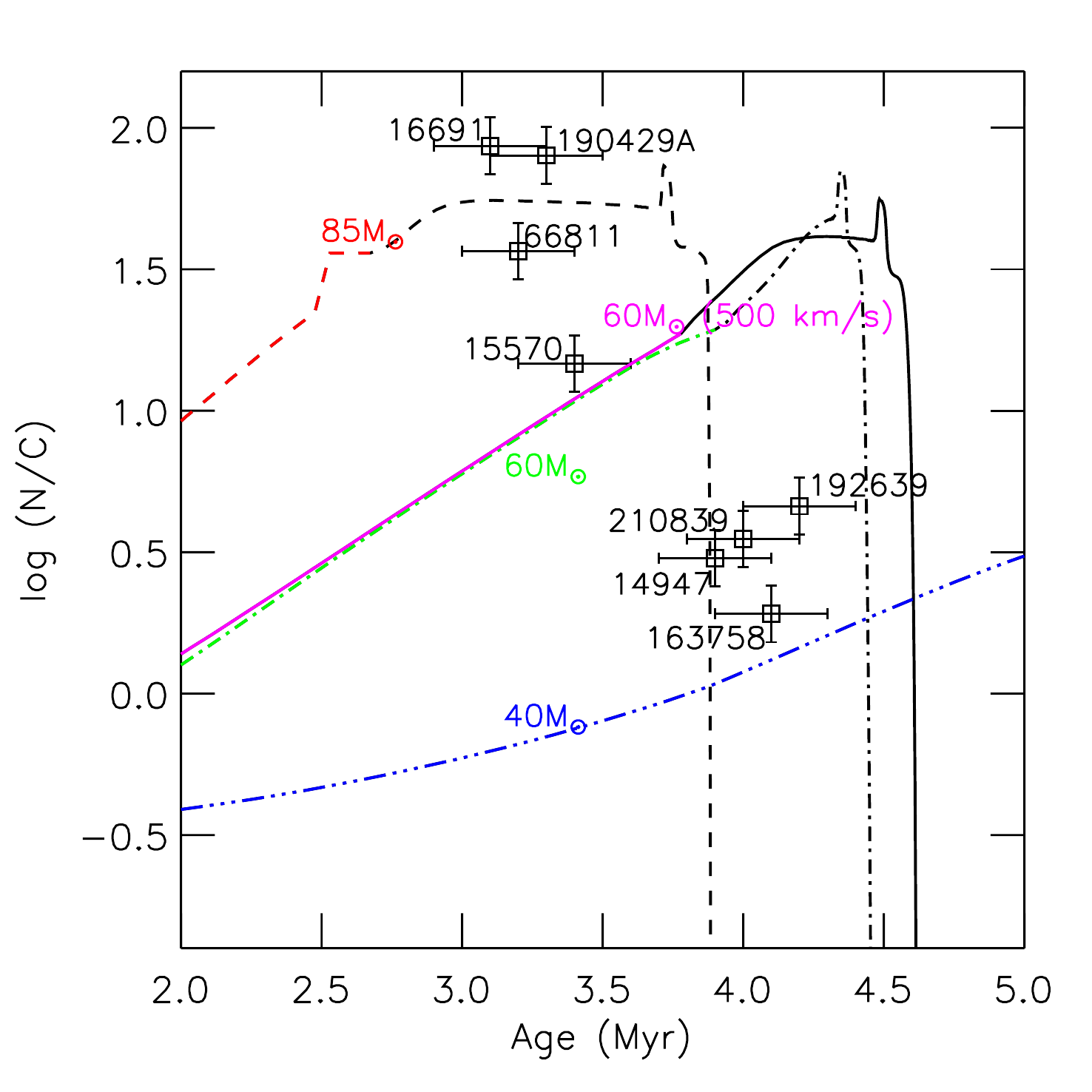}
      \caption[12cm]{Evolution of the ratio of nitrogen-to-carbon mass-fractions as a function of time in Geneva models 
      \citep{meynet03}. 
      The full (black) line is for a model with 60 \msol\ and an initial rotation velocity of 500 \kms . Overplotted in red, green, blue and pink are the parts of the tracks corresponding to the main-sequence phase
      (more exactly the part of the hydrogen burning phase where X(H) $\geq$ 0.4).}
         \label{fig5}
   \end{figure}

Figure \ref{fig5} presents a comparison between the measured N/C ratios (mass-fractions) and the predictions of the Geneva stellar evolution models, as a function of stellar age.
We can use this plot, together with the H-R diagram, to check whether a star is evolving according to the prediction of standard evolutionary tracks.
We find two different behaviors  -- one  for O4-type supergiants and the other for later-type (O5 to O7.5) supergiants. 
The measured N/C ratio of this later group behaves qualitatively and quantitatively as predicted by stellar evolution models. 
These stars are less massive and older (around 4 Myr or older) than the first group of objects between 3 and 3.5 Myr. 
In the early-type, younger, group, N/C ratios are higher than predicted for the corresponding masses. For  \object{\hdsixteen} and  \object{\hdo}, it is about ten times higher than predicted by any evolutionary model, which
additionally confirms the advanced evolutionary status of these objects.

\cite{maeder09} recently highlighted that N/C ratios depend on many variables including stellar age, mass, metallicity, rotational velocity, and multiplicity or magnetic fields and warned against 
an overinterpretation of measurements compared to theoretical predictions. However, taken at face value, these plots suggest that the global efficiency of mixing in stars more massive than 60 \msol\ is underestimated in stellar evolution models for a given rotation rate. 

In Fig. \ref{fig5} we also show the expected N/C ratio as a function of age for a 60 \msol\ model with an
initial rotation velocity of 500 \kms. The primary effect of increasing the initial rotation velocity is to maintain the N/C ratio at its maximum value up to 4.6 Myr (shift by 0.15 Myr compared to a model with 300 \kms). 
This rotation increase has no effect in the region of the plot where the O4 supergiants are located, i.e. around 3 Myr. 
Hence, a spread in rotation rates appears unable to account for the higher N/C ratios we observe. 

Two stars of our sample,  \object{\hdsixtysix} and  \object{\hdtwohundten}, have high rotation rates with \vsini\ $\ge$ 210 \kms.  \object{\hdsixteen} has a lower \vsini\  ($\approx$ 135 \kms) but presents several spectral characteristics 
usually associated with a high rotation 
rate (\niii\  and \ciii, \heii\ \lb 4686, \halpha, see Sect. \ref{hdsixt_sect} and Appendix A).  \object{\hdsixtysix} and  \object{\hdsixteen} have the strongest nitrogen enrichment of the 
sample and  \object{\hdsixteen} the lowest carbon abundance (hence the highest N/C ratio). Both stars support the idea that rotational mixing is responsible for the surface composition changes. 
On the other hand,  \object{\hdtwohundten}, which also has a high rotation rate, goes against this conclusion because it has the lowest nitrogen enrichment and marginal carbon depletion. 
The other stars have intermediate \vsini\ and no correlation emerges
with the N/C ratio. This contradictory result is reminiscent of the findings of \cite{hunter07}, though we stress that our sample is too small for a statistical interpretation. 

\begin{figure}[h]
   \centering
   \includegraphics[scale=0.6, angle=0]{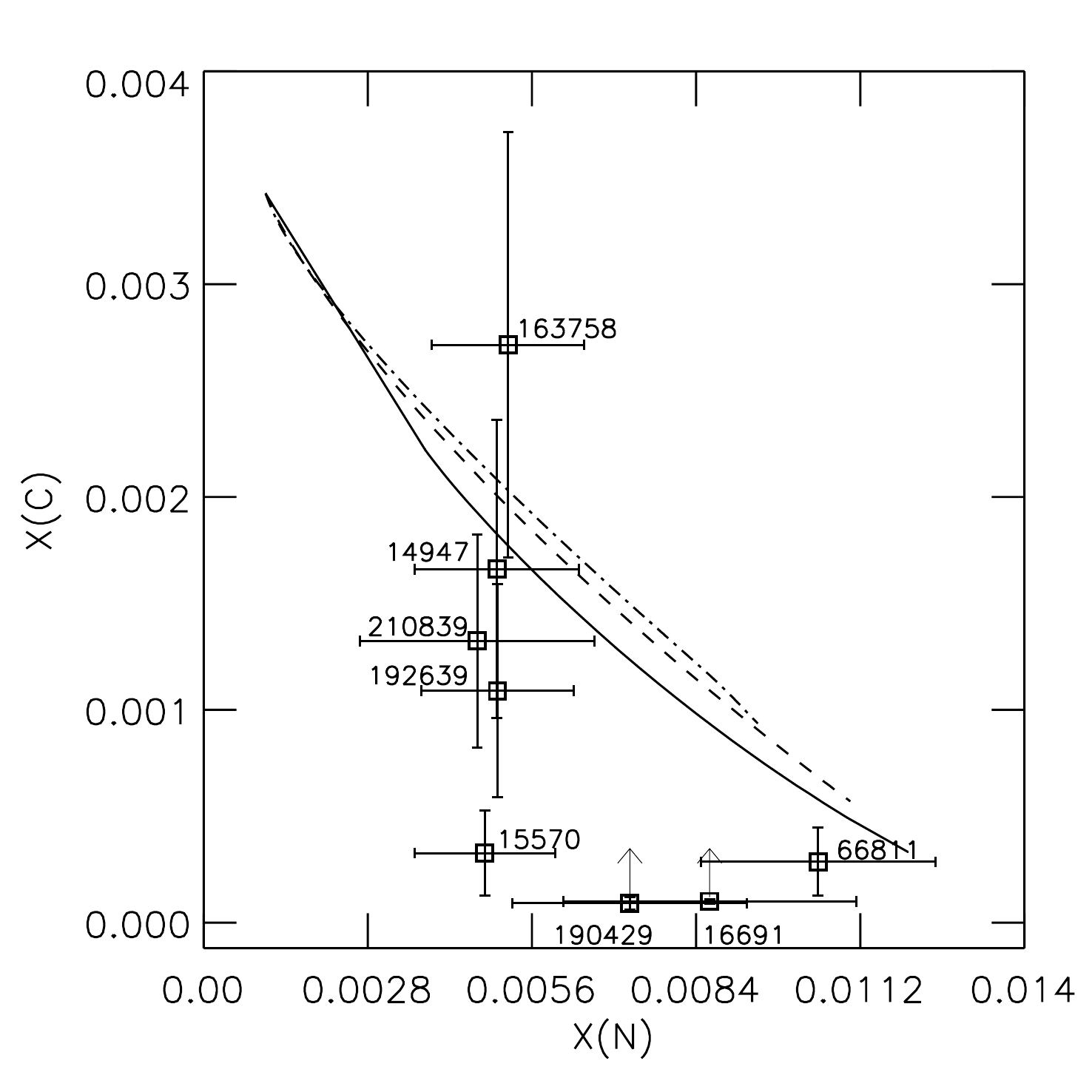}
      \caption[12cm]{Carbon mass-fraction as function of nitrogen mass-fraction, as predicted by Geneva models 
      with an initial rotation velocity of 300 \kms\  \citep{meynet03}. Solid line is for 85 \msol, dashed for 60 \msol, and dot-dashed for 40 \msol. 
      Symbols are for values derived from this study.}
         \label{fig6}
   \end{figure}

Let us now compare how the carbon and nitrogen mass-fractions relate to each other in a diagram where $X(\mathrm{C})$ is plotted as a function
of $X(\mathrm{N})$. It is expected that in the early phases of stellar evolution, $X(\mathrm{C})$ decreases, while $X(\mathrm{N})$ is built up at the expense of carbon. 
Fig.\,\ref{fig6} shows that, as expected, stars with higher nitrogen content show stronger carbon depletion. 
For the measured stellar masses, and for stars still evolving in the hydrogen burning phase, the observed range of the nitrogen mass-fractions is qualitatively compatible with the predictions of stellar evolution models.  However, quantitative differences are observed. 
Six out of eight stars cluster in a region where the carbon depletion is too strong for the measured nitrogen enrichment (or alternatively, nitrogen enrichment is too low for the measured carbon depletion). 
Once again, we observe a dichotomy in the behavior of the O4 supergiants and the O5 - O7.5 supergiants. 
The former group of stars, with masses $\approx$ 60 \msol,
shows the strongest carbon depletion, which is so extreme in two cases that it would not be reached in any model at any evolutionary phase. In this context  it is interesting that these stars
show  WN spectral features in their near-infrared spectra \citep{conti95}.
Only  \object{\hdfourteen} (and  \object{\hdonesixtythree} to a lesser extent, given its very peculiar abundances) agree qualitatively and quantitatively with the 40 - 60 \msol\ models that are expected to best describe the properties of this sample. 
This agreement probably means that both stars are actually very close to \mbox{CNO} equilibrium. Furthermore, the position of  \object{\hdonesixtythree} in Fig. \ref{fig6} is consistent with its carbon-rich status as quoted 
from observations by \cite{gomez07}. 
 
\subsubsection{\mbox{CNO}}
In massive stars, the total mass fraction of \mbox{CNO} remains roughly constant during the \mbox{CNO} cycles. In the early phases of evolution, 
the CN-equilibrium is achieved first  while the oxygen content remains about constant, and
somewhat later enters the burning cycle. The full CNO-equilibrium is achieved only after a significant amount of hydrogen has been burnt in the core. 
Changes in the oxygen surface abundance are only expected in later stages in the more massive and faster rotating models.

\begin{figure}[h]
   \centering
   \includegraphics[scale=0.6, angle=0]{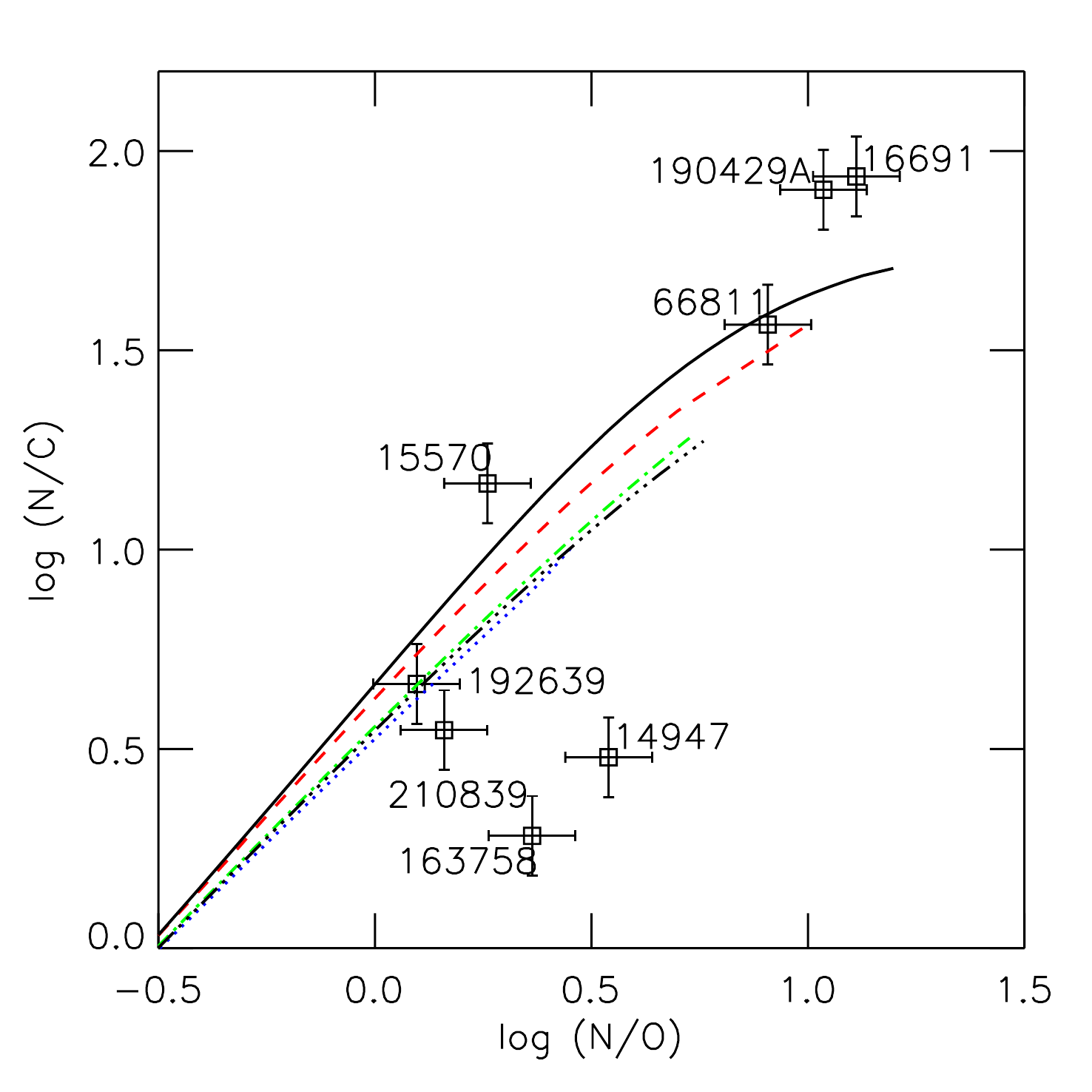}
      \caption[12cm]{N/O vs. N/C abundance ratio (by mass) measured for the sample stars and compared  to the ratios predicted in Geneva models 
      \citep{meynet03}. Solid line is for 120 \msol, dashed red line is for 85 \msol, dot-dashed green line is for 60 \msol, and dotted blue line is for 40 \msol.  
      The black dot-dot-dot-dashed line is for a model with 60 \msol and \vsini\ = 500 \kms. Only the part of evolutionary tracks corresponding to the evolutionary phase where X(H) $\geq$ 0.4 are plotted. }
         \label{fig7}
   \end{figure}

In Fig. \ref{fig7}, we compare our N/C versus N/O ratios to those predicted by stellar evolution models for different masses
and X(H) $\geq$ 0.4  \citep{meynet03}. 
This plot indicates whether the variations of N/C and N/O follow the trend expected for material processed through the \mbox{CNO} cycle.
The amplitudes of the departure from the initial abundance ratios are notoriously sensitive to differences in rotational velocities, masses, and ages.

For the O4 supergiants, N/C ratios are too high for the measured N/O ratios, compared to predictions by the evolution models. 
For three stars, their location in Fig. \ref{fig7} indicates that their N/C and N/O ratios are consistent with more 
advanced evolutionary status (WNs) than actually derived. 
The fourth star,  \object{\hdfifteen}, exhibits different properties: although  its N/C and N/O ratios are slightly higher than those predicted by evolutionary models for 60 \msol,  they
are nevertheless compatible with the its evolutionary status.
Within this group of stars, the observed behavior is qualitatively compatible with the expected evolution: the 
more massive stars are farther along the evolutionary tracks than the less massive ones, assuming that they are in the same evolutionary phase/status (as suggested by
the H-R diagram, cf. Fig. \ref{fig2}).

The opposite behavior is observed for later- and mid-type 
supergiants, whose measured N/C ratios are lower than predicted from evolution models for stars in the 
40 - 60 \msol\ range (for the measured N/O ratios). 

The case of  \object{\hdonesixtythree} is noteworthy because it presents unusual abundance patterns (see Sect. \ref{hdonesixtythree_sect}), including a super-solar
carbon abundance and a very high nitrogen content while being only mildly oxygen-depleted. Such patterns likely result from a non-standard (non solar-like) original mixture of \mbox{CNO} elements,
and the relative enhancement of the nitrogen mass-fraction depends on the initial carbon abundance available for conversion during the \mbox{CNO}-cycle. 

Figure \ref{fig6} indicates that the nitrogen enrichment in  \object{\hdfourteen} is consistent with the carbon depletion predicted for a 50 \msol\ star. 
Moreover, Fig. \ref{fig5} indicates that the measured N/C ratio is consistent with the age of the star. 
From these two results, we conclude that the measured N/C ratio is compatible with the predictions of stellar evolution models.
Therefore, the position of  \object{\hdfourteen} in Figure \ref{fig7} suggests that N/O is too high compared to evolutionary predictions, or in other words, that
the measured oxygen abundance is too low (lower than predicted). 
If the oxygen depletion is not the result of abnormal initial abundances, it could be explained if deeper layers of the star are exposed to the surface, for example 
by deeper mixing than predicted by the rotating models.  
Alternatively it could be explained by stripping of the star by winds or a binary 
companion; or by the accretion of oxygen-depleted layers from a former companion \citep{langer08, demink11}.  

We note that the total of the \mbox{CNO} abundances of the sample stars (see Table \ref{cno_tab}) shows that with a mean combined abundance 
$\epsilon\ (\Sigma \mbox{CNO})$ = 9.02, the initial \mbox{CNO} mixture for several objects is potentially different from the assumed standard cosmic abundances
adopted for the computation of the stellar evolution models by \cite{meynet03}.  Because 
there is evidence for a significant spread in the metallicity for solar-type stars in the thin disk ($\sigma$[Fe/H] = 0.191 dex, Fuhrmann 2011),
there might be a spread in the actual \fe\ abundances in O-type stars as well.  Although solar values were assumed in our study, they might not be relevant 
for all our sample stars. 

Although striking, the scatter in N/C and N/O ratios for this sample is real, as indicated by their very similar spectra (hence physical parameters) but different \mbox{CNO} line intensities. 
For the first group of stars (\object{\hdsixteen},  \object{\hdo},  \object{\hdsixtysix},  \object{\hdfifteen}), the amplitude of the deviation from the models is low enough that 
differences in the initial abundances could help reconcile the outputs from stellar evolution models and our measurements. The slope of the theoretical tracks is 
very sensitive to the adopted initial mixture \cite[see, e.g., Fig. 3 of][]{przybilla10} and a span of N/O and N/C ratios is expected.  
Including magnetic fields in stellar evolution models could also change the speed and amplitude at which abundance patterns
are built up at the surface. A recent study by \cite{meynet11} confirmed that the inclusion of magnetic braking yields
very different results for the surface abundances, which depend on the assumed rotation law inside the star. 
For stars with differential rotation, mixing is faster and stronger, but for stars with solid-body rotation, mixing is inhibited because the stars spin down rapidly. 
In this case, surface abundance ratios are lower than in models without magnetic braking. However, \cite{meynet11} 
obtained results for a 10 \msol\ star and dedicated simulations would be necessary to confirm that more massive stars with magnetic fields can still be rotating relatively 
normally and  produce significant changes in surface abundance ratios after a few Myr. 

For the second group of (later) stars, an obvious solution appeals to models with higher rotation velocities, which exhibit more mixing.  
Again, if magnetic braking plays a role, mixing would be modified (see above). 
Alternatively, (close) binarity would affect the evolution and likely the surface abundances  of the stars (from mixing and possibly mass transfer). 
Binarity would also imply different luminosities, hence evolutionary masses. In any case, comparison with tailored models for magnetic braking 
(and/or lower masses) would be required for more quantitative conclusions.  

The interplay between rotational mixing and mass-loss as currently implemented in the models might not be efficient enough to account for the 
observed the \mbox{CNO} patterns.
We note in this context that we derive mass-loss rates up to three times lower than those used in the calculations of evolution models (see Sect. \ref{wind_section}). 
The actual metal content in the Geneva evolutionary models is also higher than what is measured in the solar neighborhood (0.02 versus 0.014) by \cite{przybilla08} 
(2008), which best corresponds to the global metallicity of the stars of the present sample. 
Both these facts undermine the direct comparison of our measurements with chemical patterns from stellar evolution models for an assumed \zzsol\ = 0.02.

The overall properties of these objects, including their location in the H-R diagram and hydrogen mass-fraction (see Fig. \ref{fig2}  and Fig. \ref{fig3}), indicate that they are observed at different stages of their chemical evolution, even though they are of very similar age.
The underlying drivers of the observed differences are unclear and could be manyfold -- from enhanced mixing related to binary 
evolution in an earlier phase prior to the supergiant stage  to evolution with rotation and a magnetic field \citep{meynet11}. Recent results by \cite{sana11a} suggest
that the binary fraction among massive stars is at least as high as 50\%. Although all but one star in the present sample are believed to be single, we cannot exclude 
 the possibility that some of them might have been 
members of a binary system in the past; this scenario is especially appealing for runaway stars like  \object{\zetap} (or \object{\hdsixteen}). On the other hand, the observed properties of the only massive binary in 
our sample,  \object{\hdo}, are qualitatively similar to those of the other stars of the same spectral type, and no clear influence of the secondary in this system is found. Nevertheless, since binarity is expected 
to have significant consequences for stellar evolution, tailored evolutionary models would be needed to compare to our measurements and to draw firmer conclusions.

\subsection{Wind properties}
\label{wind_section}
\subsubsection{Mass-loss rates and filling factors}
\begin{figure}
   \centering
\center{\includegraphics[scale=0.6, angle=0]{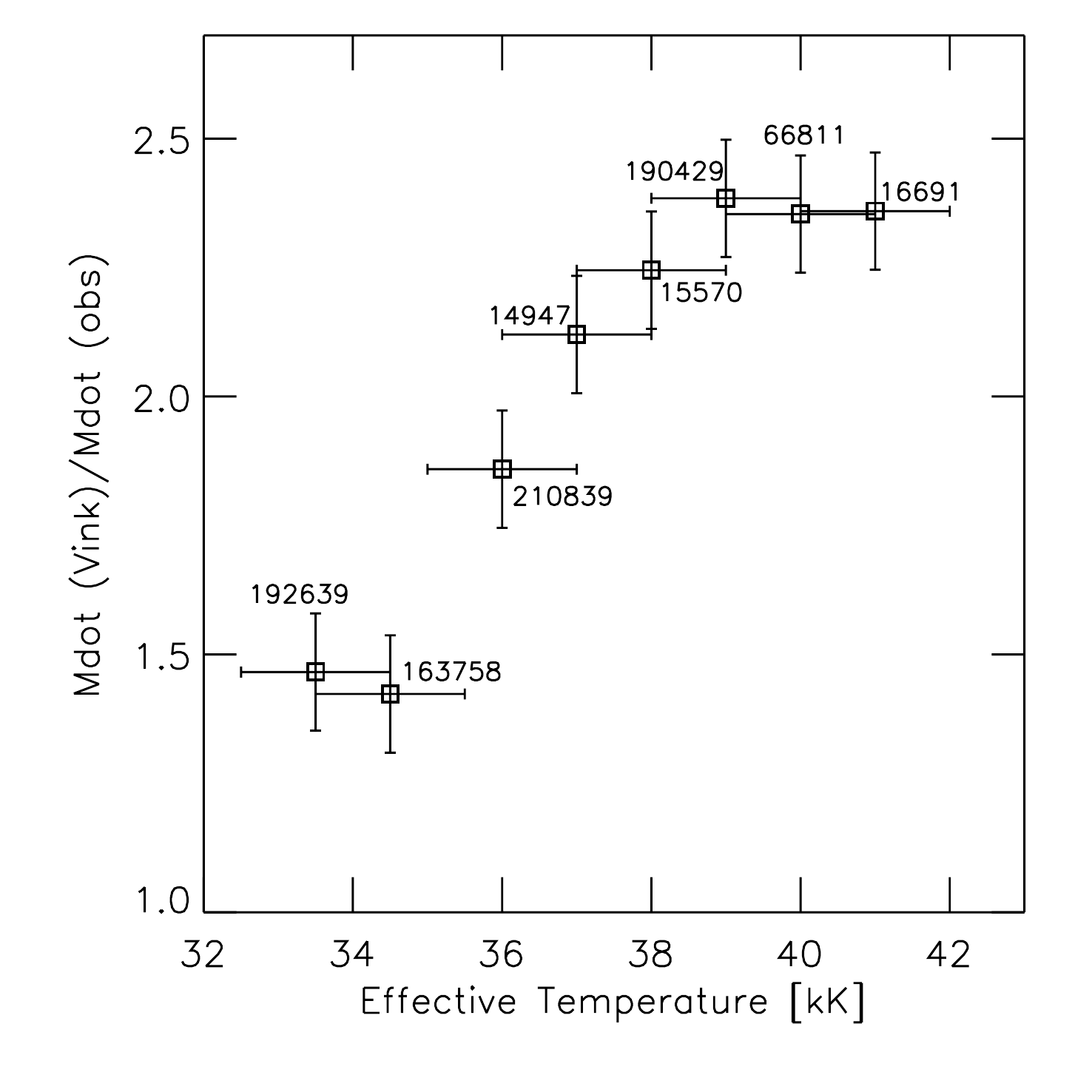}}
      \caption[12cm]{Ratio of theoretical calculations by \cite{vink00} to measured mass-loss rates as function of effective temperature. Theoretical mass-loss rates are computed for the best-fit parameters for each star. The uncertainties in the ratios are estimated using error bars for measured \mdot\ only. }
         \label{fig8}
   \end{figure}

Clumping as considered in this paper is often referred to as micro-clumping because it assumes that the clumps are optically thin.
The filling factors we derive cluster around \finf\ = 0.05 with a fairly limited dispersion $\pm 0.02$. This corresponds to mass-loss rates reduced by factors of four to six compared to the homogeneous values that would be derived from fitting \halpha.
More interestingly, for all stars our mass-loss rates are within a factor of three
of the predictions  by \cite{vink00}.
Fig. \ref{fig8} displays the ratio of predicted to measured mass-loss rates as a function of effective temperature. 
The ratios seem to increase with  \teff\ (the same trend is observed with luminosity). 
Theoretical mass-loss rates by \cite{vink00} decrease for decreasing 
effective temperature for fundamental stellar parameters typical of early-mid O-type supergiants. 
Therefore, Fig. \ref{fig8} shows that the theoretical mass-loss rates increase more rapidly with \teff\ (alternatively $L$) than measured in the present sample. 

As \teff\ increases, the maximum of the flux distribution gradually shifts to shorter wavelengths, where more lines are available to
contribute to the driving, thus increasing the mass-loss rates. 
The difference apparent in Fig. \ref{fig8} suggests that more driving is present at higher \teff\ in the theoretical
calculations. 
However, we used a very large number of species and ions for our calculations of hydrodynamics in the models, at least as many as were included in \cite{vink00}.
Since mass-loss rates are mostly set by ions of the iron group, it is unlikely that we missed important lines that would instead be accounted for in the theoretical calculations by \cite{vink00}. 

Small-scale clumping significantly alters the ionization balance in the wind \citep{hillier91, bouret05}. 
Depending on the temperature regime, dominant ions for different species might change, which could lead to a direct effect on the radiative driving. 
A recent study by \cite{muijres11} argued that the impact of clumping on the wind driving is an intricate function of the stellar temperature, the density inside the clumps and the physical size of the clumps. In the specific case of optically thin clumps, increased recombinations lead
to an increase in the mass-loss rate, as a consequence of  an increased line force. The amplitude of this effect is stronger as the clumping factor increases. 
This would be somewhat compensated for if porosity effects were included in their calculation, as a consequence of a decreased interaction between photons and the 
gas. All things considered, empirical mass-loss rates such as those we derived here  cannot be reconciled as yet to those resulting from theoretical consideration, even including 
clumping corrections.

A promising alternative comes from the work by \cite{lucy07, lucy10b, lucy10a}, who showed that turbulence in the photosphere reduces the wind driving force at the sonic point, thus strongly reducing 
the mass outflows, especially when compared to the calculations of \cite{vink00}. Calculations were made in the microturbulence framework, because the nature and cause of the actual turbulence are unknown. It is expected that microturbulence should depend on the effective temperature, hence the reduction in  mass-loss rates \citep[relative to][]{vink00} should be higher at higher \teff,
which might reduce the trend observed on Fig. \ref{fig8}.

Most likely, a proper description of stellar winds from massive stars should include optically thick clumps together with non-void interclump medium, and a non-monotonic velocity field \citep{sundqvist11}. 
This is especially important in accounting for velocity-porosity effects, which Sundqvist and collaborators conclude is especially relevant for resonance lines. 
Using a complex mixture of 1D radiation-hydrodynamics models to create 2D and 3D stochastic wind models, they computed synthetic line profiles for \lambcep, based on a Monte-Carlo radiative transfer code.
They were able to reproduce the observed profiles of \pv\ resonance lines and \halpha\ and to a lesser extent of \heii\ \lb4686. 
They derived a mass-loss rate of 1.5 \evi\ \msolyr, which is much higher than the mass-loss that they would derive assuming optically thin clumps only (4. \evii\ \msolyr), 
although this last value is obtained from the fit to the  \pv\ resonance lines only. 

The mass-loss rate we derive for  \object{\lambcep} is very close to the one quoted by \cite{sundqvist11}. Our fit to the wind lines is equally good when adopting lower-than-solar phosphorus abundance.  
Interestingly, both works fail to produce a double-peaked emission profile at \heii\ \lb4686. It turns out that this specific shape, 
(as well as the absorption trough seen in \halpha) is quite sensitive to the actual treatment of the rotation law in the wind. 
 \cite{sundqvist11} had to adopt a fairly large amount of macro-clumping to reproduce the central absorption in \halpha. 
The extent to which a more proper treatment of the fast rotation of this star would change this conclusion is unclear. 
We will present models of  \object{\lambcep} and  \object{\zetap} including a realistic treatment of rotation in a forthcoming paper, but we argue at this step that this ingredient is as important as clumping
(either optically thin or thick) to explain the observed shapes of \halpha\ and \heii\ \lb4686 and that neglecting it leads to overemphasize the effect of (large-scale) clumping. 
At the same time the volume-filling factor approach slightly overestimates the strength of \halpha\ \citep{sundqvist11}. All in all,  \object{\lambcep} might not be as representative of the influence of wind structures 
on emerging spectra as initially expected.  

Clumping was found to start close to the photosphere in all but three stars our of sample, two of which are fast rotators. For the third one, namely  \object{\hdsixteen},  we already mentioned that several emission profiles 
are characteristic of fast rotation, although not reflected by the \vsini. We realize that the need to start clumping farther out in the wind is mostly driven by the need to fit the absorption trough seen in \halpha\ in the case
of  \object{\zetap} and  \object{\lambcep}. Such a trough is not as clearly seen in  \object{\hdsixteen} (as \mdot\ is much higher) and the fit to \niv\ \lb1718 was the driver for the adopted \vcl. In any case, an accurate treatment of rotation is required for these objects to better 
investigate the actual properties of clumping.

\subsubsection{Optically thin clumps and the \pv\ problem}
We showed in previous studies \citep{bouret03, hillier03, bouret05} that the volume filling factor associated with wind clumping can be derived from
the simultaneous fit to \pv \lb\lb1118-1128, \ov\ \lb1376, \niv\ \lb1718, \heii\ \lb4686, and \halpha. The abundances for helium, nitrogen and oxygen were fixed from the analysis of photospheric lines; hence we had no other leverage but clumping and mass-loss rates to fit the wind lines listed above. The case of phosphorus is somewhat different because the FUV resonance doublet is the only available feature. 
Since the abundance of phosphorus is not expected to change over the lifetime of an individual star, 
a solar abundance should be appropriate. However, we had to use lower-than-solar phosphorus abundances, in addition to clumping, to improve the fit to \pv\ resonance doublet.
This result is reminiscent of our findings \citep[see][]{hillier03, bouret05} and although it might hint at a selective depletion of phosphorus in O supergiants 
with respect to the other metal abundances, it is also possibly related to the assumptions adopted to describe clumping in the models. 
The recent introduction of velocity porosity in radiative transfer calculations
by \cite{sundqvist11} showed that the \pv\ \lb\lb1118-1128 lines of  \object{\lambcep} can be fitted with a solar abundance of phosphorus. We note in this context that 
these lines in \object{\hdonesixtythree} do not reflect the ratio of oscillator strengths of the two lines, thus indicating that these lines are saturated. \cite{prinja10} concluded from their study of  \siiv\ \lb\lb1393-1403 
on a sample of B supergiants that velocity porosity could explain this behavior. \object{\hdonesixtythree} is therefore a very good candidate to investigate a more realistic treatment of wind inhomogeneities for  O supergiants.  

\begin{figure}[h]
\center{\includegraphics[scale=0.56, angle=0]{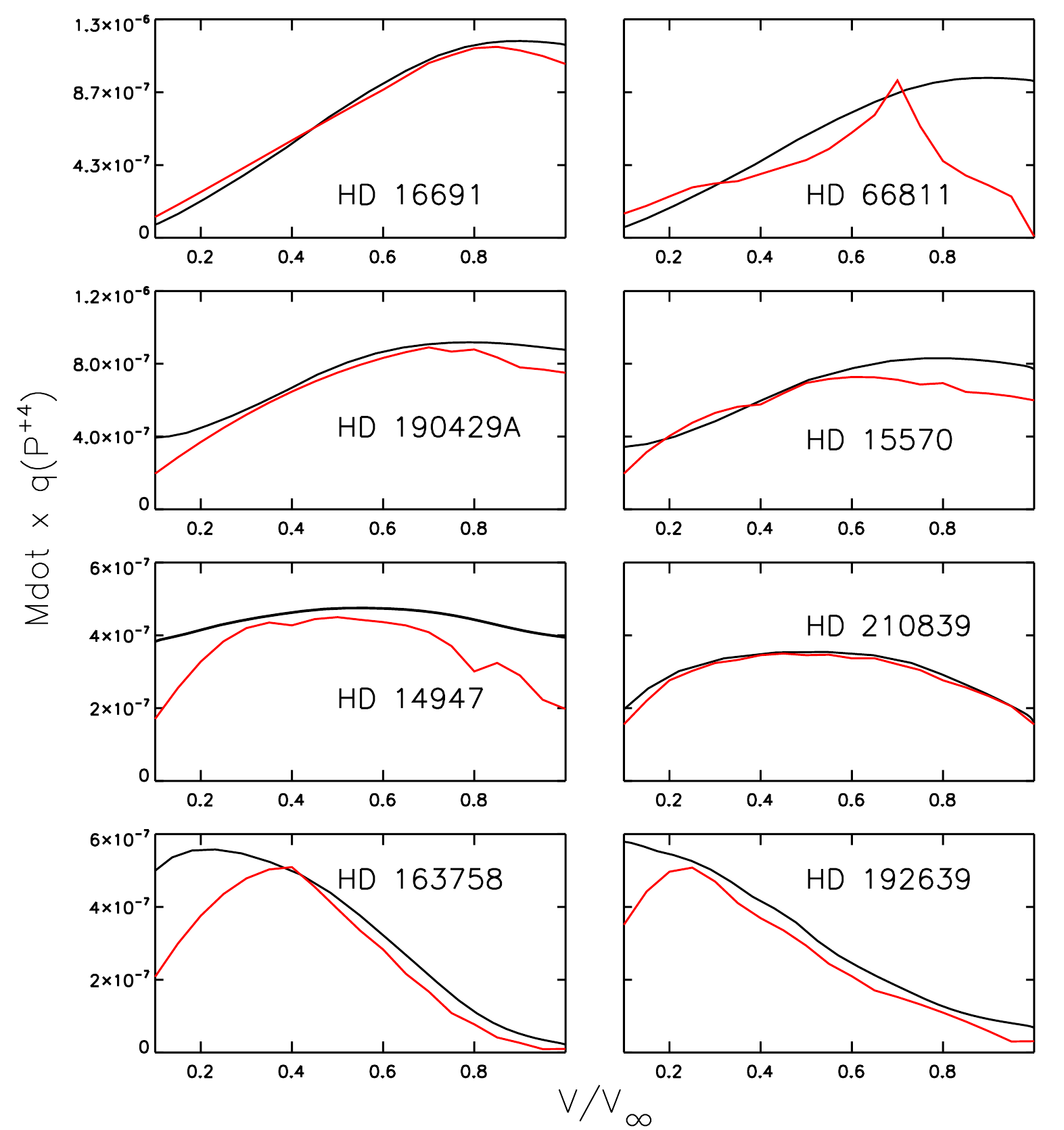}}
      \caption[12cm]{$\dot{M}{\rm q(P^{+4})}$ as a function of normalized velocity derived from the modeling with \cmfgen\ (black) or from SEI (red).}
         \label{fig10}
   \end{figure}

The primary consequence of being forced to reduce the phosphorus abundance in \cmfgen\ models 
is to change the radial optical depth of the 
resonance lines, hence their strength. 
Since the optical depth of the wind in one line is related to the ionization fraction of the corresponding 
ion ($\tau_{rad} \propto \dot{M}q_{\rm i}A_{\rm E}$ with $A_{\rm E}$ and $q_{\rm i}$ the abundance of element E and its ionization fraction 
for stage i), the need to decrease the phosphorus abundance compared to solar could be related to 
the incorrect determination of the wind ionization (i.e., over- production of the $\rm P^{+4}$ ion). 
To investigate the problem of phosphorus ionization in more details, we derived an empirical $\dot{M}{\rm q(P^{+4})}$ by profile fitting using the Sobolev with exact integrationÕÕ (SEI) method \citep{lamers87}. 
We refer to \cite{massa03} for a detailed presentation of the method. We used wind quantities such as \vinf\ and $\beta$ from the spectroscopic study (Table \ref{tab_results}). 
Figure \ref{fig10} compares the $\dot{M}{\rm q(P^{+4})}$ derived with this approach to those from the modeling with \cmfgen. 
Overall, the agreement between the quantities from both techniques is very satisfactory when a reduced phosphorus abundance is used.  In other words, in this case only, very similar optical 
depths are derived, while when $P = P_{\odot}$ is used, the optical depths from \cmfgen\ are too high.  
Although the global scaling with $P/P_{\odot} \leq 1$ improves the agreement, a clear trend for too high $\dot{M}{\rm q(P^{+4})}$ in \cmfgen\ is observed at high velocities 
(typically above v/\vinf $\geq 0.7$) for all but one star. The fit (Appendix A) shows that
the synthetic \cmfgen\ profiles are indeed too strong in absorption near the blue-edge of the absorption components of the P Cygni profiles of \pv\ lines\footnote{{ 
 \object{\zetap} shows by far the largest disagreement for empirical and theoretical $\dot{M}{\rm q(P^{+4})}$, which also translates to the poorest fit to the observations within the sample. The clear peak in the SEI $\dot{M}{\rm q(P^{+4})}$ distribution between $0.6 \leq$ v/\vinf $\leq 0.7$ 
might indicate the presence of a DAC in the wind of the star at the time of the observation}}.
This divergence between $\dot{M}{\rm q(P^{+4})}$ from \cmfgen\ and SEI is stronger for stars hotter than 36,000~K. 
It indicates that the ionization fraction of  ${\rm q(P^{+4})}$ is predicted too high in \cmfgen\ in the outer part of the winds.
The ionization fraction of $\rm P^{+4}$ in our models shows that \pv\ is a dominant ion below 0.5\vinf\ for all stars, consistent with results in \cite{fullerton06} and \cite{kritcka09}. 
In stars with \teff $\geq$ 37,000 K, \pv\ remains the dominant phosphorus ion throughout the wind. 

\cite{sundqvist11} suggested that X-rays have a significant influence on the ionization in the outer part of the winds. We performed several tests with different X-ray fluxes for these stars.  We did not find a significant influence of X-ray radiation on the ionization of phosphorus, unless very high X-ray fluxes were used, which would alter the ionization of other species as well. Such very high X-ray fluxes
provides a ``tail'' of XUV radiation that is strong enough to alter the \pv\ ionization fraction but the corresponding (emitted) \logLX\ are not consistent with the now well-defined relation \citep{sana07}.
We note that \cite{kritcka09} already concluded that X-ray radiation alone cannot change the ionization fraction of \pv\ significantly enough to avoid \mdot\ reduction when clumping is taken into account. 

On the other hand, \cite{waldron10} concluded that strong wind XUV-EUV emission line radiation could alter the ionization fraction of \pv, while remaining compatible with observational constraints.
Nevertheless, other important questions such as photoionizations from excited states, influence of other species and optical depth effects in the \heii\ continuum that might challenge their conclusion are ignored. 

Our fits to the \pv\ \lb\lb1118-1128 resonance lines are very satisfactory, given the relative simplicity of the adopted treatment of clumping. 
Again, the adopted phosphorus abundance is lower than solar but we seem to need the same P/P$_{\odot} = 0.5 \pm 0.2$ ratio for all stars, which
we can adopt as a fiducial value for subsequent modeling of other early-type supergiants.
We believe that wind parameters as derived from the present study, by fitting several wind profiles over a range of species and ionization
stages provide a realistic and robust description of the stellar winds. 
Nevertheless, more consistent descriptions of wind clumping will need to be considered in future modeling. 
The velocity porosity presented by \cite{sundqvist11} is promising, although the velocity structure of 
the hydrodynamic models do not exhibit the porosity in velocity space that is required to produce the
observed P-Cygni profile shapes, and reconcile H$\alpha$ and UV mass-loss rates. Other important questions
will need to be addressed, including a self-consistent calculation of the ionization structure, the extension to optically thick winds, 
and a realistic description (as yet lacking) of the structure of stellar winds.

\subsection{Line force consistency  check}

Since \cmfgen\ computes the full radiation field, we can accurately compute the radiative force. Consequently, we can
check whether the derived radiative force is consistent with that needed to drive the wind. For convenience we write the momentum equation as

\begin{equation}
g_{rad} =  v {dv \over dr} + {1 \over \rho}{dP_g \over dr} +g 
\label{eq_mom}
\end{equation}

\noindent
where $P_g$ is the gas pressure, $v$ is the velocity, $g=GM/r^2$, and $g_{rad}$ is the total radiative force, which is evaluated using
\begin{equation}
g_{rad}= {\chi_{f} L \over 4\pi c \rho r^2},
\end{equation}
where $\chi_{f}$ is the flux mean opacity.

\begin{figure*}
\includegraphics[scale=0.6, angle=0]{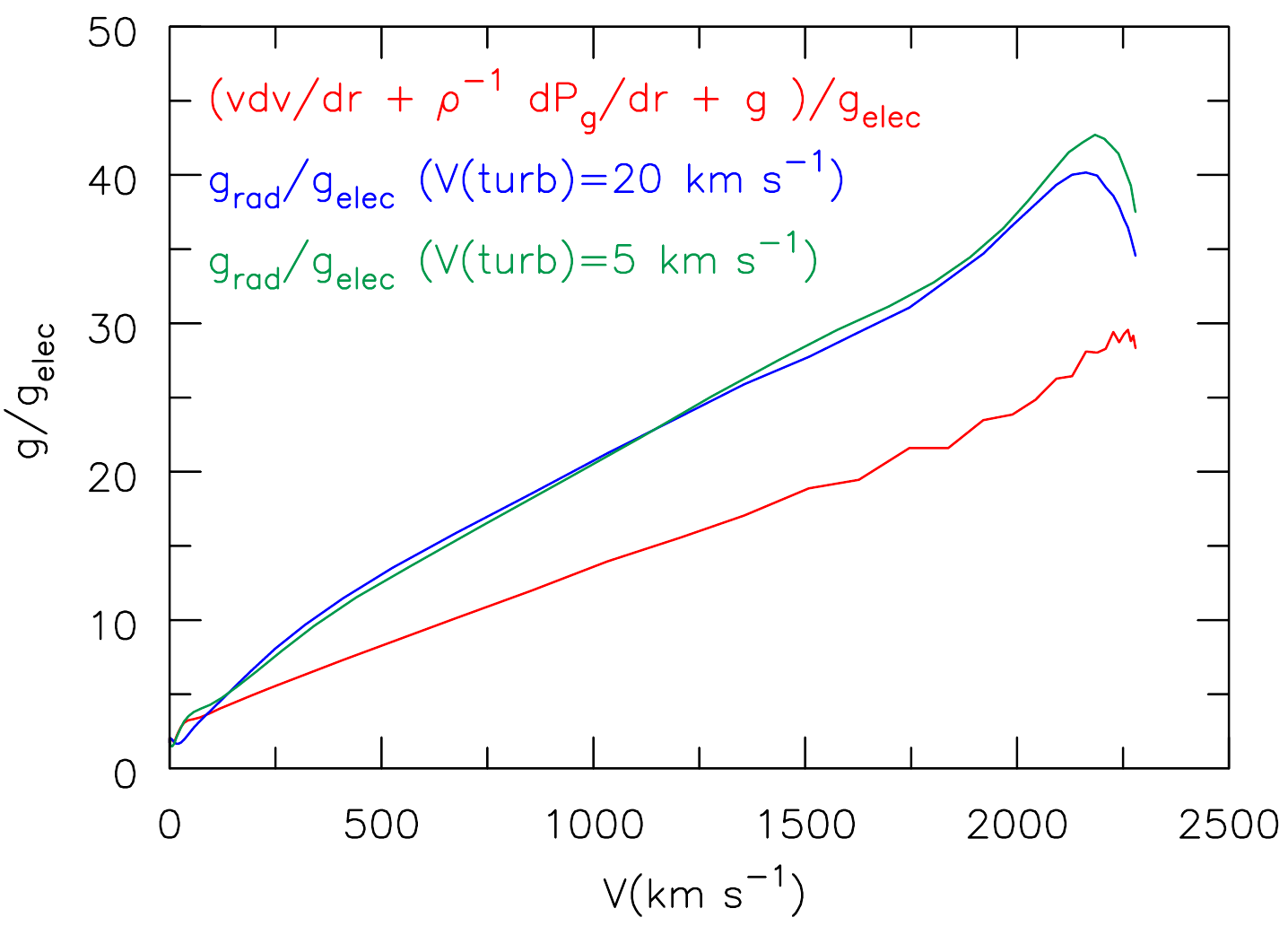}
\hoffset=4.0 \in \includegraphics[scale=0.6, angle=0]{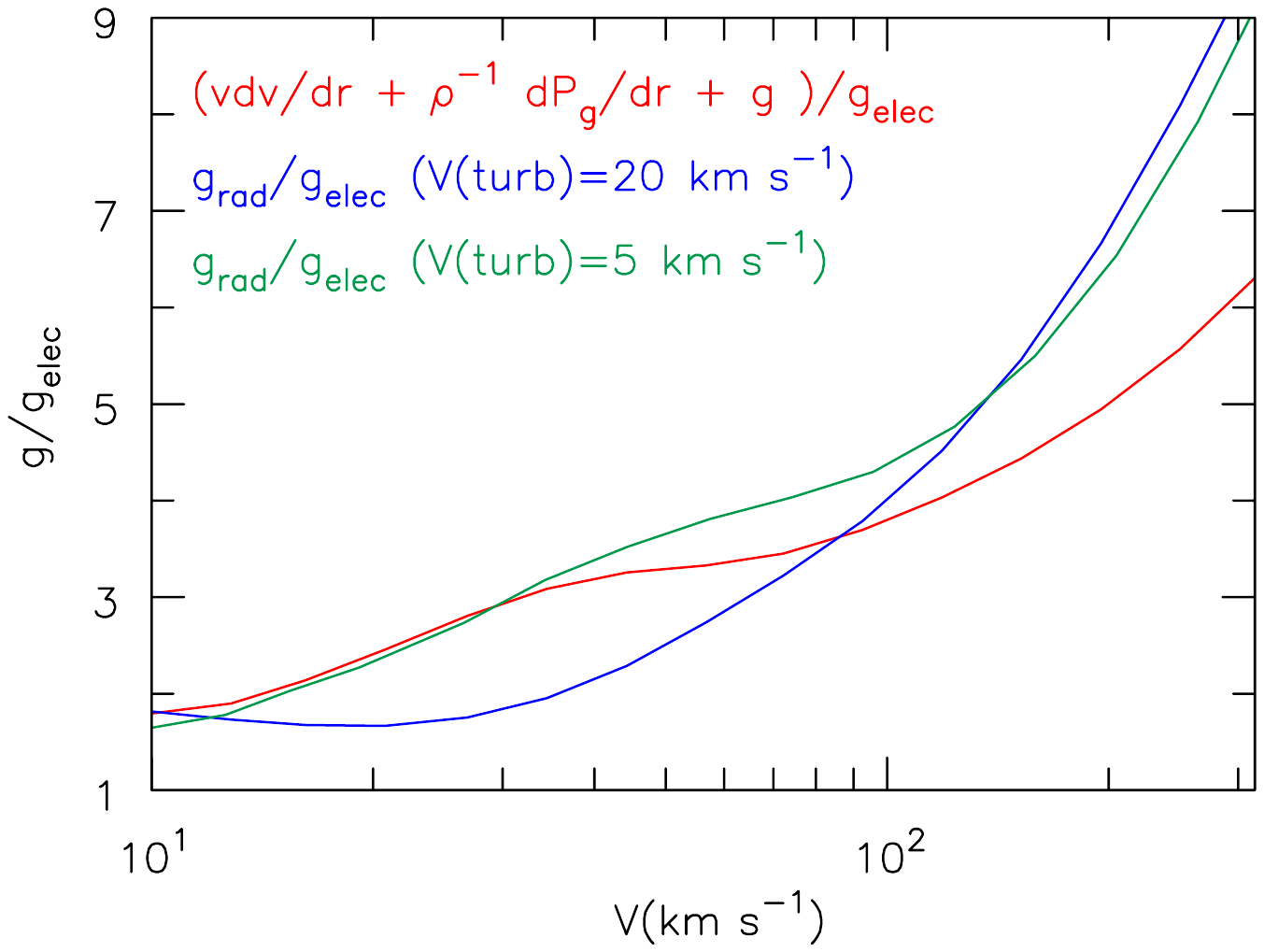}
      \caption[12cm]{Illustration comparing the radiative line force with other terms in the momentum equation.
      The model parameters are similar to those of  \object{\zetap}; \teff=41,000\,K and \msol$=1.7 \times 10^{-6}$\,\msolyr.
      The right-hand plot illustrates the sensitivity of the line force to the adopted microturbulent velocity  around the sonic point.
      Below 10\,\kms\ hydrostatic equilibrium is valid.}
         \label{fig11}
\end{figure*}

If the mass-loss rate, velocity law, and gravity in our models are consistent, the left-hand-side (LHS) of Eq. \ref{eq_mom} should be the same as the right-hand-side (RHS). In Fig.~\ref{fig11} we compare the LHS and RHS for a model similar to that found for  \object{\zetap}: \teff=41,000\,K, \mdot$=1.7 \times 10^{-6}$\,\msolyr, and \finf\  = 0.05 (which can be compared to our best-fit model of \teff=40,000\,K, \mdot$=2.0 \times 10^{-6}$\,\msolyr, and \finf\ = 0.05). As is readily apparent, our line force is too strong in most of the wind, indicating that the velocity law is incorrect, or more likely, the mass-loss rate is too low. The same conclusion is reached if we use the parameters found for the best-fit \object{\zetap} model. 

The simplest means of obtaining better consistency between the LHS and RHS would be to increase the mass-loss rate, and the simplest way of maintaining ``consistency'' with the spectral fit would be to simultaneously increase the volume filling factor $f$.
As expected, tests show that a moderate increase in the mass-loss rate to  $2.7  \times 10^{-6}$\,\msolyr\ and \finf\  = 0.1 yields a much better agreement. As discussed previously,  a slightly higher $f$ value might give consistent spectral fits if the effects of velocity porosity were taken into account.

Because we can perform this consistency check, we could, in principle, use the wind dynamics to derive an additional constraint on \mdot. However, there are several difficulties involved. First, it is very difficult to obtain agreement between the LHS and RHS around the sonic point ($v \sim 20$\,\kms) -- this is not surprising, since the critical point lies near the sonic point. We can only obtain a moderate agreement if we adopt a low microturbulent velocity of 5\,\kms, a value which is inconsistent with that required to fit the Fe lines in the UV, and with that generally inferred for O supergiants. 
Understanding the nature of this microturbulence is of crucial importance for understanding radiation driven winds. Moreover, we found it easier to achieve a fit when $\beta <1$ in this region, a value more in line with the predictions of radiation-driven wind theory. Second, there is increasing evidence that winds are clumped and dynamic - not steady-state as assumed by Eq. \ref{eq_mom}. Therefore, for full consistency, we need to run hydrodynamic simulations, and since  \object{\zetap} is a rapid rotator, this may necessitate 3D simulations. Third, there is the problem of velocity porosity, which will tend to reduce the line force for a given mass-loss rate and clumping. Fourth, we need to simultaneously reduce the abundances, because mass-loss rates are predicted to scale as Z$^{0.85}$\citep{vink01}. A systematic error in the abundances by a factor of 1.5 (which is not unrealistic) will cause a $\sim$ 40\% error in \mdot. 
Consequently, while we can conclude that the spectroscopically derived mass-loss rates, and the mass-loss rate inferred from the line force, are consistent 
to within a factor of 2 for  \object{\zetap}, it is almost impossible to obtain better consistency at the present time. It is also worth noting that it is necessary to include Ne, S, Ar, Cl etc. in the model calculations. While they do not have a major influence on the observed spectrum, they do significantly enhance the radiative line force in the wind.

In general, the derived spectroscopic mass-loss rates, and those inferred from the dynamics for O supergiant agree reasonably well. The same is not true for dwarf O stars, where the derived CMFGEN line force indicates a much higher mass-loss rate than that derived spectroscopically from the weak UV P~Cygni profiles \citep[e.g.,][]{martins12}.

\section{Conclusions}
\label{conclu_sect}
We have modeled the FUV-UV and optical spectra of a sample of eight O-type supergiants to investigate their surface abundances and wind properties. 
The results that we obtained can be summarized as follows:
\begin{itemize}
\item Supergiants with spectral types between O4 to O7.5 order along a well-defined evolutionary sequence, in terms of ages and masses, from younger and more massive to older stars with lower initial masses. O4 supergiants cluster around the 3 Myr isochrone and are more massive than 60 \msol, while from O5 to O7.5, stars have masses in the range 50 - 40 \msol\  and are 4$\pm 0.3$ Myr old. 
No mass-discrepancy is observed since spectroscopic and evolutionary masses agree within the uncertainties (and in any case within 20\%). 
Despite the high signal-to-noise ratios that can be achieved, the use of echelle spectra compromises the rectification of spectra, especially in the vicinity of the 
broad \hyd\ lines that are the primary diagnostics of \logg.
Improving \logg\ determinations is crucial for determining stellar masses by spectroscopic methods.
\item All stars exhibit significant enrichment of nitrogen, as expected from their evolutionary status. Carbon is depleted with respect to an initial abundance assumed to be solar-like for all but one star, namely  \object{\hdonesixtythree}. This star turned out to be carbon-rich (twice solar at least) as expected from its spectral classification. If depleted 
by the amount expected on the basis of its evolutionary status, its initial carbon abundance was likely
very high. All other stars have either a nitrogen enrichment that is too low for the measured carbon depletion, or alternatively too strong a carbon depletion for the measured nitrogen enrichment. 
\item The observed carbon and nitrogen mass-fractions are compatible with those expected from the models, for the measured stellar masses. On the other hand, the N/C ratios as a function of age are inconsistent with the theoretical predictions for the four earliest (O4 spectral type) stars of the sample. The efficiency of rotational mixing as a function of age is questioned for these stars. 
Other mechanisms might be needed to explain the observed patterns. 
A solution could be that the observed metallicity gradient within the galactic disk should be taken into account rather than assuming standard solar values for metals. Alternatively, a modified efficiency of mixing may be needed in stellar evolution models. 
\item Mass-loss rates derived using clumped models are lower than theoretical predictions by factors that range from 1.5 to 2.5. 
This difference is actually very reasonable given the known uncertainty on both estimates. 
The corresponding filling-factors associated with small-scale clumping are 0.05$\pm 0.02$. Clumping is found to start close to the photosphere for all but three stars, two of which are fast rotators, namely  \object{\lambcep} and  \object{\zetap}. Although its \vsini\ is smaller,  \object{\hdsixteen} behaves like these two fast rotators; several optical lines indicate that this star could indeed be a fast rotator as well, seen at a higher inclination. 
\item  The ratios of theoretical to observed mass-loss rates increase with luminosity (or \teff), which suggests that too much driving might be present in the models for the hotter and more massive stars. We found no 
convincing explanation for this behavior at this point. The need for a sub-solar phosphorus abundance to fit the \pv\ resonance line was investigated. Overall, empirical ionization fractions derived within the SEI 
framework agree well with the theoretical fractions from \cmfgen. Some remaining discrepancies are observed, especially in the outer regions of the winds (higher radial velocities). \pv\ is predicted
to be overabundant by \cmfgen\ in these circumstances. In any case, sub-solar phosphorus abundance are needed to artificially decrease the radial optical depth of the \pv\ lines. Although this
problem is directly related to the micro-clumping framework we work with, the needed abundance reduction is fairly constant and could be adopted as default for future work.             
 \end{itemize}

To reach a more complete, more quantitative, understanding of the observed trends of the photospheric and wind properties for Galactic supergiants, we will extend our analysis to later spectral types 
(down to the end of the O-type stars sequence). In forthcoming papers,  we will also explore the different technical problems we met during the present analysis, such as the formation of excited lines
of \mbox{CNO} elements that are fundamental for an abundance analysis;  and the extent to which incorporating rotation improves fits to key diagnostic features in the 
spectra of massive stars.
 
\begin{acknowledgements}
We thank the referee, Sergio Sim{\'o}n-D{\'{\i}}az, for useful comments and suggestions.
We also thank Selma de Mink for fruitful discussions about stellar evolution and abundances.
This research has made use of the SIMBAD database, operated at CDS, Strasbourg, France.
We want to acknowledge the valuable help of Hugues Sana with the \feros\ data of \zetap. 
STScI is operated by the Association of Universities for Research in Astronomy, Inc., under NASA contract NAS5-26555. 
Support for MAST for non-HST data is provided by the NASA Office of Space Science via grant NAG5-7584 and by other grants and contracts.
J.-C. Bouret is indebted to George Sonneborn for his invitation to work at NASA/GSFC during the completion of this work. We thank the French 
Agence Nationale de la Recherche (ANR) for financial support. 
D. J. Hillier acknowledges support from STScI theory grant HST-AR-11756.01.A and from NASA ADP Grant:  NNG04GC81G.

\end{acknowledgements}

\bibliographystyle{aa}
\bibliography{article}

\newpage
\pagestyle{empty}

\begin{appendix}
\section{Best fits}
\label{bestfit_sect}
In this appendix we present our best-fit models to the optical and UV spectra for the eight stars in our sample. 
The wavelength range between 1200 and 1225 \AA\ was not used in the spectral analysis since it suffers from a strong interstellar 
Lyman absorption. Nevertheless, we take this interstellar absorption into account by adding the effects of a representative hydrogen column density to our synthetic spectra in our final plots.

\begin{figure*}
\includegraphics[scale=0.51,angle=0]{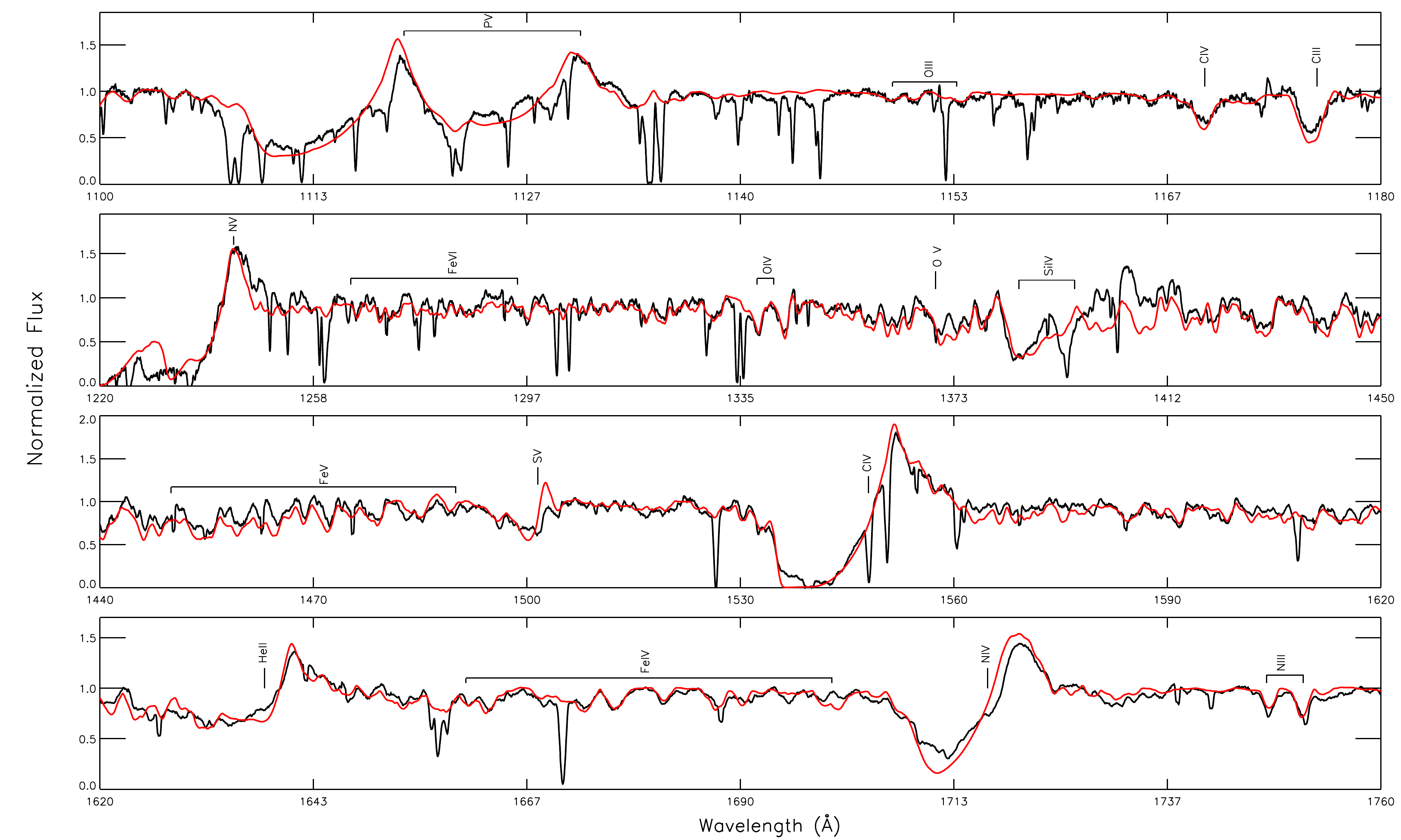}
      \caption[12cm]{Best-fit model for  \object{\hdo} (red line) compared to \fuse\ and \iue\ spectra (black line)}
         \label{Fig_hd190}
   \end{figure*}

\begin{figure*}
\includegraphics[scale=0.51, angle=0]{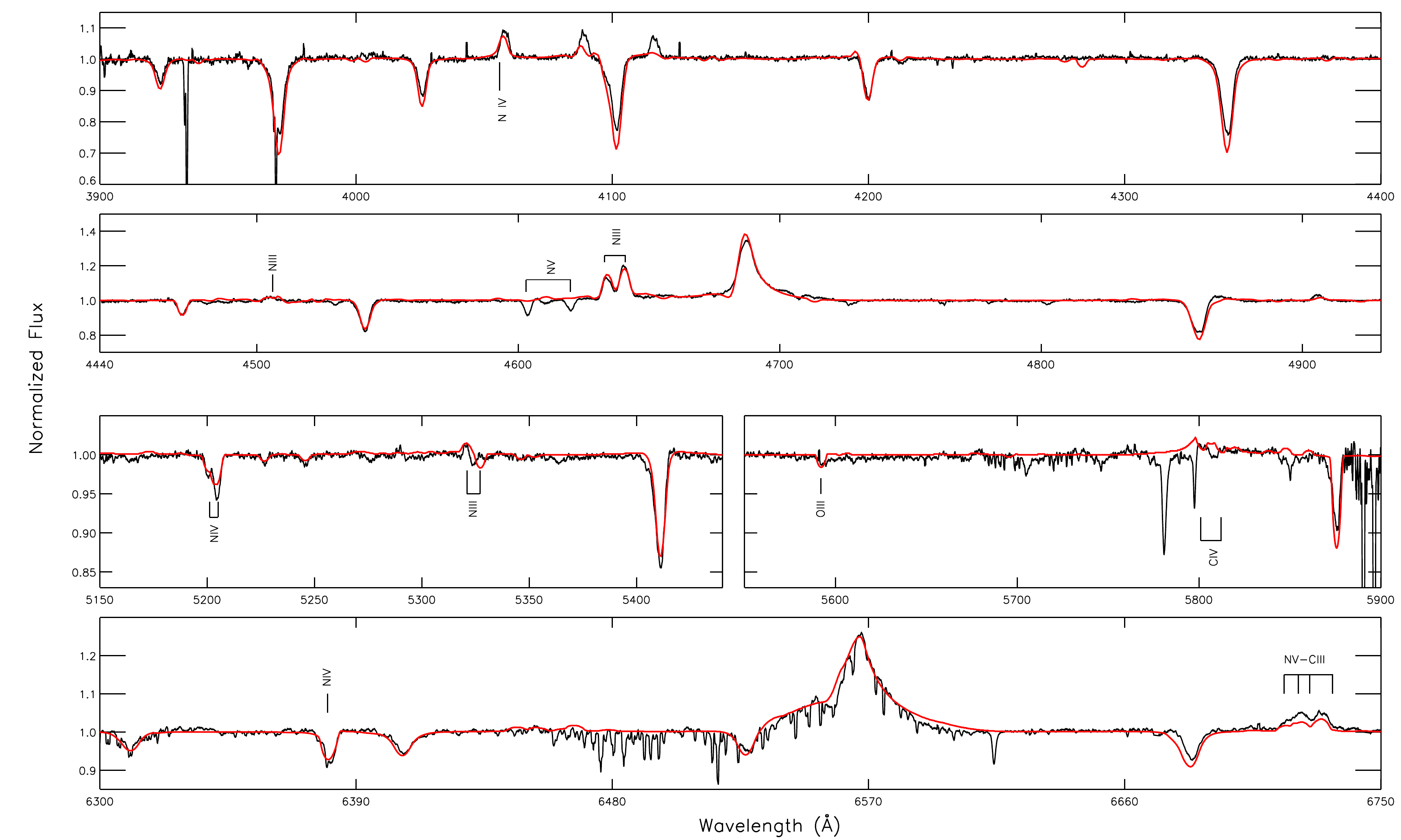}
      \caption[12cm]{Best-fit model for  \object{\hdo} (red line) compared to the \elodie\ spectrum (black line)}
         \label{Fig_hd190}
   \end{figure*}
   
\begin{figure*}
\includegraphics[scale=0.51, angle=0]{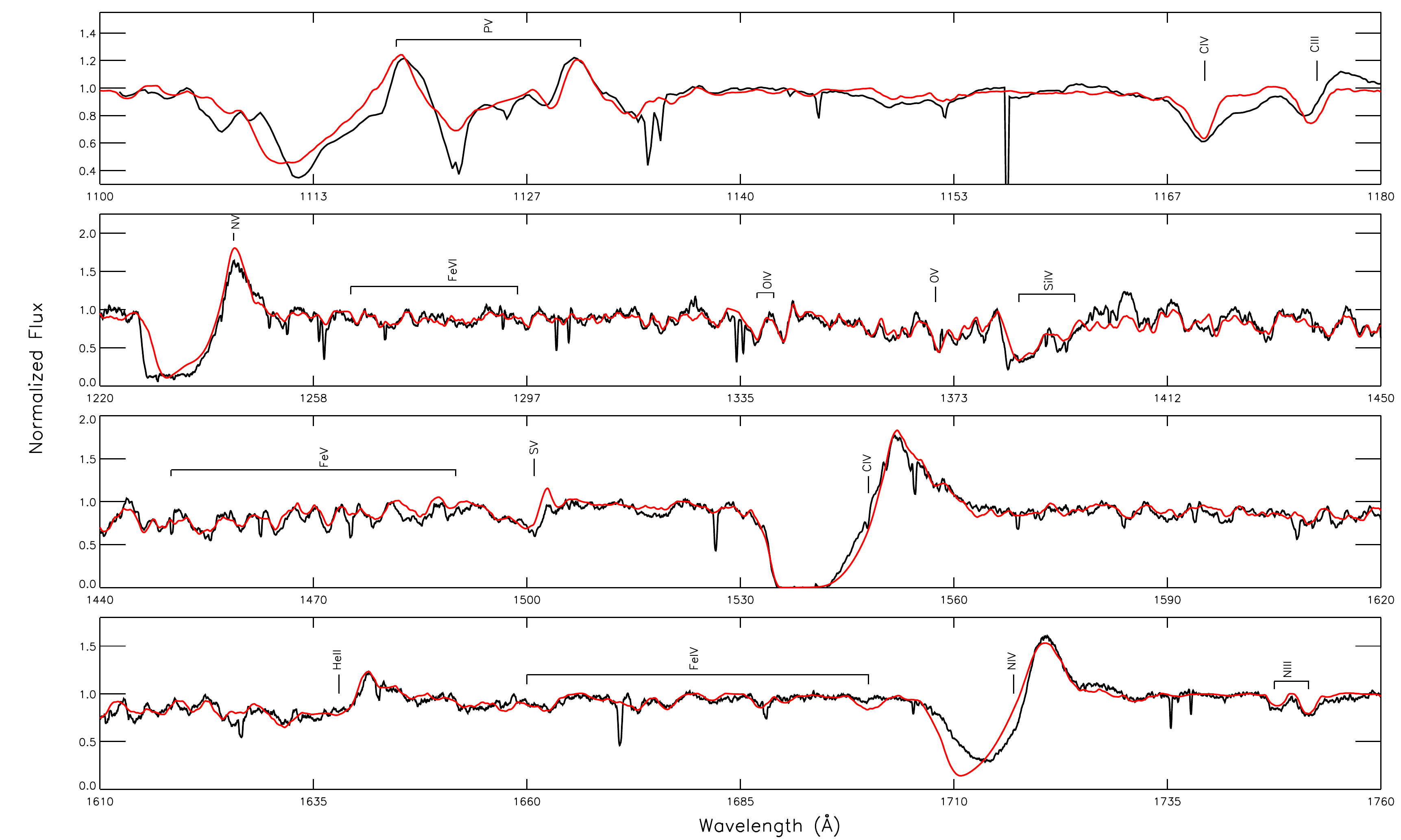}
      \caption[12cm]{Best-fit model for  \object{\hdsixtysix} (red line) compared to \copernicus\ and \iue\ spectra (black line)}
         \label{Fig_hh66}
   \end{figure*}

 \begin{figure*}
   \centering
\includegraphics[scale=0.51, angle=0]{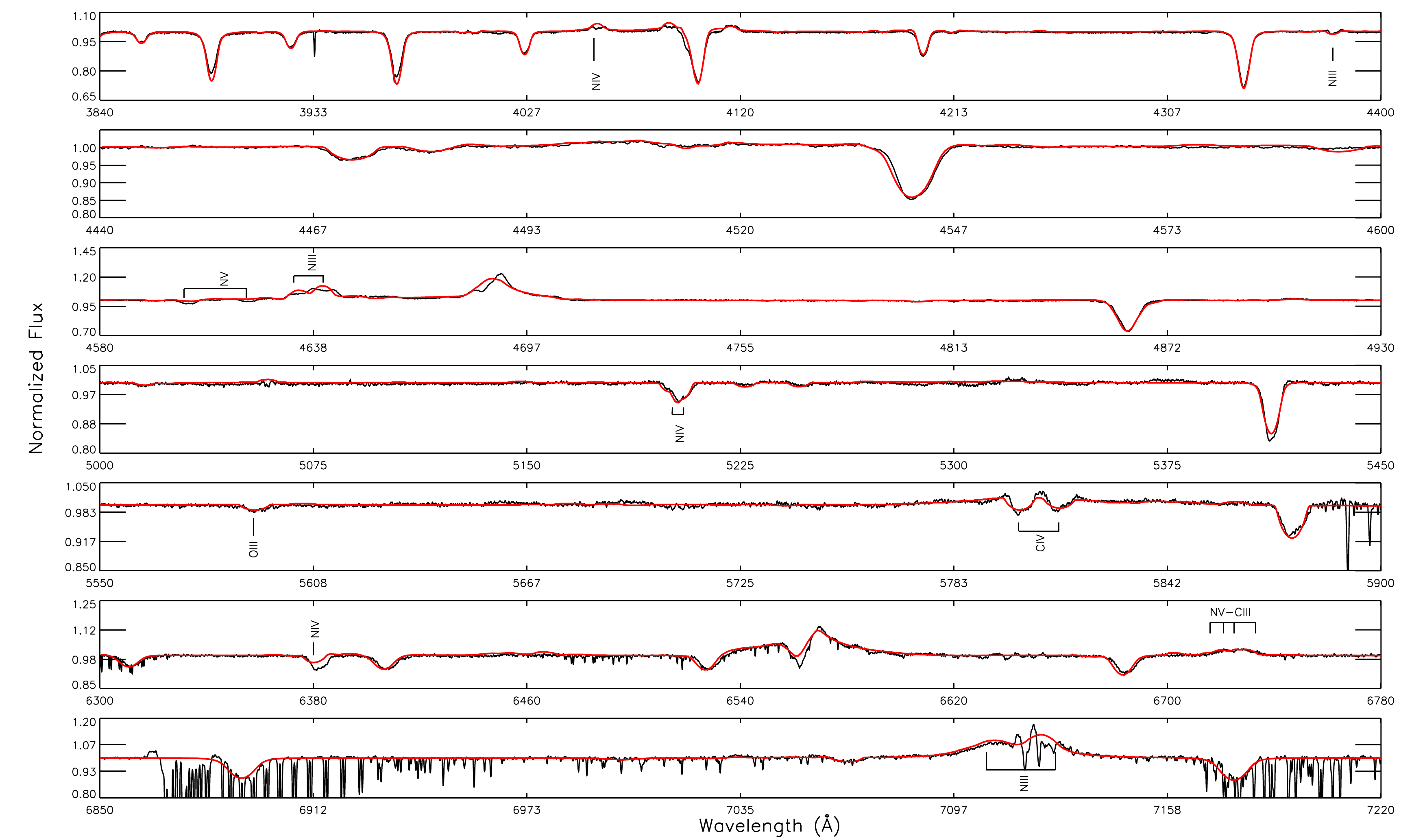}
      \caption[12cm]{Best-fit model for  \object{\hdsixtysix} (red line) compared to the \feros\ spectrum (black line)}
         \label{Fig_hd66}
   \end{figure*}


\begin{figure*}
\includegraphics[scale=0.51, angle=0]{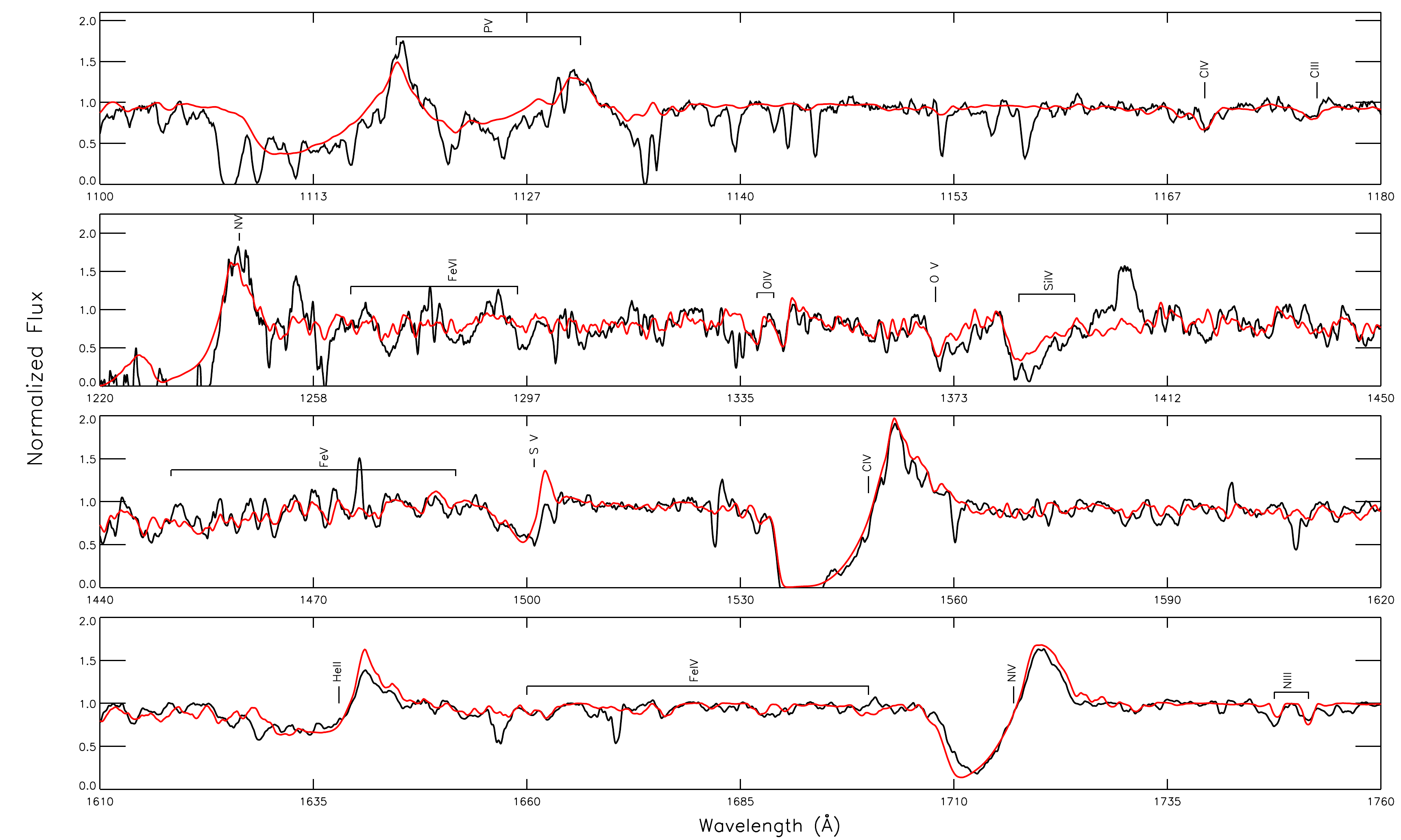}
      \caption[12cm]{Best-fit model for  \object{\hdsixteen} (red line) compared to \fuse\ and \iue\ spectra (black line)}
         \label{Fig_hd16}
   \end{figure*}
   
\begin{figure*}
\includegraphics[scale=0.51, angle=0]{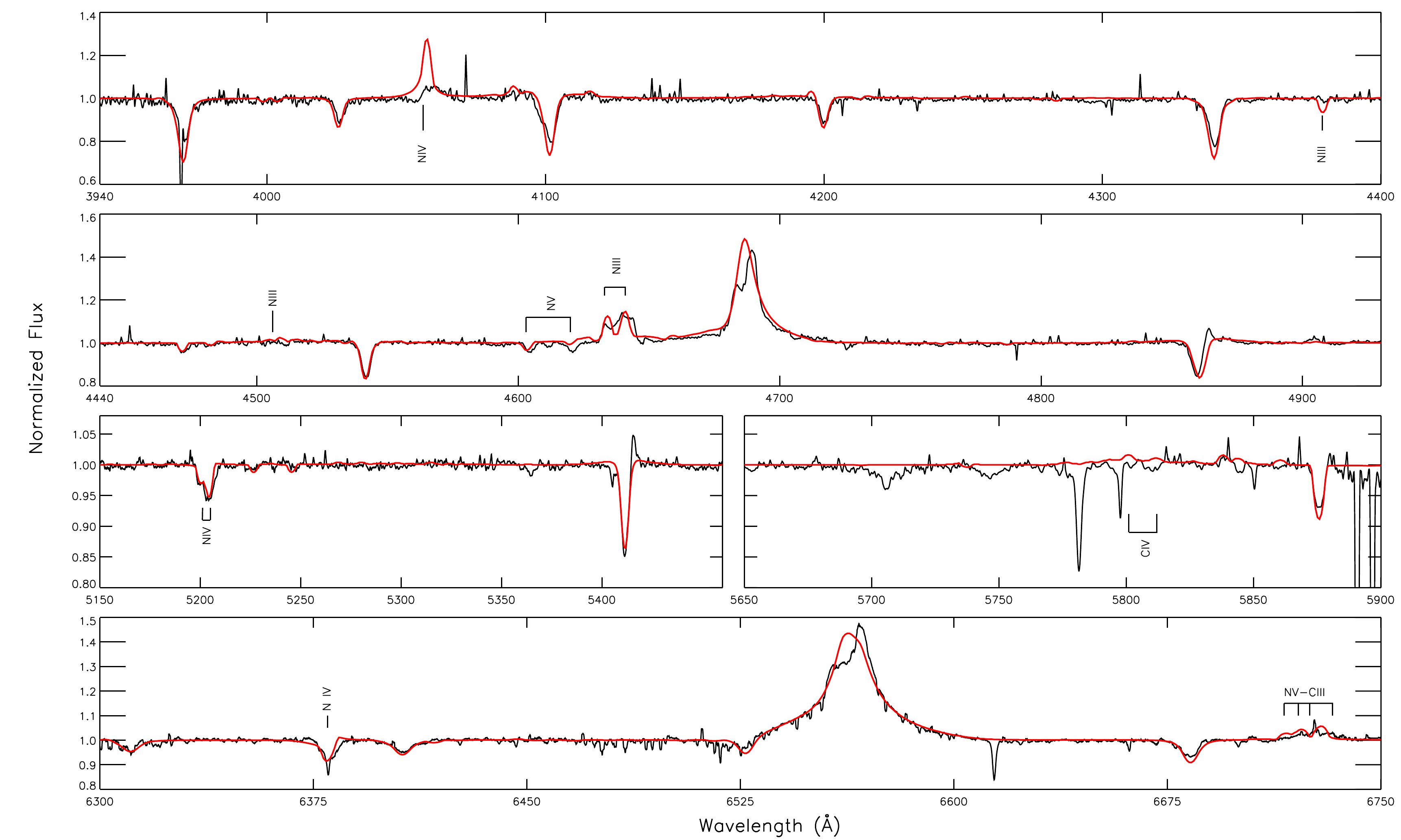}
      \caption[12cm]{Best-fit model for  \object{\hdsixteen} (red line) compared to the \elodie\ spectrum (black line)}
         \label{Fig_hd16}
   \end{figure*}   


\begin{figure*}
\includegraphics[scale=0.51, angle=0]{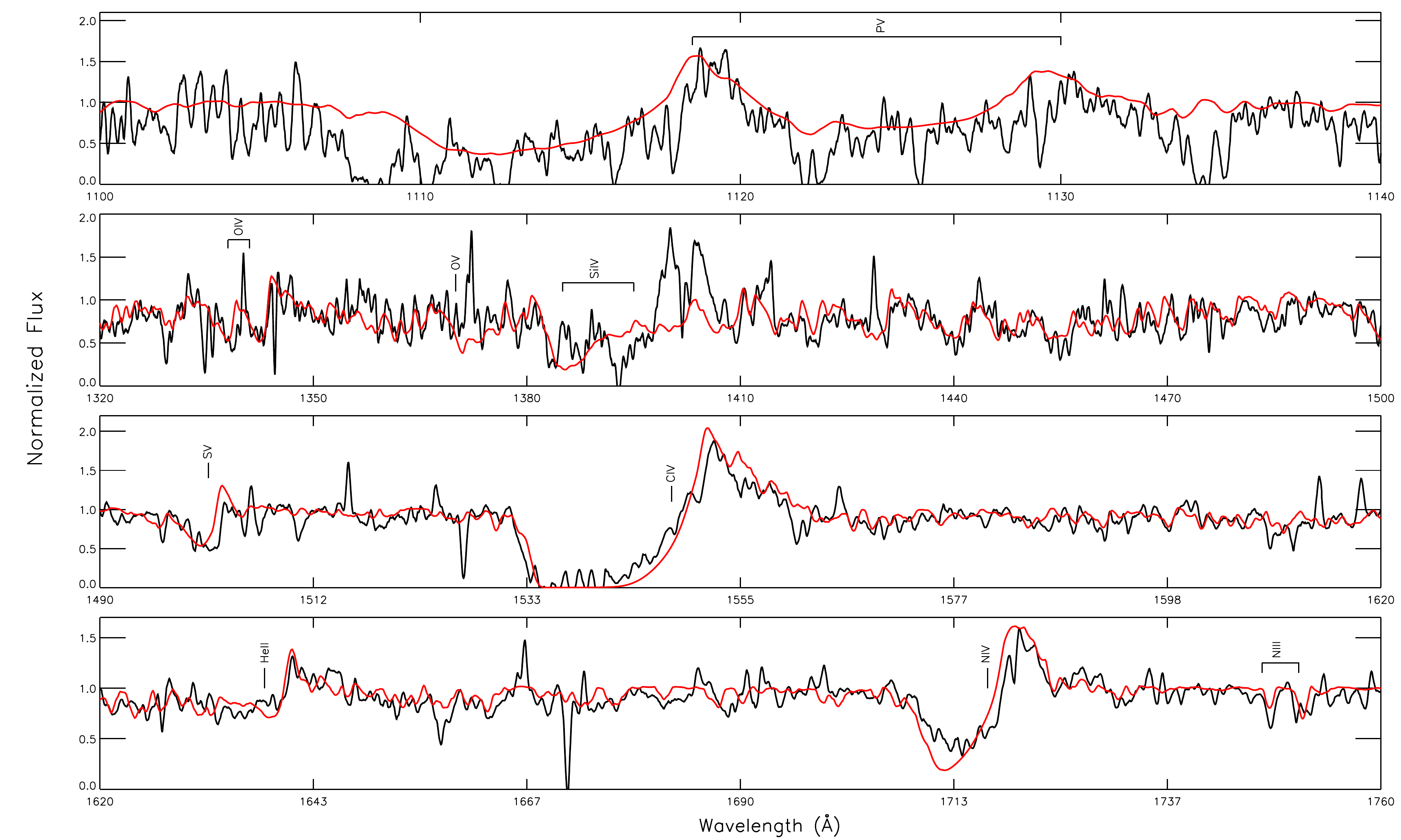}
      \caption[12cm]{Best-fit model for  \object{\hdfifteen} (red line) compared to \fuse\ and \iue\ spectra (black line)}
         \label{Fig_hd15}
   \end{figure*}
   
\begin{figure*}
\includegraphics[scale=0.51, angle=0]{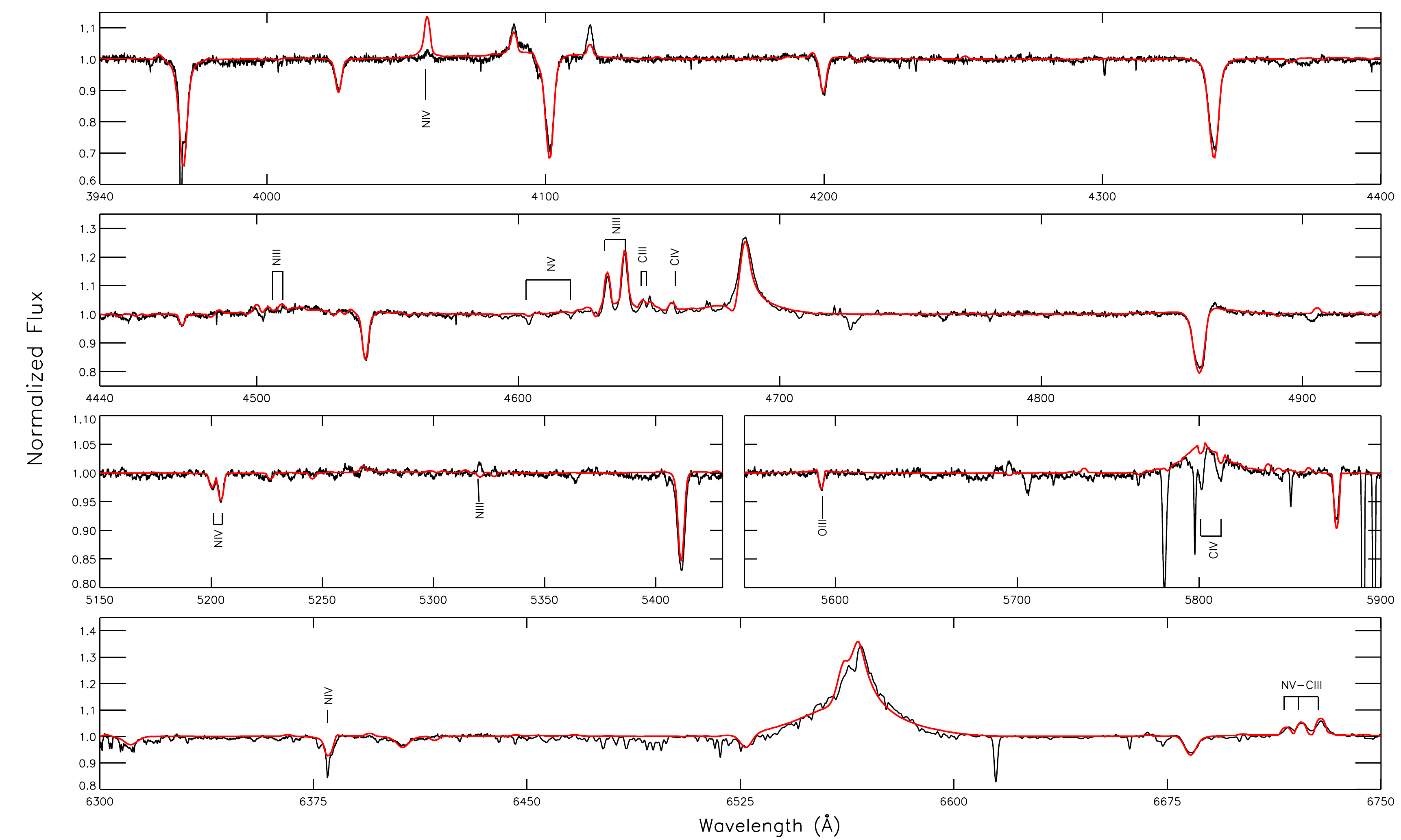}
      \caption[12cm]{Best-fit model for  \object{\hdfifteen} (red line) compared to the \elodie\ spectrum (black line)}
         \label{Fig_hd15}
   \end{figure*}
%
%

\begin{figure*}
\includegraphics[scale=0.51, angle=0]{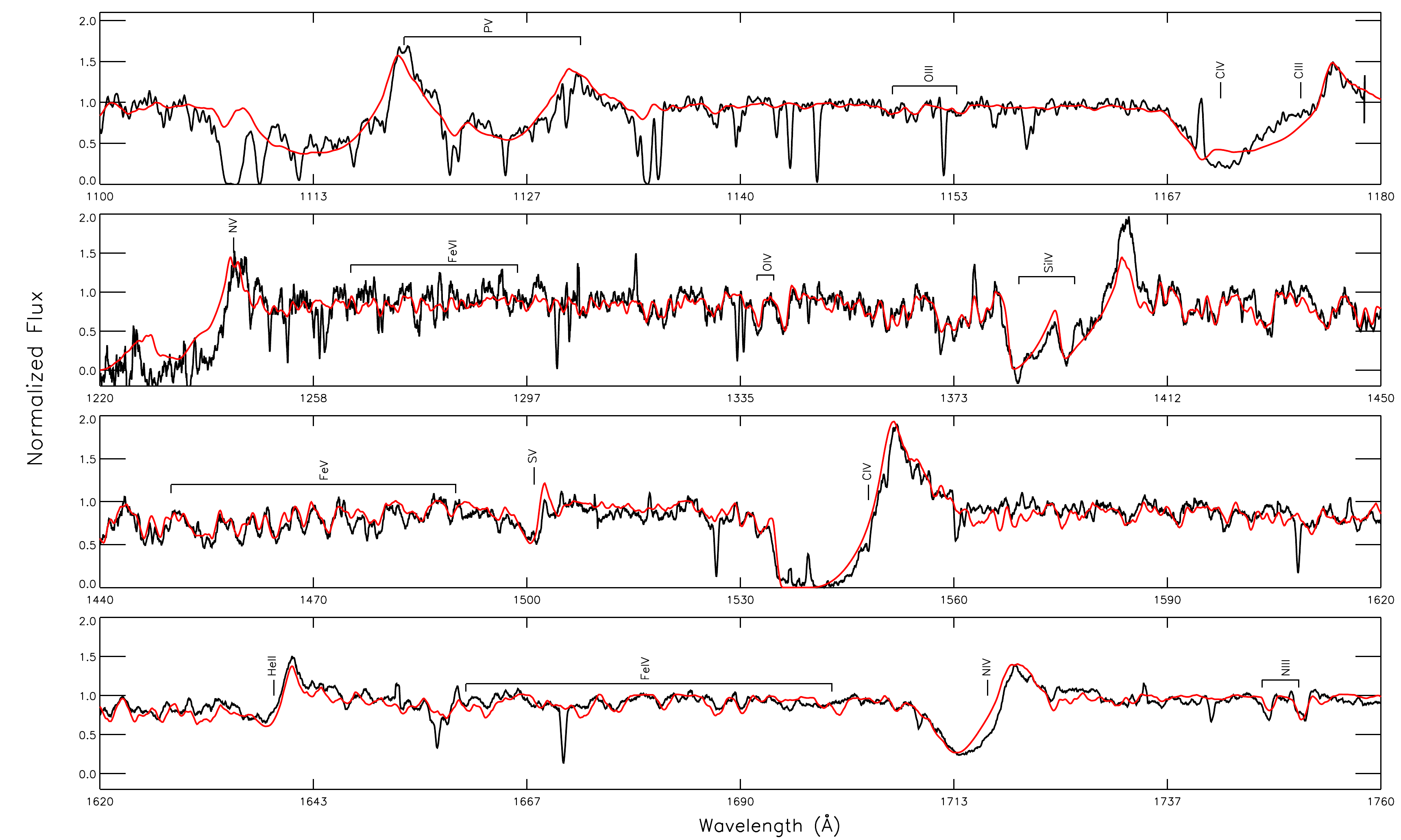}
      \caption[12cm]{Best-fit model for  \object{\hdfourteen} (red line) compared to \fuse\ and \iue\ spectra (black line)}
         \label{Fig_hd14}
   \end{figure*}
   
   \begin{figure*}
\includegraphics[scale=0.51, angle=0]{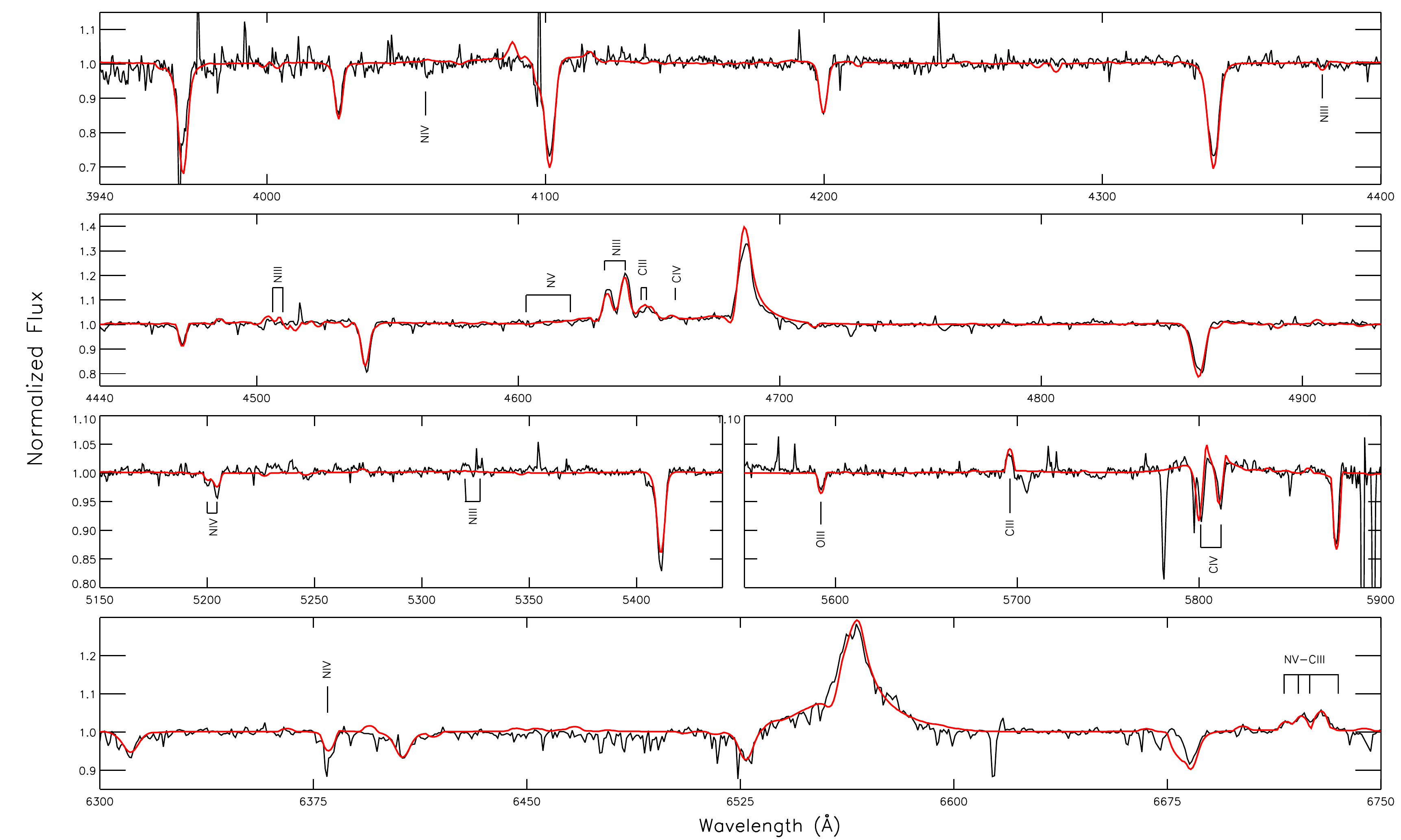}
      \caption[12cm]{Best-fit model for  \object{\hdfourteen} (red line) compared to the \elodie\ spectrum (black line)}
         \label{Fig_hd14}
   \end{figure*}



\begin{figure*}
\includegraphics[scale=0.51, angle=0]{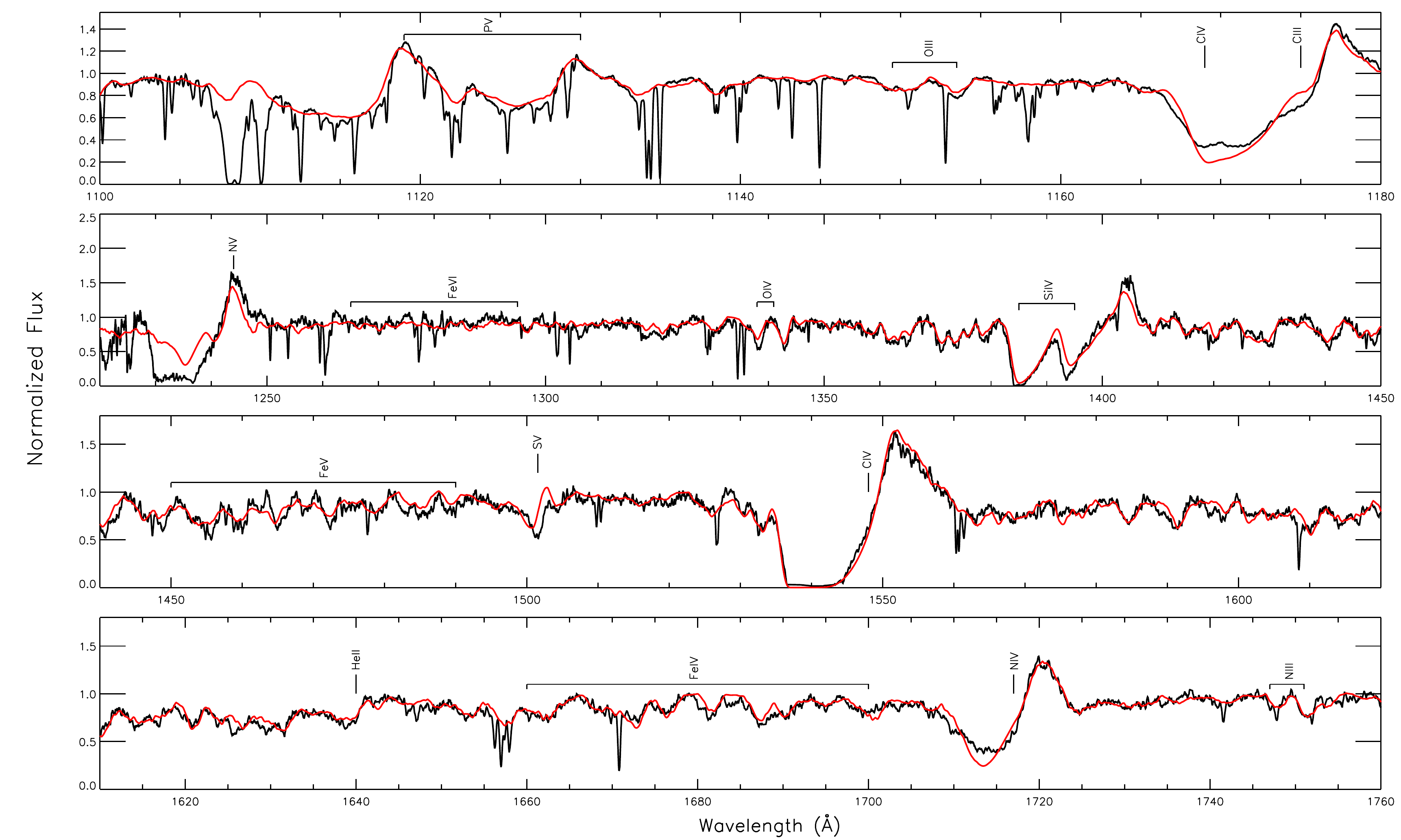}
      \caption[12cm]{Best-fit model for  \object{\hdtwohundten} (red line) compared to \fuse\ and \iue\ spectra (black line)}
         \label{Fig_hd210}
   \end{figure*}
   
\begin{figure*}
\includegraphics[scale=0.51, angle=0]{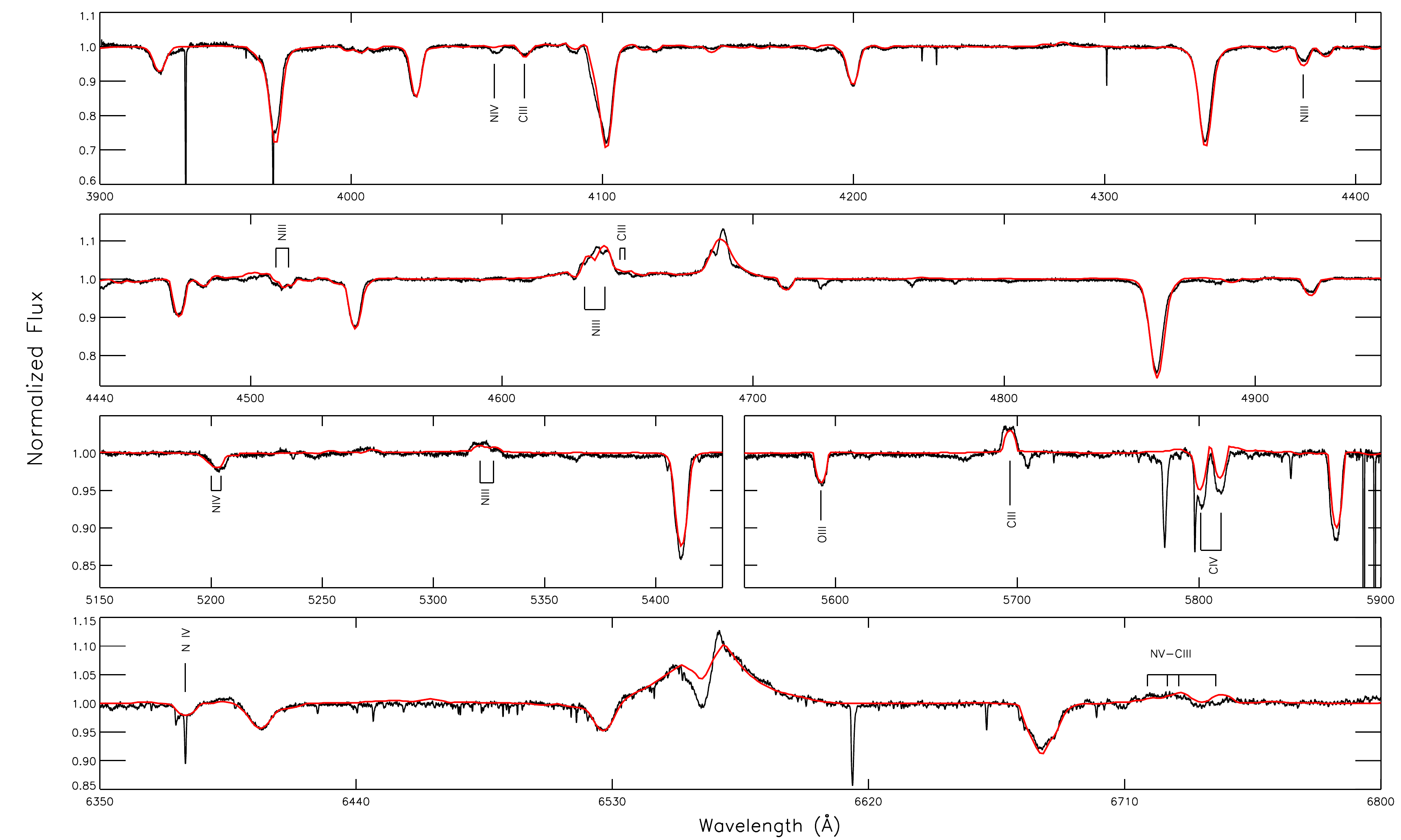}
      \caption[12cm]{Best-fit model for  \object{\hdtwohundten} (red line) compared to the \narval\ spectrum (black line)}
         \label{Fig_hd210}
   \end{figure*}



\begin{figure*}
\includegraphics[scale=0.51, angle=0]{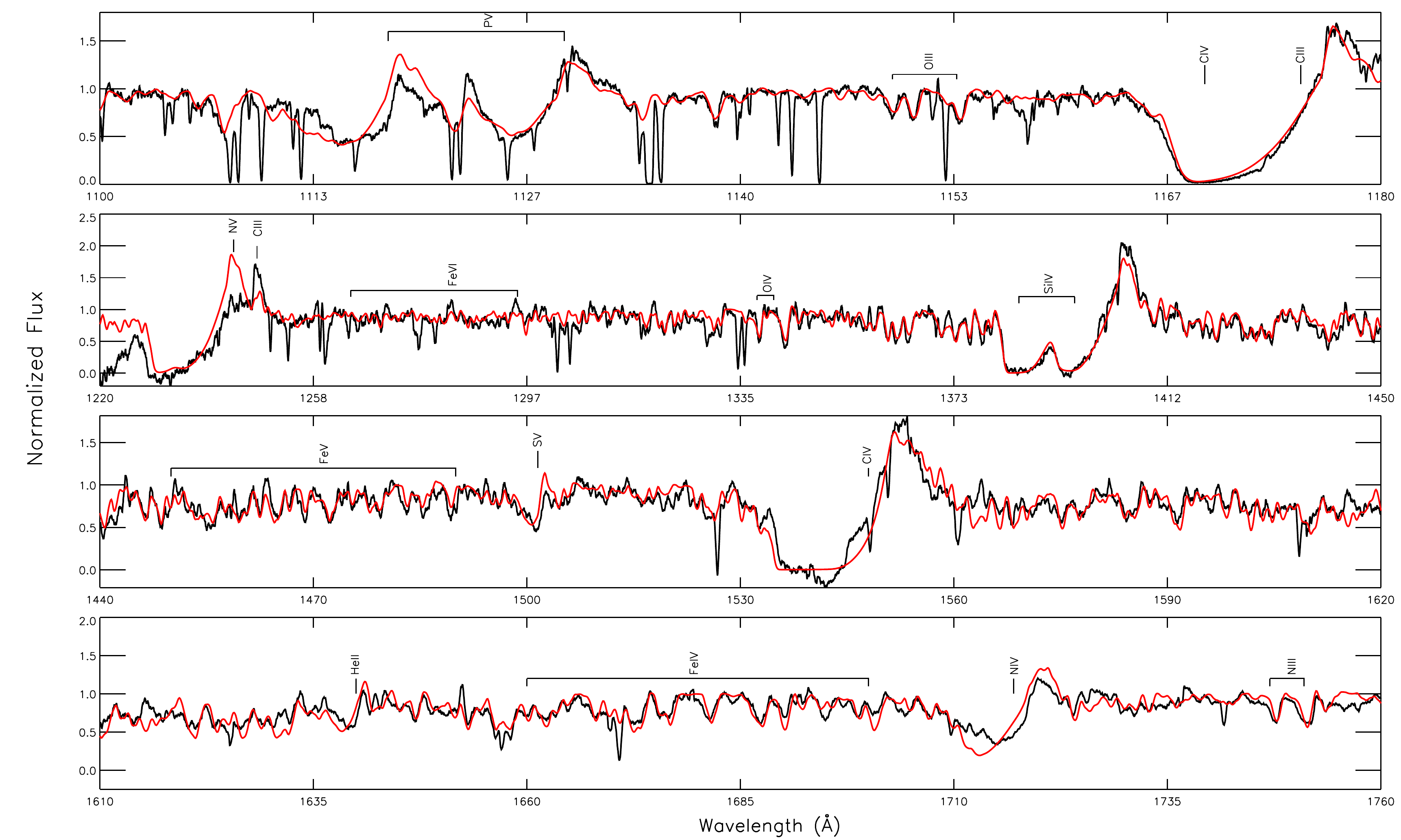}
      \caption[12cm]{Best-fit model for \object{ \hdonesixtythree} (red line) compared to \fuse\ and \iue\ spectra (black line)}
         \label{Fig_hd163}
   \end{figure*}
   
\begin{figure*}
\includegraphics[scale=0.51, angle=0]{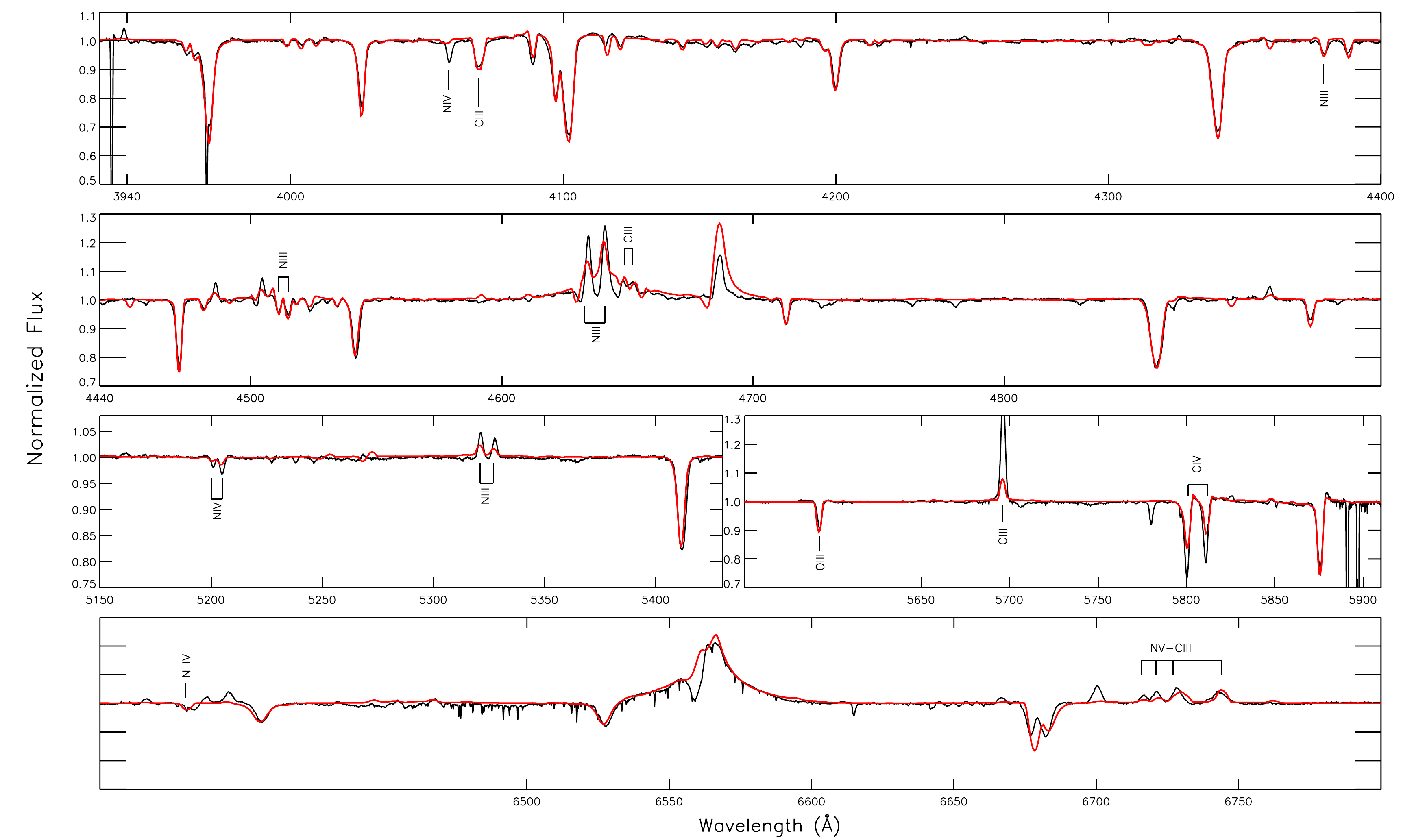}
      \caption[12cm]{Best-fit model for  \object{\hdonesixtythree} (red line) compared to the \uves\ spectrum (black line)}
         \label{Fig_hd163}
   \end{figure*}
%


\begin{figure*}
\includegraphics[scale=0.51, angle=0]{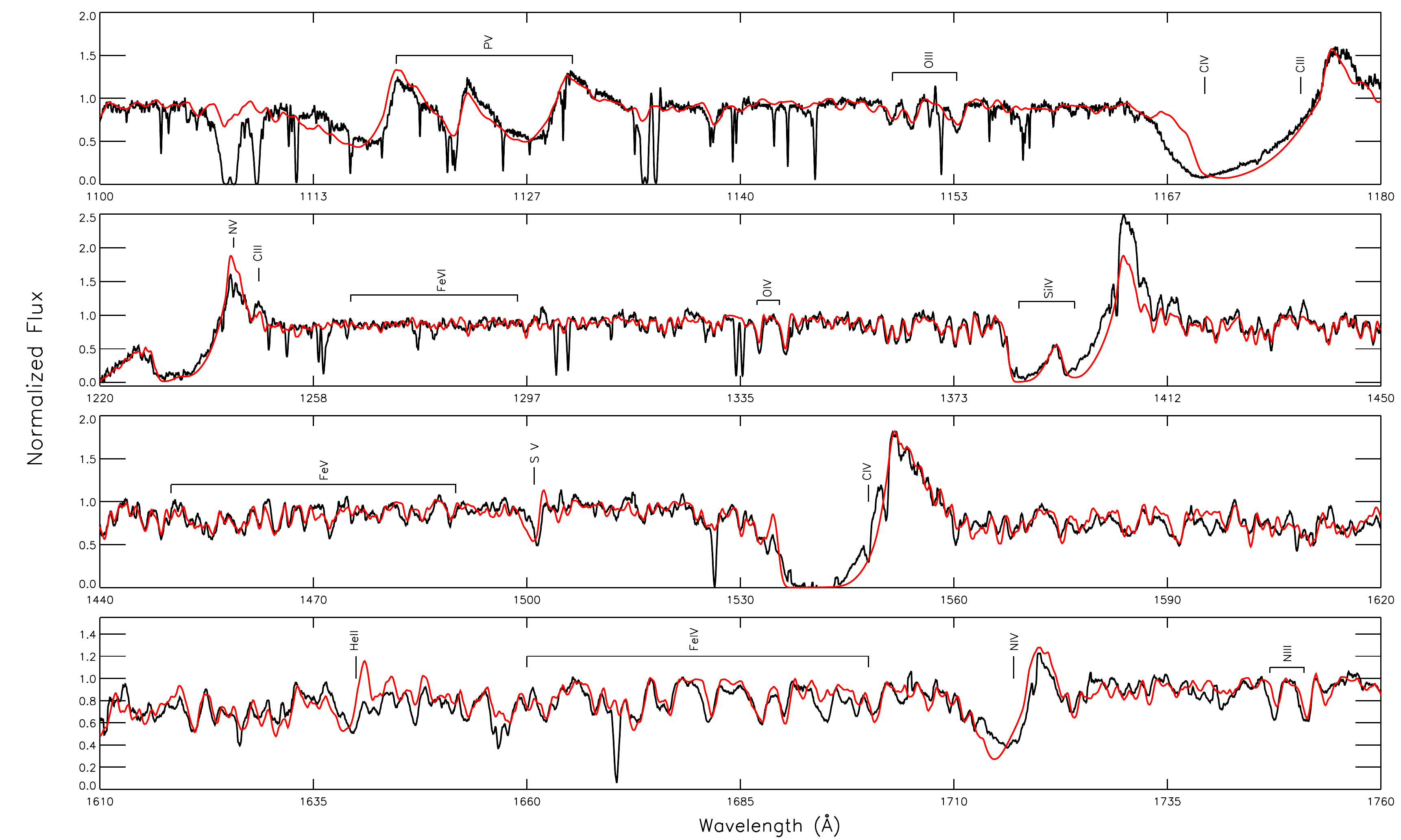}
      \caption[12cm]{Best-fit model for  \object{\hdoneninetwo} (red line) compared to \fuse\ and \iue\ spectra (black line)}
         \label{Fig_hd192}
   \end{figure*}
   
\begin{figure*}
\includegraphics[scale=0.51, angle=0]{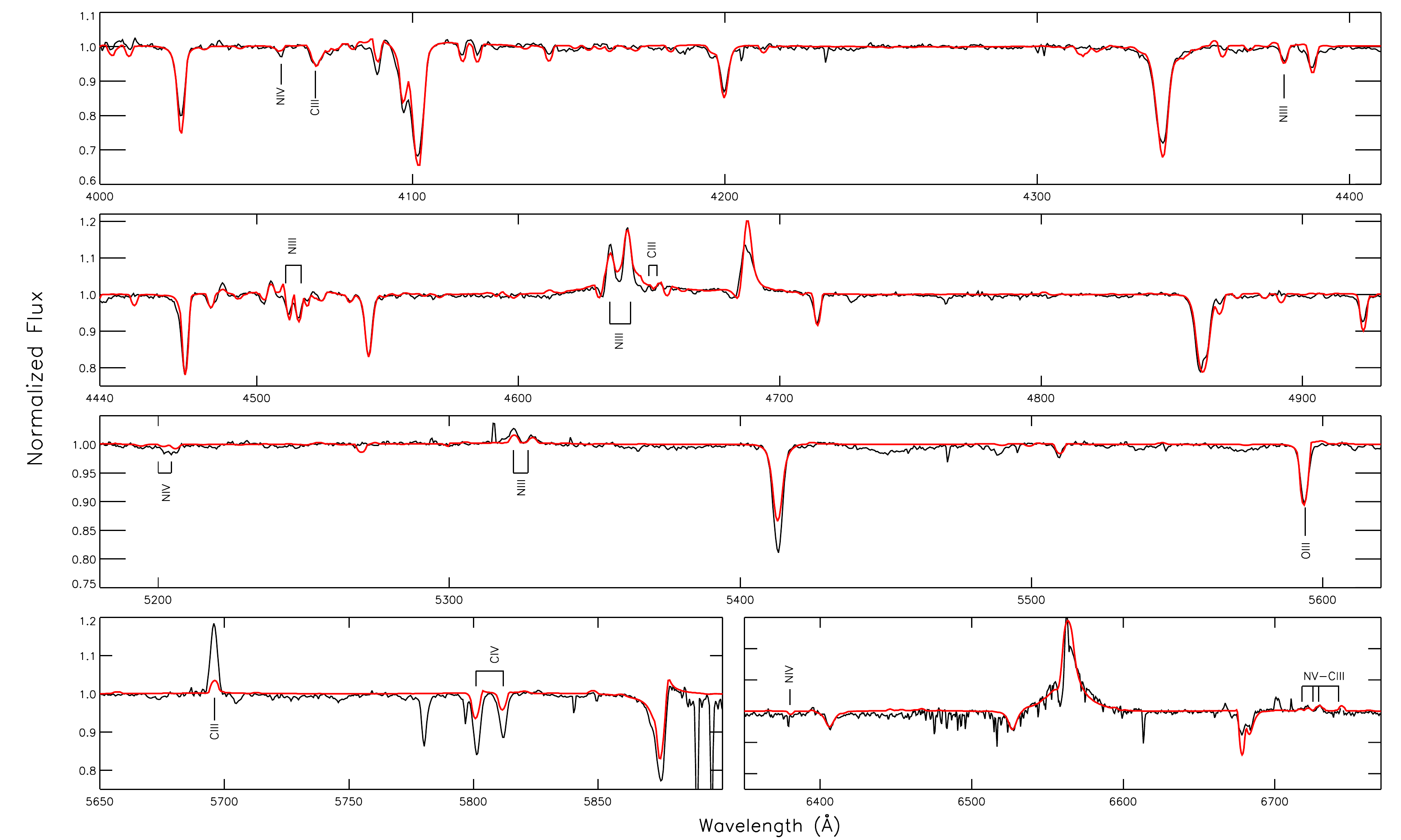}
      \caption[12cm]{Best-fit model for  \object{\hdoneninetwo} (red line) compared to the \elodie\ spectrum (black line)}
         \label{Fig_hd192}
   \end{figure*}

\end{appendix}


%
\end{document}